\documentclass[epj]{svjour}
\usepackage[british]{babel}
\usepackage{epsfig}
\usepackage{appendix}
\usepackage{subfig}
\usepackage[T1]{fontenc}

\usepackage[switch,columnwise]{lineno}

\usepackage[space]{cite}
\usepackage{amsmath}
\usepackage{multirow}

\newcommand\ba{\begin{eqnarray}}
\newcommand\ea{\end{eqnarray}}
\newcommand\nn{\nonumber}
\newcommand{\be}{\begin{equation}}
\newcommand{\ee}{\end{equation}}

\usepackage{xspace}
\newcommand{\PANDA}{$\overline{\textrm{P}}\textrm{ANDA}$\xspace}

\begin{document}
\raggedbottom

\hugehead
\title{Feasibility studies of time-like proton electromagnetic form factors at \PANDA at FAIR}
\author{{\large The \PANDA Collaboration} \\ \\
B.~Singh \inst{1}
\and W.~Erni \inst{2}
\and B.~Krusche \inst{2}
\and M.~Steinacher \inst{2}
\and N.~Walford \inst{2}
\and B.~Liu \inst{3}
\and H.~Liu \inst{3}
\and Z.~Liu \inst{3}
\and X.~Shen \inst{3}
\and C.~Wang \inst{3}
% ...
}
\institute{Aligarth Muslim University, Physics Department, Aligarth India
\and Universit\"at Basel, Basel Switzerland
\and Institute of High Energy Physics, Chinese Academy of Sciences, Beijing China
% ...
}

\date{Received: date / Revised version: date}

\abstract{Simulation results for future measurements of electromagnetic proton form factors at \PANDA (FAIR) within the PandaRoot software framework are reported. The statistical precision with which the proton form factors can be determined is estimated. The signal channel $\bar p p \to e^+ e^-$ is studied on the basis of two different but consistent procedures. The suppression of the main background channel, \textit{i.e.} $\bar p p \to \pi^+ \pi^-$, is studied. Furthermore, the background versus signal efficiency, statistical and systematical uncertainties on the extracted proton form factors are evaluated using two different procedures. The results are consistent with those of a previous simulation study using an older, simplified framework. However, a slightly better precision is achieved in the PandaRoot study in a large range of momentum transfer, assuming the nominal beam conditions and detector performance.
\PACS{
       {25.43.+t}{Antiproton-induced reactions} \and
       {13.40.Gp}{Electromagnetic form factors}
}
}

\maketitle

\section{Introduction}
\label{Sec:Introduction}

The \PANDA \cite{Lutz:2009ff} experiment at FAIR (Facility for Antiproton and Ion Research, at Darmstadt, Germany) will detect the products of the annihilation reactions induced by a high-intensity antiproton beam with momenta from 1.5 to 15 GeV/$c$. The comprehensive physics program includes charmonium spectroscopy, search for hybrids and glueballs, search for charm and strangeness in nuclei, baryon spectroscopy and hyperon physics, as well as nucleon structure studies \cite{Wiedner:2011mf}. Here we focus on the extraction of time-like (TL) proton electromagnetic form factors (FFs) through the measurement of the angular distribution of the produced electron (positron) in the annihilation of proton-antiproton into an electron-positron pair.

Electromagnetic FFs are fundamental quantities, which describe the intrinsic electric and magnetic distributions of hadrons. Assuming parity and time invariance, a hadron with spin $S$ is described by $2S + 1$ independent FFs. Protons and neutrons (spin 1/2 particles) are thus characterized by two FFs: the electric $G_E$ and the magnetic $G_M$. In the TL region, electromagnetic FFs have been associated with the time evolution of these distributions \cite{Kuraev:2011vq}.

Theoretically, the FFs enter in the parameterization of the proton electromagnetic current. They are experimentally accessible through measurements of differential and total cross sections for elastic $ep$ scattering in the space-like (SL) region and $\bar p p \leftrightarrow e^+e^-$ in the TL region. It is assumed that the interaction occurs through the exchange of one photon, which carries a momentum transfer squared $q^2$. In the TL region, this corresponds to the total energy squared $s$. 

Space-like FFs have been rigorously studied since the 1960's \cite{Hofstadter:1956qs}. However, the polarization transfer method \cite{Akhiezer:1968ek,Akhiezer:1974em} that was used for the first time in 1998 gave rise to new questions in the field. Recent access to high precision measurements over a large kinematic range further contributed to the new interest \cite{Perdrisat:2006hj}. Elastic $e^- p \rightarrow e^- p$ data from the JLab-GEp collaboration \cite{Jones:1999rz,Punjabi:2005wq,Gayou:2001qd,Puckett:2010ac}, covering a range of momentum transfer squared up to $Q^2=-q^2 \simeq $ 8.5~(GeV/$c$)$^2$, showed that the electric and magnetic distributions inside the proton are not the same. This is in contrast to what was previously reported: the ratio of the electric and the magnetic FF, $\mu_p G_{E}/G_{M}$ ($\mu_p$ is the proton magnetic moment) decreases almost linearly from unity as the momentum transfer squared increases, approaching zero.

In the TL region, the precision of the proton FF measurements has been limited by the achievable luminosity of the $e^+e^-$ colliders and $\bar p p$ annihilation experiments. Attempts have been made at LEAR \cite{Bardin:1994am}, BABAR \cite{Lees:2013uta} and more recently at BESIII \cite{Ablikim:2015vga}. The obtained FF ratios show a different tendency, being somehow inconsistent in the limit of combined systematical and statistical uncertainties which definitely calls for more precise experiments. The results of LEAR and BABAR disagree with each other with a significance up to 3$\sigma$, while the BESIII measurements have large total uncertainties.

The \PANDA experiment, designed with an average peak luminosity of ${\cal L}=2\cdot10^{32}$ cm$^{-2}$s$^{-1}$ in the so-called \textit{high luminosity mode} (with ${\cal L}\sim10^{31}$ cm$^{-2}$s$^{-1}$ available at the beginning of operation), will bring new information in two respects: the precision measurement of the angular distribution for the individual determination of FFs, and the measurement of the integrated cross section for the extraction of a generalized FF up to larger values of $s$. These data are expected to set a stringent test of nucleon models. In particular, the high $s$-region brings information on analyticity properties of FFs and on the asymptotic $q^2$ behavior predicted by perturbative Quantum ChromoDynamics (pQCD) \cite{Kivel:2012zz}.

The FAIR facility and the \PANDA experiment are under construction in Darmstadt (Germany). Simulations of the different physics processes have been performed or are in progress. The feasibility of the FFs measurement with the \PANDA detector, as suggested in Ref.~\cite{TomasiGustafsson:2008gq}, has been investigated in Ref.~\cite{Sudol:2009vc}. Since the latter paper was published, much progress has been made in the development of a new simulation framework (PandaRoot, see Ref.~\cite{SpataroPandaRoot}) with a much more realistic detector model and more elaborated reconstruction algorithms. This is the motivation for reinvestigating this channel. In addition, a lot of progress has been made recently regarding the design of the detector. Prototypes of sub-detectors have been built. A series of performance tests has been carried out, which provided data for improvements in the design. The Technical Design Reports of most of the detectors are available. Although the development of the simulation and analysis software is still ongoing in parallel with the detector construction, a realistic description of the sub-detectors and new algorithms for the tracking, digitization, and particle identification (PID) has been implemented. A realistic magnetic field map, calculated with TOSCA software \cite{tosca}, is part of the simulation.

In the PandaRoot version\footnote{version 25544} used for this work, the description of most of the sub-detectors has been complemented with the passive materials, beam pipe, magnet yoke, etc. Moreover, during recent years GEANT4 \cite{Agostinelli:2002hh} has undergone continuous improvements, concerning in particular the shape of the electromagnetic shower.

The aim of this paper is to present new simulation results, based on a recent PandaRoot version, in order to check the validity of the previously made assumptions and to confirm the conclusions regarding the feasibility of the $e^+e^-$ detection at a sufficient level of precision. In addition, new and efficient analysis tools have been developed for the extraction of the physical information, which will be applied also to the treatment of the experimental data.
 
The paper is organized as follows. The kinematics of the reactions of interest (signal and background) and the evaluation of the counting rates are described in Section \ref{basics}. The detector is briefly described in Section \ref{experiment}. The standard chain of the full simulation with PandaRoot and the procedure to identify and analyze the signal and the background channels are described in Section \ref{experiment}. In Section \ref{sec:analysis} we present the results in terms of the proton FF ratio $\mbox{R}={|G_{E}|}/{|G_{M}|}$, individual FFs $|G_E|$ and $|G_M|$, the angular asymmetry ${\cal{A}}$ and the effective FF. In Section \ref{sec:systematic} several sources of systematic uncertainties are discussed. Finally, in Section \ref{sec:competitiveness} the competitiveness of the \PANDA with respect to existing and planned experiments is discussed. The Conclusions section contains a summary and final remarks.

\section{Basic formalism}
\label{basics}
Let us consider the reactions:
\be
\bar p(p_1)+ p (p_2) \to \ell^-(k_1)+ \ell^+(k_2),~ \ell=e, \mu, \pi,
\label{eq:eqreac}
\ee
where the four-momenta of the particles are written within parentheses. In the center-of-mass (c.m.) system, the four-momenta are:
\ba
p_{1}&=&(E, \vec p),~p_{2}=(E, -\vec p),\nn\\
k_1&=&(E, \vec k), k_2=(E, - \vec k),~ \vec p\cdot \vec k= p k \cos\theta,
\ea
$\theta$ is the angle between the negative emitted particle and the antiproton beam.

The cylindrical symmetry around the beam axis of the unpolarized binary reaction enforces an isotropic distribution in the azimuthal angle $\phi$. These reactions are two-body final state processes. The final state particles are emitted back to back in the c.m. system, and each of them, having equal mass, carries half of the total energy of the system, $E=\sqrt{s}/2$, where the invariant $s$ is $s=q^2=(p_{1}+p_2)^2=(k_1 + k_2)^2$.

All leptons in the final state ($e$, $\mu$, $\tau$) contain the same information on the electromagnetic hadron structure. However, the experimental requirements for their detection are peculiar for each particle species. In this work we focus on the electron-positron pair production, denoted the \textit{signal reaction}, and on the charged pion pair production, denoted the \textit{background reaction}. The cross section of hadron production is expected to be much larger than that of leptons: for charged pions it is $\sim10^6$ times larger than for $e^+e^-$ production \cite{Zichichi:1962ni,Eisenhandler:1975kx,VandeWiele:2010kz}. The signal and the background reactions have very similar kinematics because the mass of the electron is sufficiently close to the pion mass in the energy scale of the \PANDA experiment (for the antiproton beam momentum in the laboratory system $p_{lab}=3.3$ GeV/$c$ the range of the electron laboratory momentum is 0.58-3.82 GeV/$c$, while for the pions it is 0.92-3.5 GeV/$c$). Therefore, the kinematics plays a minor role in the electron (positron)/pion separation. The discrimination between electrons and pions requires high performance PID detectors and precise momentum measurement. For example, the information from the electromagnetic shower induced by different charged particles in an electromagnetic calorimeter plays an important role for electron identification. The kinematic selection suppresses contributions from hadronic channels with more than two particles in the final states, as well as events with secondary particles originating from an interaction of primary particles with the detector material. A kinematic selection is also very efficient in suppressing neutral pions, as discussed in Refs. \cite{Sudol:2009vc,MoraEspi12}. Note that the cross section of neutral pion pair production, $\pi^0\pi^0$, is ten times smaller than that of $\pi^+\pi^-$.

\subsection{The signal reaction}
\begin{figure}
  \resizebox{0.5\textwidth}{!}{ \includegraphics{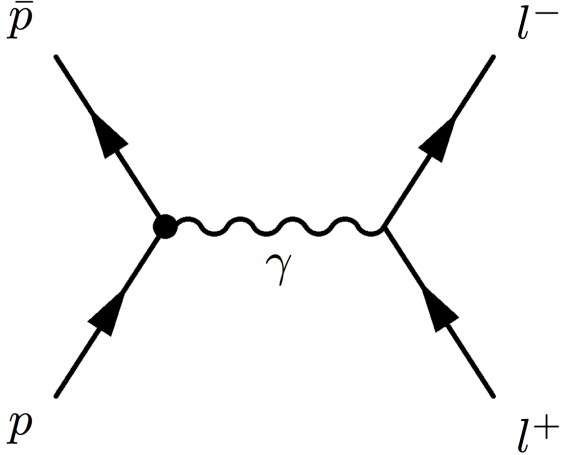} }
  \caption{Tree-level contributing diagram to $\bar p p \rightarrow l^+ l^- $.}
  \label{fig:one}
\end{figure}

The expression of the hadron electromagnetic current for the $\bar p p $ annihilation into two leptons is derived assuming one-photon exchange. The diagram which contributes to the tree-level amplitude is shown in Fig.~\ref{fig:one}. The internal structure of the hadrons is then parameterized in terms of two FFs, which are complex functions of $q^2$, the four momentum squared of the virtual photon. For the case of unpolarized particles the differential cross section has the form \cite{Zichichi:1962ni}: 
\ba
\label{eq:eq1}
\frac{d\sigma}{d\cos\theta } &= & \frac{\pi\alpha^2}{2\beta s} \left [(1+\cos^2\theta)|G_M|^2+\displaystyle\frac{1}{\tau}\sin^2\theta|G_E|^2 \right ],
\label{eq:eqdsigma}
\ea
where $\beta=\sqrt{1-1/\tau}$, $\tau=s/(4m^2)$, $\alpha$ is the electromagnetic fine-structure constant, and $m$ is the proton mass. This formula can also be written in equivalent form as \cite{TomasiGustafsson:2001za}: 
\be
\displaystyle\frac{d\sigma}{d\cos\theta}=\sigma_0\left [ 1+{\cal A} \cos^2\theta \right ],
\label{eq:eqsth}
\ee
where $\sigma_0$ is the value of the differential cross section at 
$\theta=\pi/2$ and ${\cal A}$ is the angular asymmetry which lies in the range $-1\le {\cal A}\le 1$, and can be written as a function of the FF ratio as:
\ba
\sigma_0&=&\frac{\pi\alpha^2}{2\beta s} \left (|G_M|^2+ \frac{1}{\tau}|G_E|^2\right )
\nn\\
{\cal A}&=&\displaystyle\frac{\tau|G_M|^2-|G_E|^2}{\tau|G_M|^2+|G_E|^2}=
\displaystyle\frac{\tau-\mbox{R}^2}{\tau+\mbox{R}^2},
\label{eq:eqsa}
\ea
where $\mbox{R}=|G_E|/|G_M|$.

The fit function defined in Eq.~(\ref {eq:eqsth}) can be reduced to a linear function (instead of a quadratic one) where $\sigma_0$ and ${\cal A}$ are the parameters to be extracted from the experimental angular distribution.  In the case of $\mbox{R}=0$, the minimization procedure based on MINUIT has problems converging, while the asymmetry $\cal{A}$ varies smoothly in the considered $q^2$ interval. Therefore, it is expected to reduce instabilities and correlations in the fitting procedure.
The angular range where the measurement can be performed is usually restricted to $|\cos\theta|\le\bar c$, with $\bar c=\cos\theta_{max}$.

The integrated cross section, $\sigma_{int}$, is:
\ba
\sigma_{int}&=&\int_{-\bar c}^{\bar c}\displaystyle\frac{d\sigma}{d\cos\theta}d\cos\theta=2\sigma_0\,\bar c\, \left(1+\frac{\cal A}{3}\,{\bar c}^2\right) \label{eq:eqsint}\\
&=&
\displaystyle\frac{\pi\alpha^2}{\beta s}\bar c\left [ \left(1+\displaystyle\frac {\bar c^2}{3}\right )|G_M|^2+ \displaystyle\frac{1}{\tau}\left(1-\displaystyle\frac{\bar c^2}{3}\right )|G_E|^2\right ].
\nn
\ea 
The total cross section, $\sigma_{tot}$, corresponds to $\bar c=1$:
\ba
\sigma_{tot}&=&2\sigma_0\left (1+\frac{\cal A}{3}\right)=\frac{2\pi\alpha^2}{3\beta s}\left [2 |G_M|^2 +\frac {|G_E|^2}{\tau}\right ]\\
&=&
\frac{2\pi\alpha^2|G_M|^2}{3\beta s}\left [2 +\frac {\mbox{R}^2}{\tau}\right ].
\nn
\label{eq:stot}
\ea
Knowing the total cross section, one can define an effective FF as:
\be
|F_p|^2=\displaystyle\frac{3\beta s \sigma_{tot}}
{2\pi\alpha^2 \left(2+\displaystyle\frac{1}{\tau}\right)},
\label{eq:Fp}
\ee
or from the integrated cross section, as:
\be
|F_p|^2=\displaystyle\frac{\beta s}{\pi\alpha^2}
 \displaystyle\frac{\sigma_{int}}{
\bar c\left [\left (1+ \displaystyle\frac{\bar c^2}{3}\right ) +
\displaystyle\frac{1}{\tau}\left (1- \displaystyle\frac{\bar c^2}{3}\right ) \right ]
},
\label{eq:Fp1}
\ee
which is equivalent to the value extracted from cross section measurements, assuming $|G_E|=|G_M|$.

\begin{figure*}[t]
\begin{center}
\subfloat[\label{fig:Data_a}]{\resizebox{0.45\textwidth}{!}{ \includegraphics{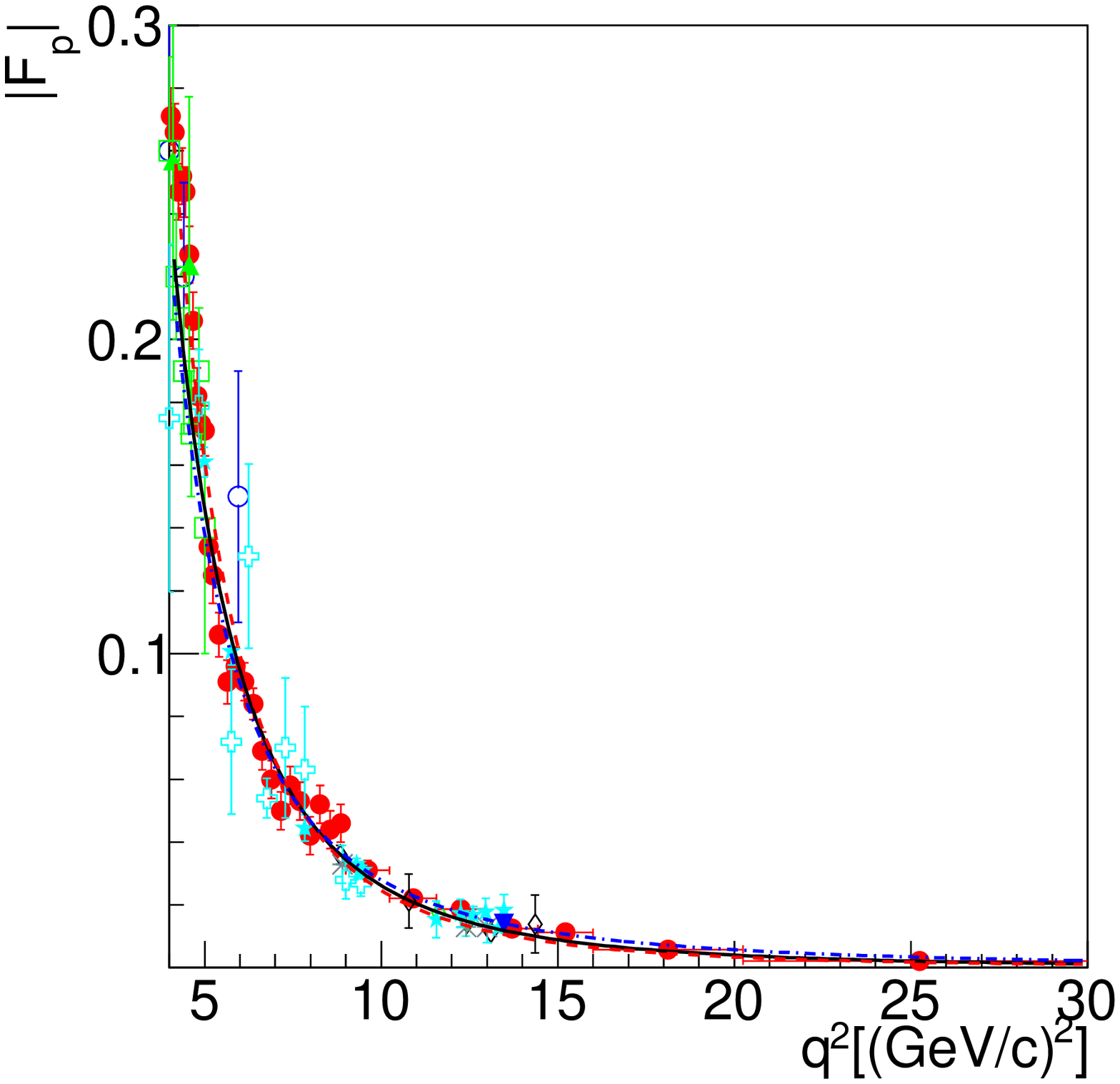} } }
\subfloat[\label{fig:Data_b}]{\resizebox{0.45\textwidth}{!}{ \includegraphics{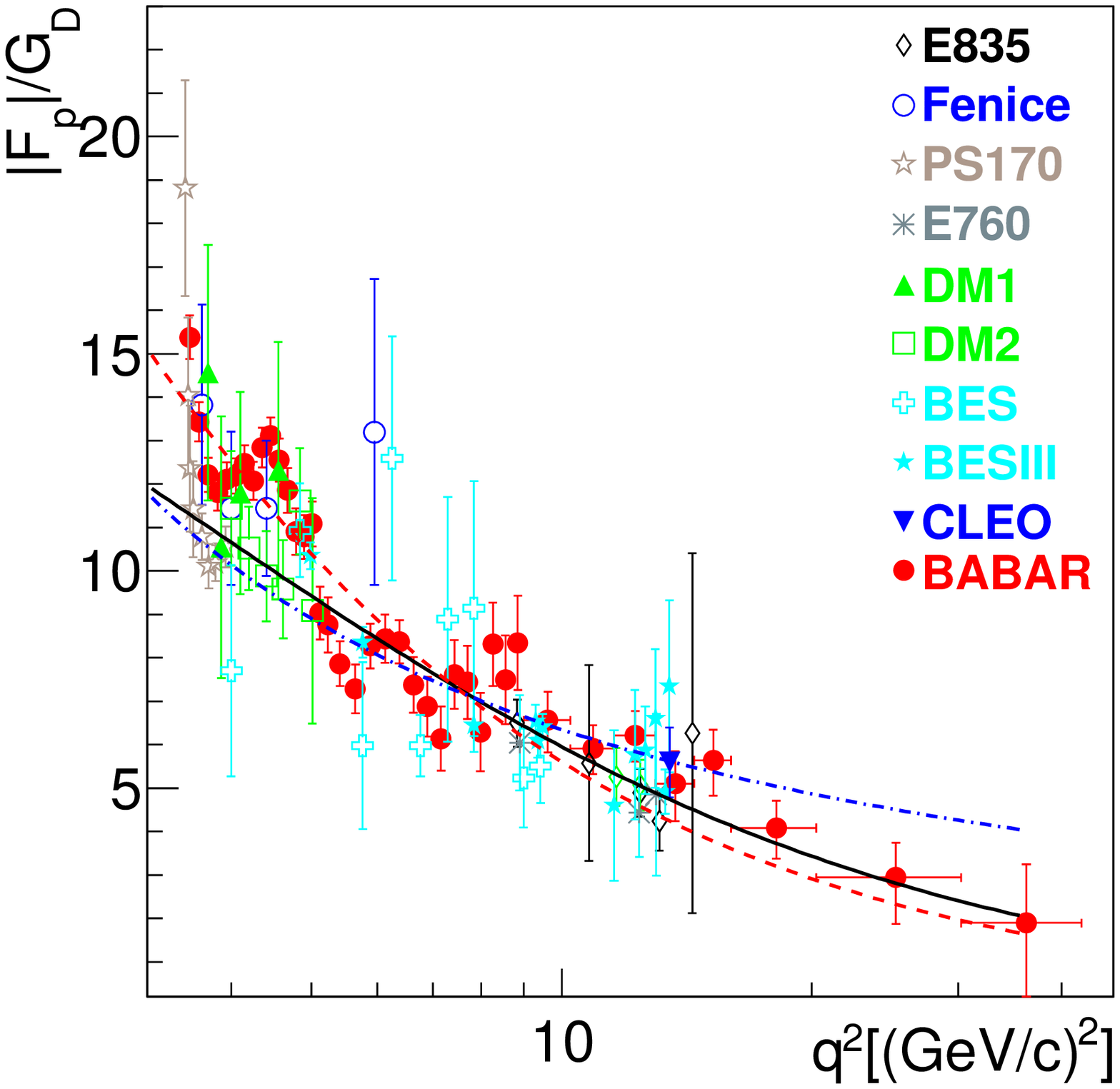} } }
\caption{$q^2$ dependence of the world data for $\bar p p\to e^+ e^-$ and $e^+ e^-\to \bar p p$. The effective proton TL FF, $|F_P|$, is extracted from the annihilation cross sections assuming $|G_E|=|G_M|$: E835 \protect\cite{Andreotti:2003bt,Ambrogiani:1999bh}, Fenice \protect\cite{Antonelli:1998fv}, PS170 \protect\cite{Bardin:1994am}, E760 \protect\cite{Armstrong:1992wq}, DM1 \protect\cite{Delcourt:1979ed}, DM2 \protect\cite{Bisello:1983at,Bisello:1990rf}, BES \protect\cite{Ablikim:2005nn}, BESIII \protect\cite{Ablikim:2015vga}, CLEO \protect\cite{Pedlar:2005sj}, BABAR \protect\cite{Lees:2013rqd,Lees:2013ebn}. Different parameterizations are shown based on Eq.~(\protect\ref{eq:eqff}) (solid black and dashed red lines) and from Eq.~(\protect\ref{eq:eqqcdbis}) (dash-dotted blue line), as described in the text.}
\label{fig:Data}
\end{center}
\end{figure*}

Literature offers several parameterizations of the proton FFs \cite{Belushkin:2006qa,Denig:2012by,Pacetti:2015iqa,Haidenbauer:2014kja,Lorenz:2015pba,Bianconi:2015vva}. The world data are illustrated in Fig.~\ref{fig:Data}. In Ref.~\cite{Sudol:2009vc} two parameterizations were considered. Cross section parameters are extracted from experimental data of the integrated cross section. BABAR data \cite{Lees:2013rqd,Lees:2013ebn} suggest a steeper decrease with $s$, and show a strong energy dependence near threshold \cite{Haidenbauer:2006dm}.

The Quantum ChromoDynamics (QCD) inspired parameterization of $|G_{E,M}|$ is based on an analytical extension of the dipole formula from the SL to the TL region and corrected to avoid 'ghost' poles in $\alpha_s$ (the strong interaction running constant) \cite{Shirkov:1997wi}:
\be
|G_{E,M}^{QCD}|=\frac{A^{QCD}}{q^4\left [\log^2(q^2/\Lambda_{QCD}^2)+\pi^2 \right ]}.
\label{eq:eqqcdbis}
\ee
The parameter $A^{QCD}=89.45~ (\mbox{GeV/$c$})^4$ is obtained from a fit to the experimental data, and $\Lambda_{QCD}=0.3$ (GeV/$c$) is the QCD cut-off parameter. It is shown in Fig.~\ref{fig:Data} as a dash-dotted blue line.

The existing data on the TL effective FF are well reproduced by the function proposed in Ref.~\cite{TomasiGustafsson:2001za}:
\ba
|G_{E,M}|&=&\frac{A}
{1+q^2/m_a^2} G_D,\nn\\
G_D&=&(1+q^2/q_0^2)^{-2},
\label{eq:eqff}
\ea
where the numerator $A$ is a constant extracted from the fit to the TL data. It is illustrated by a solid black line with the nominal parameters $A=22.5$, $m_a^2=3.6$ (GeV/$c$)$^2$, and $q_0^2=0.71$ (GeV/$c$)$^2$. Note that an updated global fit with a data set including 85 points (starting from $s=4$ GeV$^2$) gives $A($fit$)=71.5$  and $m_a^2($fit$)=0.85$ (GeV/$c$)$^2$, with a value of $\chi^2/\mbox{NDF}=1.4$ (dashed red line), overestimating the low energy data. These parameterizations reproduce reasonably well the data in the considered kinematic region. In our calculations, we chose the parameterization from Eq.~(\ref{eq:eqff}) with nominal parameters.

The expected count rates for an ideal detector are reported in Table~\ref{table:kincount}, assuming $\mbox{R}=|G_E|/|G_M|=1$ and using the parameterization from Eq.~(\ref{eq:eqff}), the angular range $|\cos\theta|\le 0.8$ and Eq.~(\ref{eq:eqdsigma}). Due to the \PANDA detector acceptance, the electron identification efficiency becomes very low above $|\cos\theta| = 0.8$ (see Section \ref{sec:analysis}). For each reported kinematic point $N_{int}(e^+e^-)$ an integrated luminosity of 2 fb$^{-1}$ is assumed. This corresponds to four months of measurement with 100\% efficiency at the maximum luminosity of ${\cal L}= 2\cdot 10^{32}$ cm$^{-2}${s$^{-1}$}. In the table, we also list the cross sections and expected number of counts $N_{int}(\pi^+\pi^-)$ of the dominant background channel, \textit{i.e.} $\bar p p \to \pi^+ \pi^-$. 
\begin{table*}
	\begin{center}
	\begin{tabular}{ l l l l l l l }
	\hline
	$p_{lab}$ & $s$ & $\sigma_{int} (e^+ e^-)$ & $N_{int}(e^+e^-)$ & $\sigma_{int} (\pi^+ \pi^-)$ & $N_{int}(\pi^+\pi^-)$ & $\displaystyle\frac{\sigma_{int}(\pi^+ \pi^-)}{\sigma_{int}(e^+ e^-)}\times 10^{-6}$\\
 	\ [GeV/$c$] & [GeV$^2$] & [pb] & & [$\mu$b] & &  \\
	\hline
	1.70 & 5.40        & 415                 & 830$\cdot 10^{3}$ & 101  & 202$\cdot 10^{9}$ & 0.24 \\
	2.78 & 7.27        & 55.6                & 111$\cdot 10^{3}$ & 13.1 & 262$\cdot 10^{8}$ & 0.24 \\
	3.30 & 8.21        & 24.8                & 496$\cdot 10^{2}$ & 2.96 & 592$\cdot 10^{7}$ & 0.12 \\
	4.90 & 11.12       & 3.25                & 6503              & 0.56 & 111$\cdot 10^{7}$ & 0.17 \\
	5.90 & 12.97       & 1.16                & 2328              & 0.23 & 455$\cdot 10^{6}$ & 0.20 \\
	6.40 & 13.90       & 0.73                & 1465              & 0.15 & 302$\cdot 10^{6}$ & 0.21 \\
	7.90 & ${^*}$16.69 & 0.21                & 428               & 0.05 & 101$\cdot 10^{6}$ & 0.24 \\
	10.9 & ${^*}$22.29 & 0.03                & 61                & 0.01 & 205$\cdot 10^{5}$ & 0.34 \\
	12.9 & $^{*}$26.03 & 0.01                & 21                &      &                   &      \\
	13.9 & $^{*}$27.90 & 0.66$\cdot 10^{-2}$ & 13                &      &                   &      \\
	\hline
	\end{tabular}
	\caption[]{Integrated cross section $\sigma_{int}$ for the range $|\cos\theta|\leq 0.8$ and number of counts $N_{int}$ for $\bar p p \to e^+ e^-$. The prediction was made according to the parameterization as in Ref.~\cite{Sudol:2009vc}. The corresponding values for the $\bar p p \to \pi^+ \pi^-$ channel are also listed. A 100$\%$ data taking efficiency and an integrated luminosity ${\cal{L}}=2$ fb$^{-1}$ were assumed for each beam momentum value, which corresponds to four months of data taking. For the $s$-values  marked with an '*' the full simulation has not been performed and the numbers are given for future references. The last value, $s=27.9$ GeV$^2$, is the upper kinematic limit for which this process could be measured at \PANDA.}
	\label{table:kincount}
	\end{center}
\end{table*}

As already mentioned, TL FFs are complex functions. However, the unpolarized cross section contains only the moduli squared of the FFs. An experiment with a polarized antiproton beam and/or polarized proton target would allow access to the phase difference of the proton FFs (Ref.~\cite{Dbeyssi:2013}). The feasibility of implementing a transversely polarized proton target in \PANDA is under investigation.

\subsection{The background reaction}
\label{sec:The background reaction}
In order to estimate the $\pi^+\pi^-$ background in the interesting kinematic range, phenomenological parameterizations for the differential cross sections and a new generator have been developed (see Ref.~\cite{Zambrana-note:2014}). 

The difficulties for a consistent physical description are related to different aspects:
\begin{itemize} 
\item The dominant reaction mechanism changes with energy and angle \cite{Dover:1992vj}.
\item At low energy, the angular distribution of the final state pions is measured \cite{Eisenhandler:1975kx}. A baryon exchange model was developed in Ref.~\cite{Moussallam:1984zj}, restricted to $p_{lab}<1$ GeV/$c$.
\item At high energy ($p_{lab}\geq 5$ GeV/$c$),  a lack of statistics does not allow us to better constrain the parameters \cite{Eide:1973tb,Buran:1976wc}, to discriminate among models.
\item Model independent considerations based on crossing symmetry or T-invariance, which help to connect the relevant reactions, in general can not be considered as predictive \cite{Stein:1977en}. 
\end{itemize}

As a consequence, the generator utilizes two different parameterizations: in the low energy region, $0.79 \le p_{lab}\le $ 2.43 GeV/$c$, the Legendre polynomial parameters up to the order of ten have been fitted to the data from Ref.~\cite{Eisenhandler:1975kx}. In the high energy region, $5\le p_{lab}< 12$ GeV/$c$, the Regge inspired parameterization from Ref.~\cite{VandeWiele:2010kz}, previously tuned to data from Refs. \cite{Eide:1973tb,Buran:1976wc,White:1994tj,Armstrong:1986ng}, was applied. In the intermediate region, $2.43\le p_{lab}< $ 5 GeV/$c$, where no data exist and the validity of the model is questionable, a soft interpolation is applied.

\begin{figure}
\begin{center}
\resizebox{0.50\textwidth}{!}{
\includegraphics{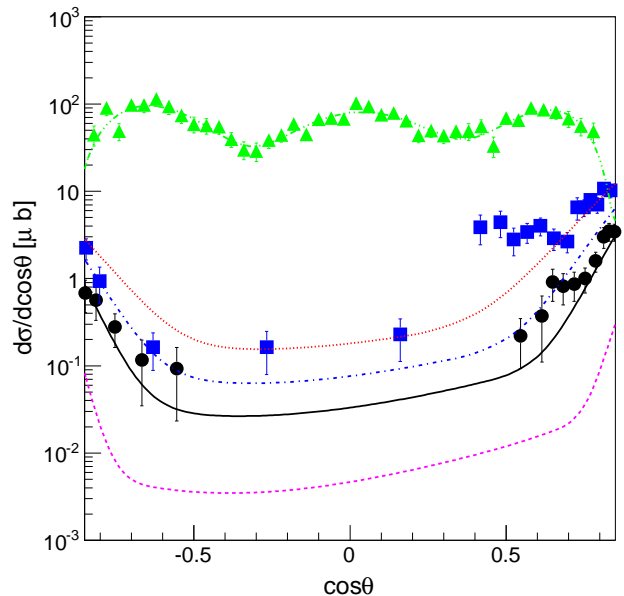}} 
\caption{Data and modeling of the $\pi^-$ angular distributions from the reaction $\bar p p\to \pi^+ \pi^-$, as a function of $\cos\theta$, for different values of the beam momentum: $p_{lab}=1.7$ GeV/$c$ \protect\cite{Eisenhandler:1975kx} (green triangles and dash-triple dotted line); $p_{lab}=5$ GeV/$c$ \protect\cite{Eide:1973tb} (blue full squares and dash-dotted line); $p_{lab}=6.21$ GeV/$c$ \protect\cite{Buran:1976wc} (black full circles and solid line). The results of the generator \cite{Zambrana-note:2014} are also given at $p_{lab}=3$ GeV/$c$ (red dotted line) and $p_{lab}=10$ GeV/$c$ (magenta dashed line).
}
\label{fig:pi_ang}
\end{center}
\end{figure}

Differential cross sections of the $\bar p p\to \pi^+\pi^-$ reaction are displayed in Fig.~\ref{fig:pi_ang}, for different $p_{lab}$. The functions used in the pion generator are shown in comparison to the data sample.

\begin{figure}
\begin{center}
\resizebox{0.50\textwidth}{!}{
\includegraphics{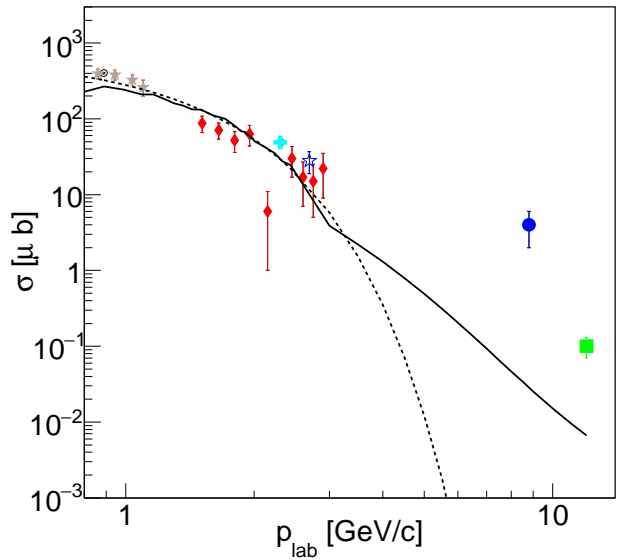}}
\caption{Total cross section for the reaction $\bar p p\to \pi^+\pi^-$, as a function of the beam momentum in the laboratory reference frame, $p_{\rm lab}$. The selected data are from Refs: \protect\cite{Bardin:1994am} (black open circle), \protect \cite{Ward:1980cw} (blue full circle), \protect\cite{Chen:1977ik} (cyan full cross), \protect\cite{Eastman:1973va} (red full diamonds), \protect\cite{Mandelkern:1972ba} (brown full stars), and \protect\cite{Domingo:1967gra} (black open star). The result from Ref.~\protect\cite{Armstrong:1986ng} (green full square) corresponds to total backward cross section and has to be considered as a lower limit. The solid line is the result from the generator \cite{Zambrana-note:2014}. The dashed line is the result of the compilation from Ref.~\protect\cite{Dbeyssi:2013}.}
\label{fig:pi_tot}
\end{center}
\end{figure}

The total cross section is shown in Fig.~\ref{fig:pi_tot} as a function of the antiproton momentum. The lack of data around $p_{lab}=4$ GeV/$c$ does not constrain the parameters and they could therefore not be fixed to a precise value in the generator. The cross section measured at $p_{lab}=12$ GeV/$c$ from Ref.~\cite{Armstrong:1986ng} should be considered as a lower limit. For comparison, the parameterization from the compilation in Ref.~\protect\cite{Dbeyssi:2013} is also shown. This parameterization reproduces the data.

The expected count rates for the background channel, as well as the total signal-to-background cross section ratio, are reported in Table~\ref{table:kincount}. Within the range $|\cos\theta|\le 0.8$, the $\bar{p}p$ rate of annihilation with the subsequent production of two charged pions is about five to six orders of magnitudes larger than that of the production of a lepton pair.

\section{The \PANDA experiment}
\label{experiment}

\subsection{The \PANDA detector}

The \PANDA experiment will offer a broad physics program thanks to the large acceptance, high resolution and tracking capability and excellent neutral and charged PID in a high rate environment. The average interaction rate is expected to reach $2\times10^7$ s$^{-1}$. The structure and the components of the detector have been optimized following the experience gained in high energy experiments. A detailed overview of the \PANDA detector and its performance can be found in Ref.~\cite{Lutz:2009ff}. In the following, we outline characteristics of the detectors which play an important role in the FF measurements.

\begin{figure*}[t]
\begin{center}
\resizebox{0.90\textwidth}{!}{
 \includegraphics{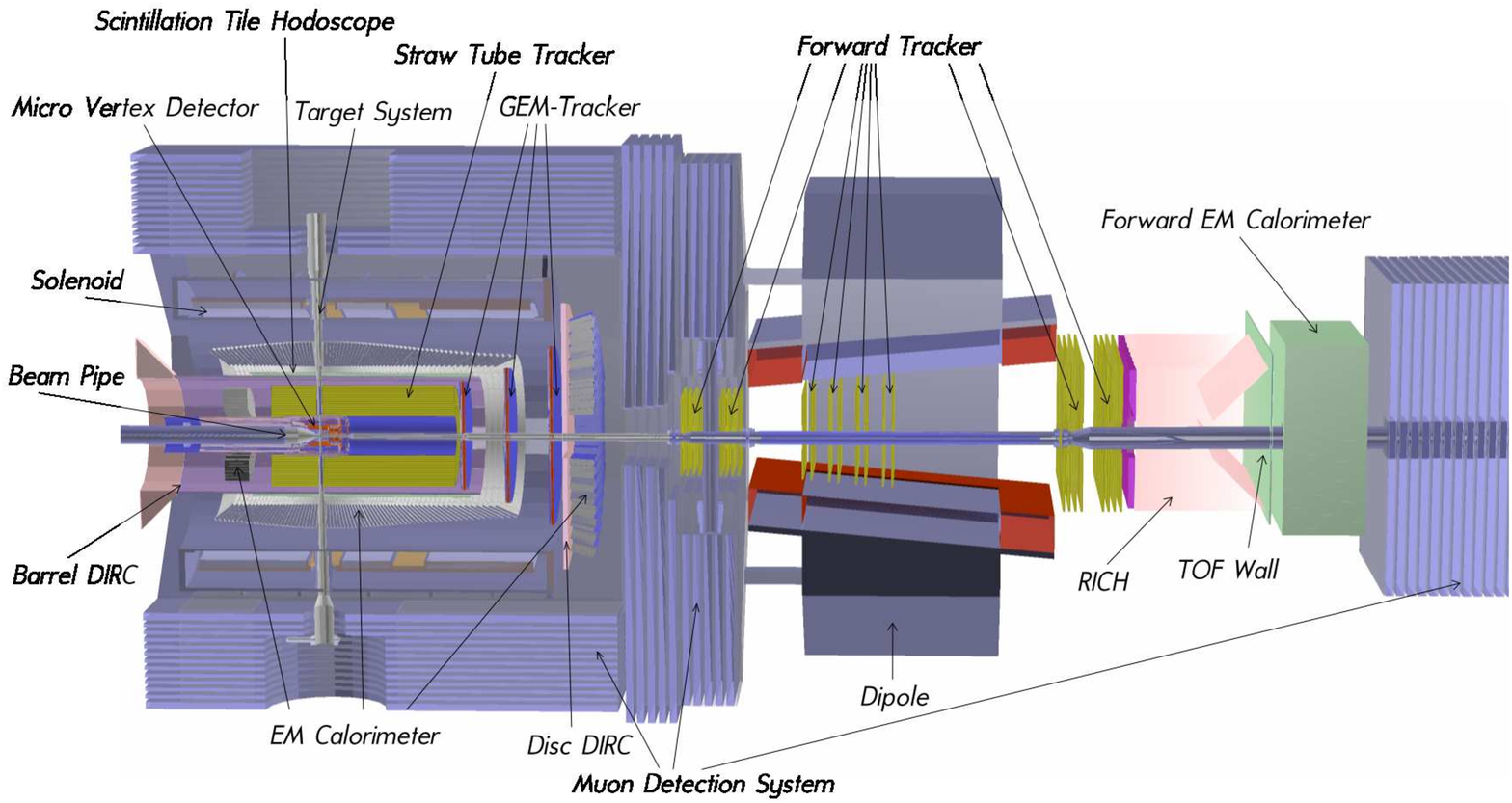}}
\caption{View of the \PANDA detector.}
\label{fig:panda}
\end{center}
\end{figure*}

An overall picture of the \PANDA detector is shown in Fig.~\ref{fig:panda}. The size of the detector is about 13 m along the beam direction. \PANDA is a compact detector with two magnets: a central solenoid \cite{Erni:2009pt} and a forward dipole. The (pellet or jet) target is surrounded by a number of detectors. The target spectrometer consists of the Micro Vertex Detector (MVD) \cite{Erni:2012kva} and the Straw Tube Tracker (STT) \cite{Erni:2013ita} to ensure a precise vertex finding as well as a spatial reconstruction of the trajectories of charged particles. In addition, both sub-detector systems are able to measure the specific energy loss to support the particle identification. The Detection of Internally Reflected Cerenkov light (DIRC) is used for particle identification at polar angles between $22^\circ$ and $140^\circ$, and momenta up to 5~GeV/$c$ \cite{Merle:2014wra}.

A time-of-flight (TOF) detector comprised of small plastic scintillator tiles (SciTil) is employed for precise time measurements to avoid event mixing at high collision rates and particle identification \cite{Gruber2016104}.

The barrel is completed by an electromagnetic calorimeter (EMC), consisting of lead tungstate (PbWO$_4$) crystals, to assure an efficient photon detection from 0.01~GeV to 14.6~GeV \cite{Erni:2008uqa} with an energy resolution better than 2\%. Besides the cylindrical barrel (11,360 crystals), a forward endcap (3856 crystals) and a backward endcap (ca. 600 crystals) are added \cite{Kavatsyuk:2011zz}. 

Particles emitted at polar angles smaller than 22$^\circ$ are detected by three planar stations of Gas Electron Multipliers (GEM) downstream of the target \cite{Bohmer:2011zz}. The achieved momentum resolution is expected to be $\Delta p/p\simeq 1.5\%$ at 1 GeV/$c$. The muon identification is performed by Iarocci proportional tubes and strips, in the gap behind the EMC and in between the layers of the laminated solenoid flux return, with forward polar angular coverage up to $60^\circ$ \cite{MuonTDR}. The Barrel DIRC is used for PID for particles with momenta of 0.8 GeV/$c$ up to about 5 GeV/$c$, at polar angles between 22$^\circ$ and 140$^\circ$ \cite{Schwarz:2011zzc}.

A time-stamp based data acquisition system, capable of a fast continuous readout, followed by an intelligent software trigger is under development.

\subsection{Simulation and analysis software}

The offline software for the \PANDA detector simulation and event reconstruction is PandaRoot, which is developed within the framework for the future FAIR experiments, FairRoot \cite{fairroot}. It is mainly based on the object oriented data analysis framework ROOT \cite{Brun:1997pa}, and utilizes different transport models such as Geant4 \cite{Agostinelli:2002hh}, which is used in the present simulations. Different reconstruction algorithms for tracking and PID are under development and optimization in order to achieve the requirements of the experiment.

\begin{figure}[th!]
	\begin{center}
	\resizebox{0.48\textwidth}{!}{
	\includegraphics{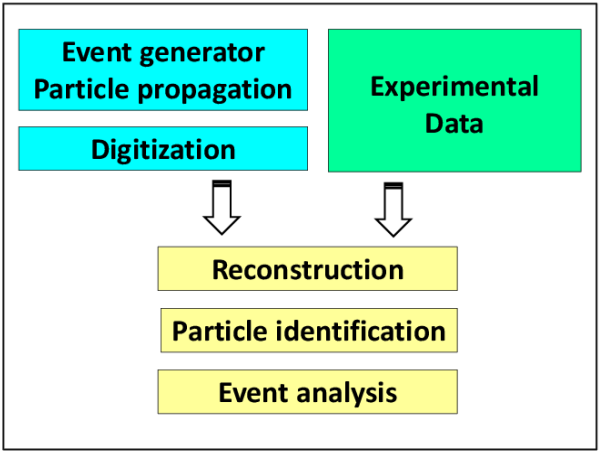}}
	\caption{Standard analysis chain in PandaRoot.}
	\label{fig:SchemeAcquisition}
	\end{center}
\end{figure}

A schematic view of the simulation and data analysis chain is shown in Fig.~\ref{fig:SchemeAcquisition}.

\subsection{Generated events}

\begin{figure*}
  \begin{center}
  \subfloat[\label{fig:pipi_cos_theta17}]{\resizebox{0.33\textwidth}{!}{ \includegraphics{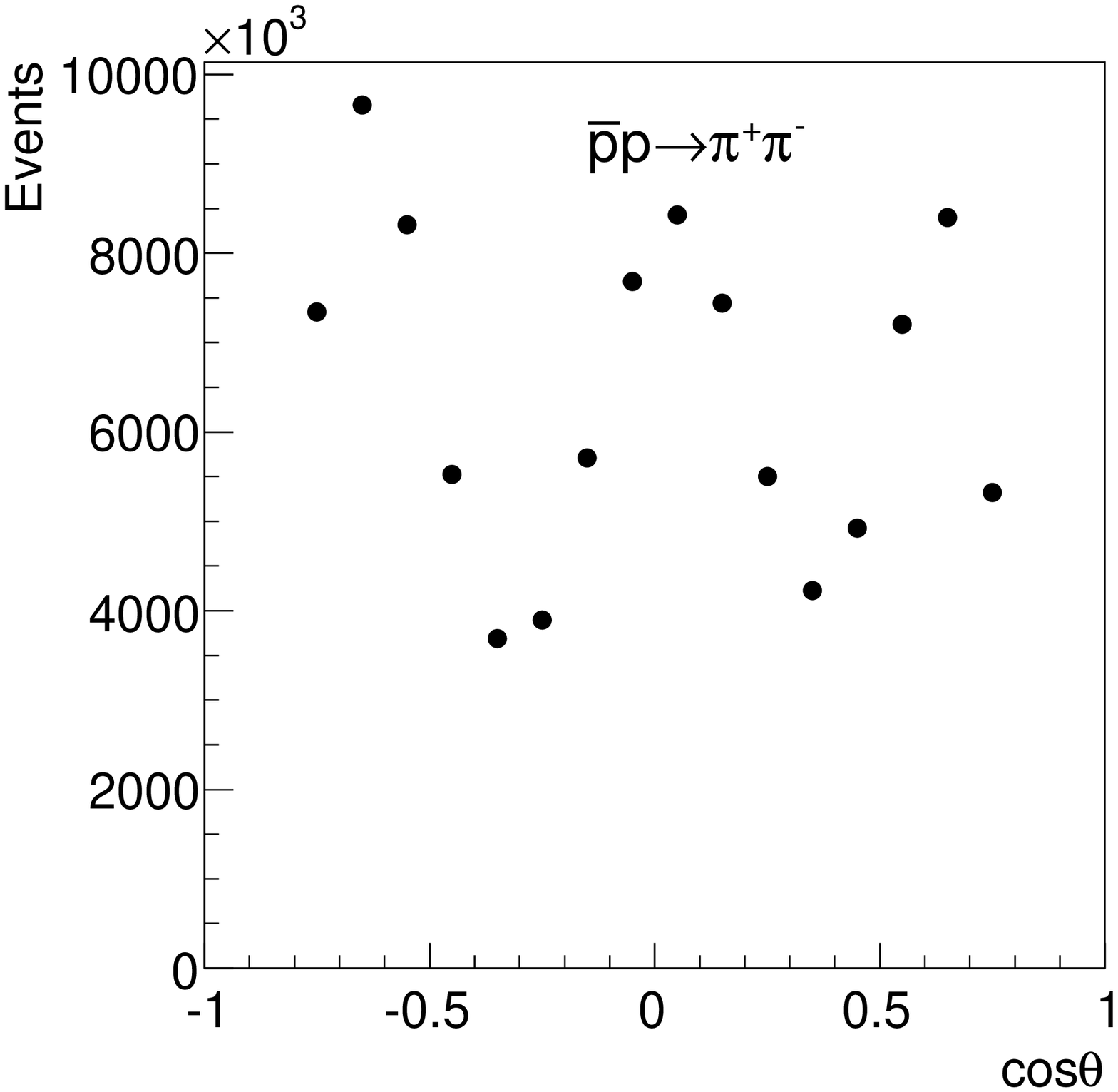} } }
  \subfloat[\label{fig:pipi_cos_theta33}]{\resizebox{0.33\textwidth}{!}{ \includegraphics{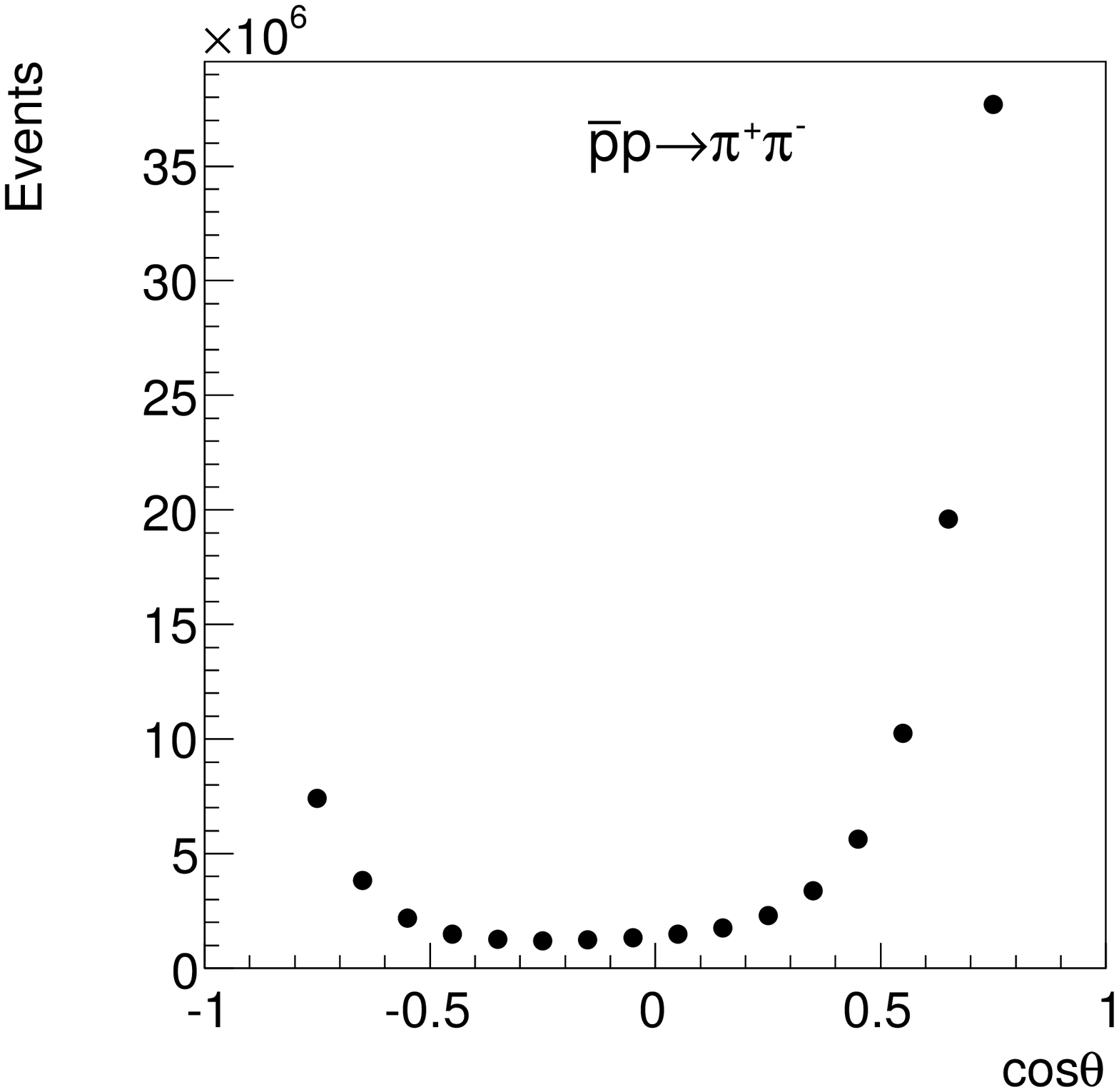} } }
  \subfloat[\label{fig:pipi_cos_theta64}]{\resizebox{0.33\textwidth}{!}{ \includegraphics{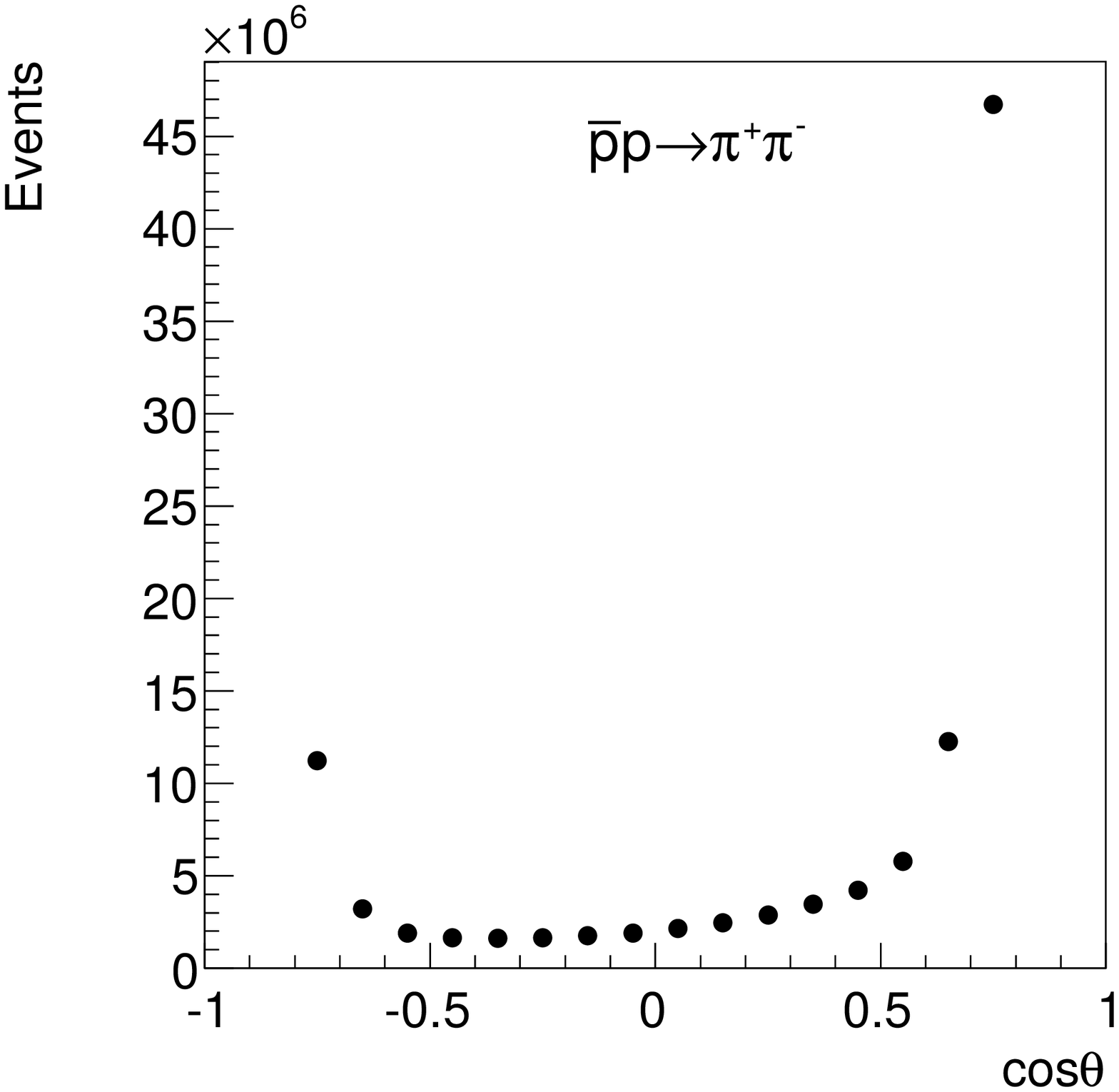} } } \\
  \caption{Angular distribution of $\pi^-$ from the $\bar p p \to \pi^+ \pi^-$ events generated using the model from Ref.~\cite{Zambrana-note:2014} for different $p_{lab}$ values: (a) $1.7$ GeV/$c$, (b) $3.3$ GeV/$c$, and (c) $6.4$ GeV/$c$.}
  \label{fig:pipi_cos_theta}
  \end{center}
\end{figure*}

The signal and background events can be produced by different event generators according to the physics case. As mentioned in the previous section, the generator from Ref.~\cite{Zambrana-note:2014} was used for the $ \bar p p \to \pi^+ \pi^-$ background simulation. Taking into account the ratio of cross sections $\sigma(\bar p p\to \pi^+ \pi^-)/ \sigma(\bar p p\to e^+ e^-) \simeq 10^{6}$, in order to make a reliable proton FF measurement, we need to achieve a background rejection factor on the order of $10^{8}$. Monte Carlo angular distributions of the $\pi^-$ mesons are shown in Fig.~\ref{fig:pipi_cos_theta} for three incident antiproton beam momenta $p_{lab}=1.7$, 3.3, and 6.4 GeV/$c$.

The EvtGen generator~\cite{Lange:2001uf} was used to generate the $\bar p p \to e^+e^-$ signal channel. The generated data were used to determine the efficiency of the signal channel with high precision. The final state leptons are produced according to two models of the angular distribution implemented in EvtGen (see Section \ref{sec:analysis}).

\subsection{PID and kinematic variable reconstruction}

In order to separate the signal from the background, a number of criteria have been applied to the reconstructed events. For this purpose, the raw output of the PID and tracking sub-detectors as EMC, STT, MVD, and DIRC have been used. In this section the most relevant reconstructed variables for the signal selection are described in detail.

\begin{figure}[h!]
	\begin{center}
	\resizebox{0.50\textwidth}{!}{
	\includegraphics{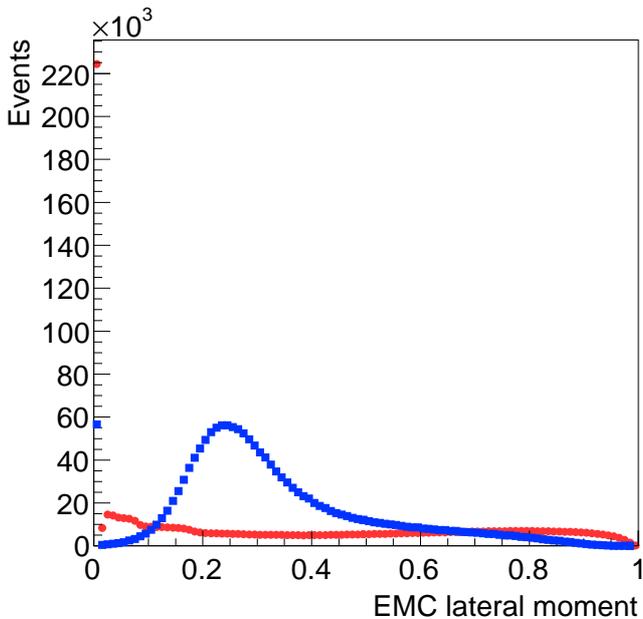}}
	\caption{EMC lateral moment for the signal (blue squares) and background (red circles) events for $p_{lab}=3.3$ GeV/$c$.}
	\label{fig:emc_lateral_moment}
	\end{center}
\end{figure}

In hadronic showers, most of the energy is typically contained in two to three crystals, while electromagnetic showers spread out over greater distances. The group of affected crystals is called a cluster. Typically, hadron showers have smaller lateral moment (LM)\cite{Erni:2008uqa} than electromagnetic showers as shown in Fig.~\ref{fig:emc_lateral_moment}. Therefore, a cut on the EMC LM is applied for the signal-background separation.

The center of a cluster is the crystal that has the highest energy among all crystals in the shower. This energy (E1) is also used for the PID.

\begin{figure*}[th!]
  \begin{center}
  \subfloat[\label{fig:epem_ep}]{\resizebox{0.4\textwidth}{!}{ \includegraphics{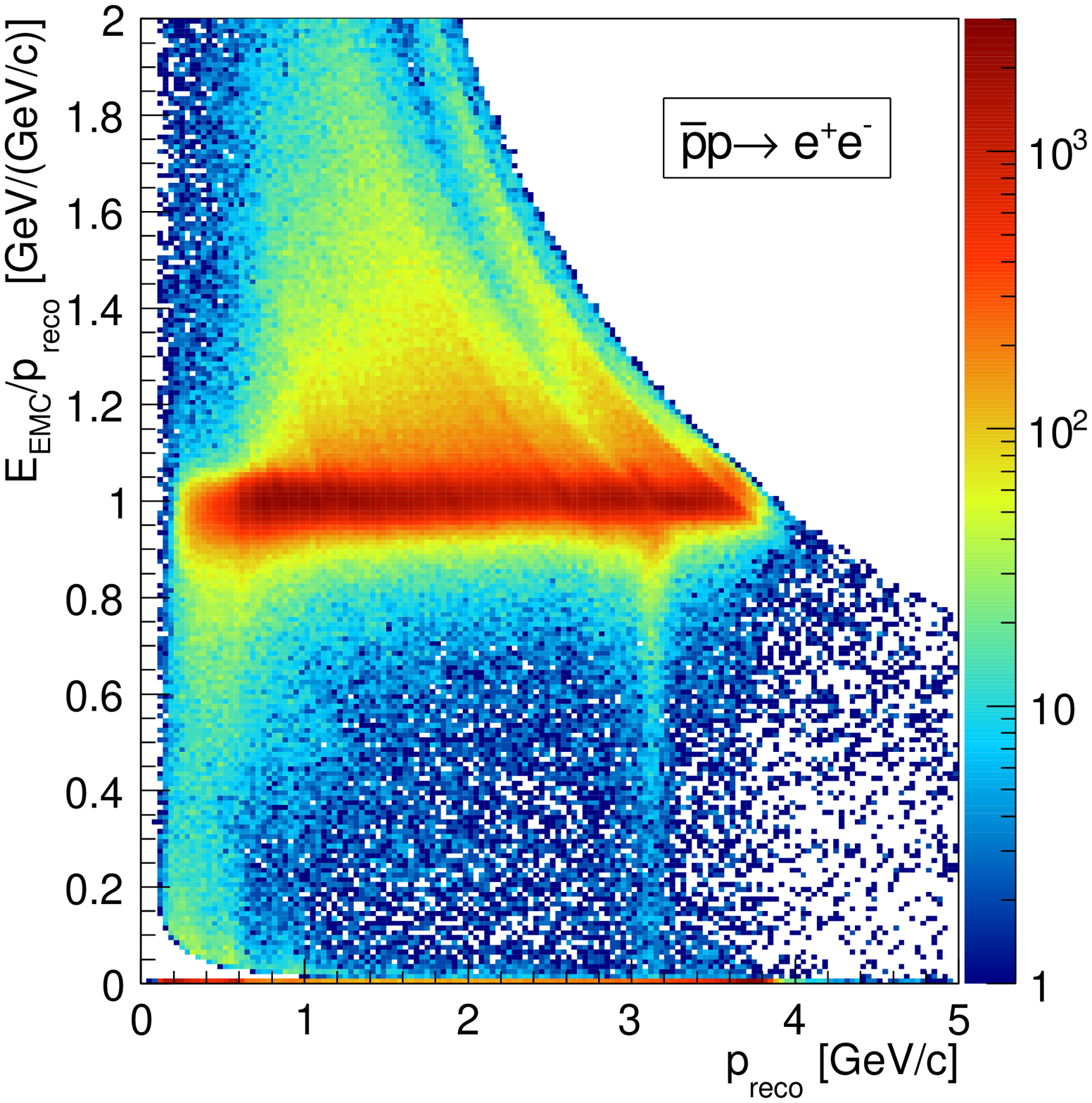} } }
  \subfloat[\label{fig:pipi_ep}]{\resizebox{0.4\textwidth}{!}{ \includegraphics{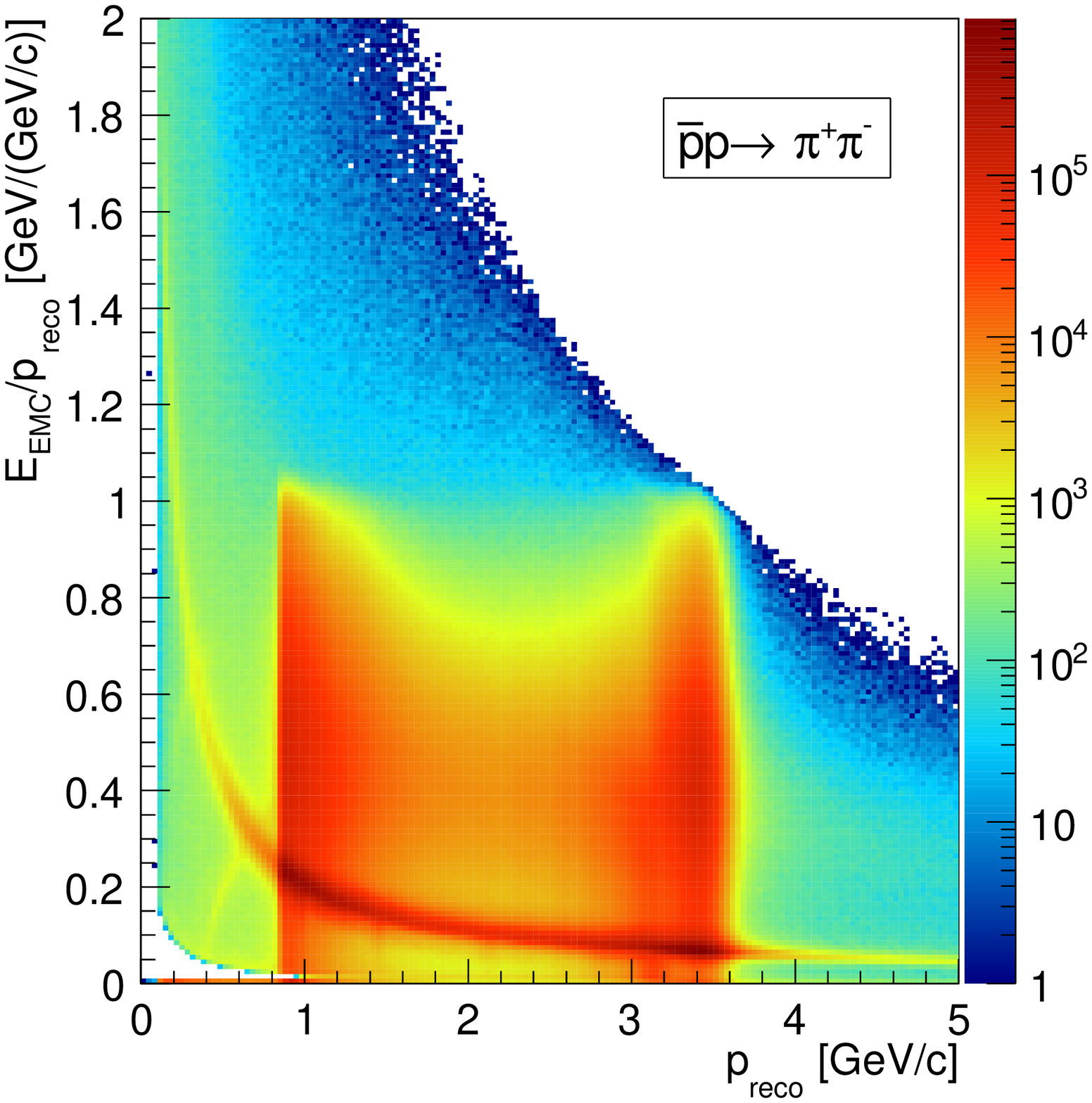} } } \\
  \subfloat[\label{fig:epem_stt}]{\resizebox{0.4\textwidth}{!}{ \includegraphics{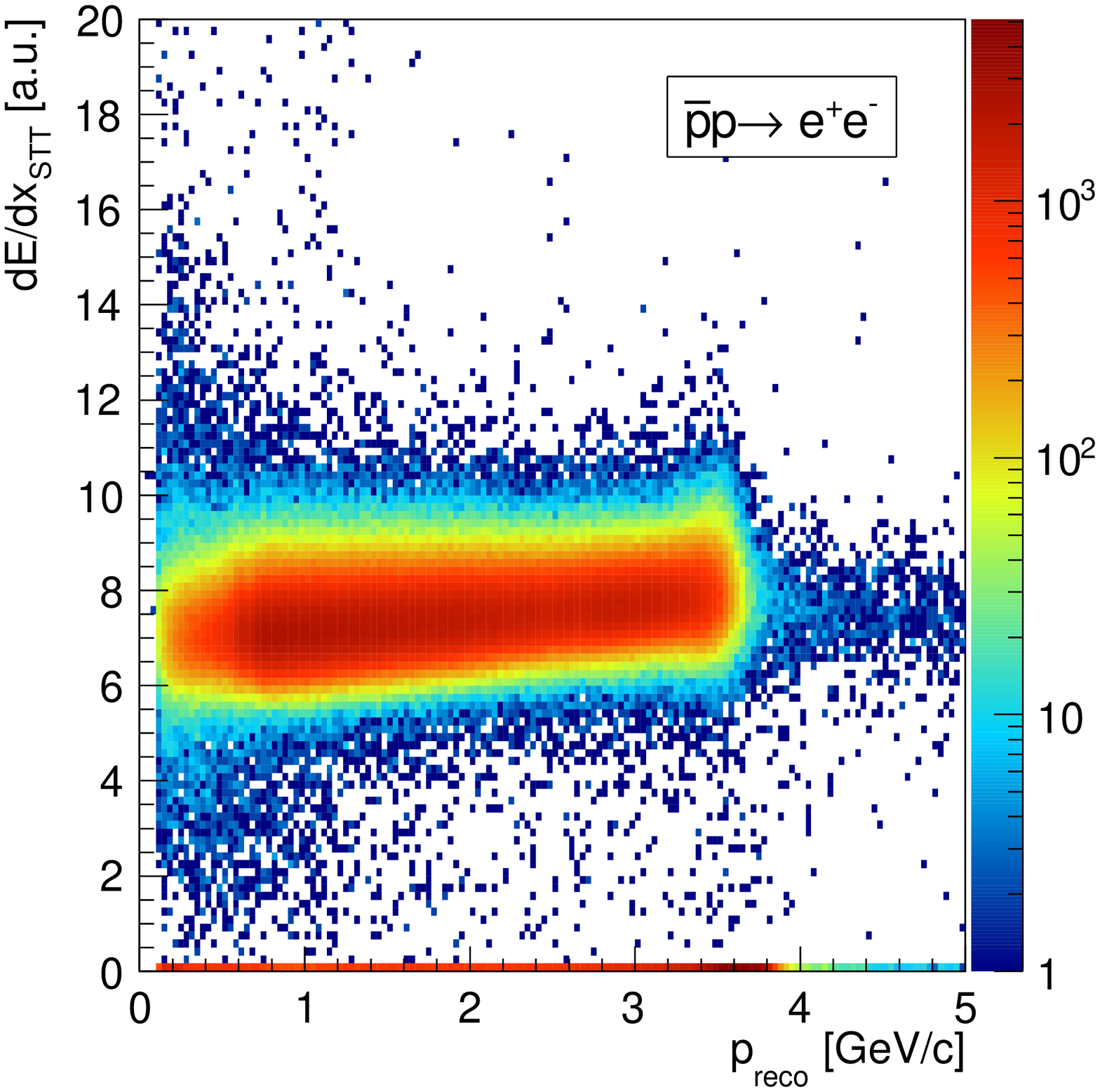} } }
  \subfloat[\label{fig:pipi_stt}]{\resizebox{0.4\textwidth}{!}{ \includegraphics{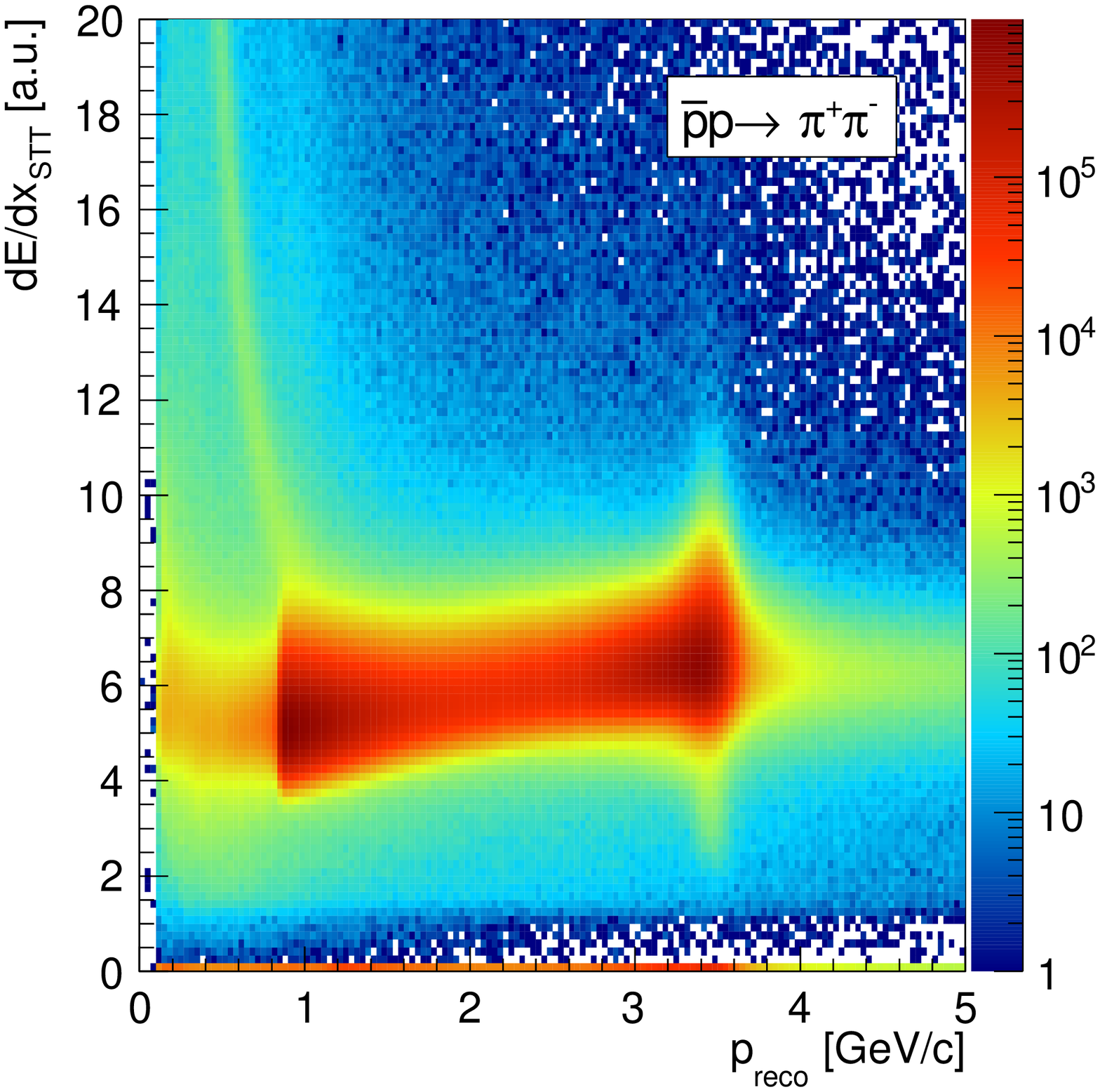} } } \\  
\caption{Detector response to the signal (left column) and the background (right column): (a,b) is the ratio of the energy deposited in the EMC to the reconstructed momentum; (c,d) is the energy loss per unit of length in the STT  as a function of momentum at $p_{lab}=3.3$ GeV/$c$.}
  \label{fig:ep_vs_p}
  \end{center}
\end{figure*}

The ratio $E_{\text{EMC}}/p_{\text{reco}}$ of the shower energy deposited in the calorimeter to the reconstructed momentum of the track associated with the shower is another standard variable for electron selection (Fig.~\ref{fig:ep_vs_p}). Due to the very low electron mass, the $E_{\text{EMC}}/p_{\text{reco}}$ ratio is close to unity for the signal (Fig.~\ref{fig:epem_ep}). The discontinuities that appear in the plot are due to the transition regions between the different parts of the EMC. For the background (Fig.~\ref{fig:pipi_ep}), the distribution shows a double structure: one narrow peak at low $E_{\text{EMC}}/p_{\text{reco}}$ values, which is due to the energy loss by ionization, and another one around $E_{\text{EMC}}/p_{\text{reco}}=0.4$ corresponding to hadronic interactions. The tail of the latter extends to much higher values, resulting in background under the electron peak.

Figures~\ref{fig:epem_stt} and ~\ref{fig:pipi_stt} show the momentum dependence of the energy loss per unit length for signal and background, respectively. Although the energy loss, denoted $dE/dx_{STT}$, shows overlapping patterns for electrons and pions, a cut on the deposited energy in the STT can be applied in order to partially suppress the pion background.

\subsubsection{PID probabilities} 

Using the raw output of the EMC, STT, MVD and DIRC detectors, the probabilities of a reconstructed particle being an electron or positron have been calculated. The probabilities can be calculated for each detector individually (PID$_s$) or as a combination of all of them (PID$_c$).

\begin{figure*}
	\centering
	\resizebox{0.9\textwidth}{!}{
	\includegraphics{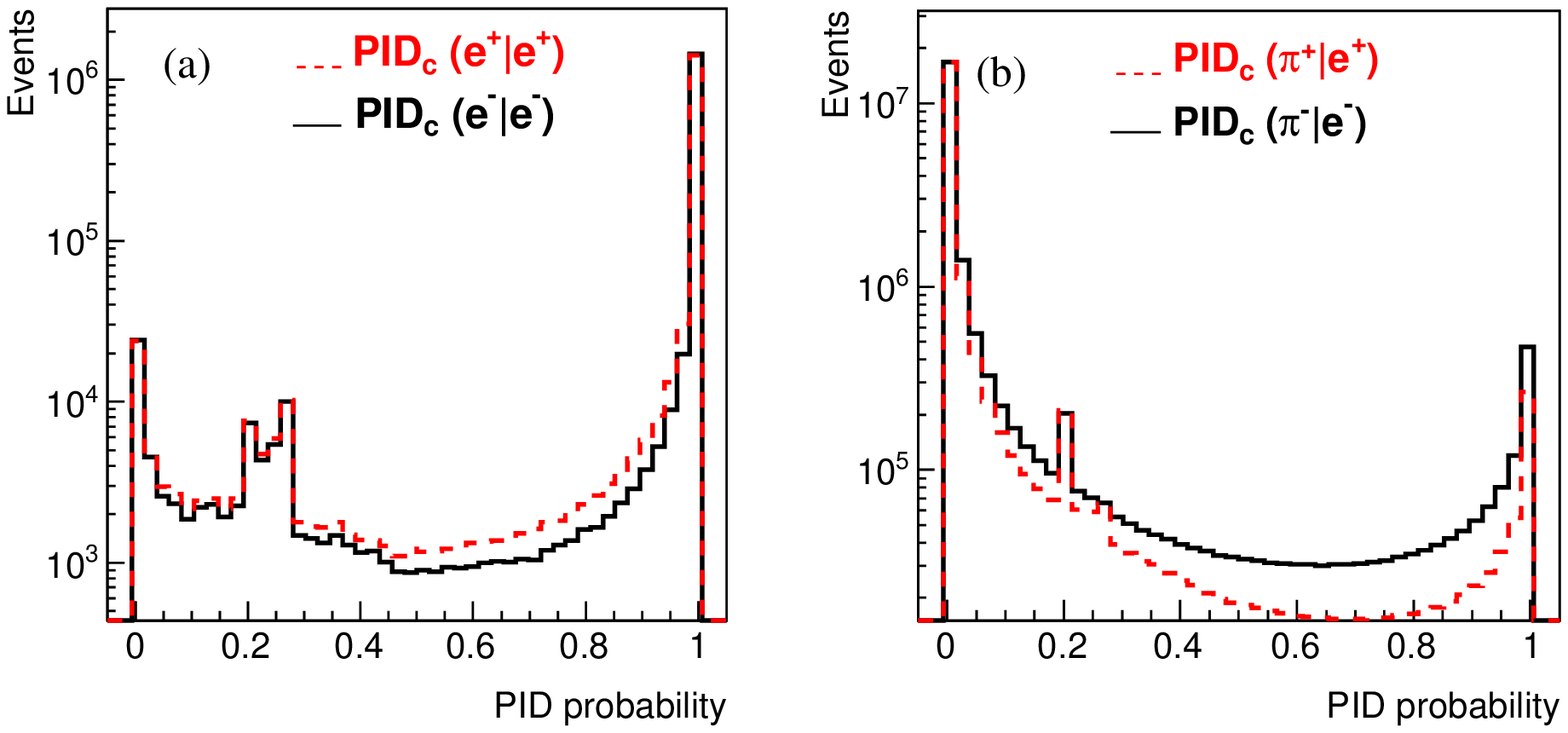}}
	\caption{(a) Total PID probability distribution for a $e^+$ to be identified as a $e^+$ PID$_c(e^+|e^+)$ (dashed red line) and for an $e^-$ to be identified as an $e^-$ PID$_c(e^-|e^-)$ (solid black line). (b) Total PID probability distribution for a $\pi^+$ to be identified as a $e^+$ PID$_c(\pi^+|e^+)$ (dashed red line) and for a $\pi^-$ to be identified as an $e^-$ PID$_c(\pi^-|e^-)$ (solid black line). $p_{lab}=3.3$ GeV/$c$.}
	\label{pidmerge}
\end{figure*}

In this work, both types of probabilities PID$_s$ and PID$_c$ have been used in order to increase the signal efficiency and the background suppression factor. The distributions of PID$_c$ are shown in Fig.~\ref{pidmerge}. For the signal, the distributions of the PID$_c$ (Fig.~\ref{pidmerge}a) have a maximum at PID$_c=1$, where the generated electron and positron events are well identified. The peak at 0.2 is related to events for which no definitive type of particle was assigned. In this case, the probability splits equally into the five particle hypotheses ($e$, $\mu$, $K$, $\pi$, and $p$). The same explanation holds for the highly populated region around PID$_c=0.3$, where two or three particle types have the same behavior in some detectors.
For the generated background events, PID$_c$ distributions are shown in Fig.~\ref{pidmerge}b. As expected, the distributions of PID$_c$ all have a maximum at zero.

\subsubsection{Kinematic variables}

The signal and background reactions are two-body final state processes. The electrons or pions are emitted back to back in the c.m. system. Since all final state particles are detected, their total energy is equal to that of the $\bar{p}p$ system.

\begin{figure*}
  \begin{center}
  \subfloat[\label{fig:epem_theta_diff}]{\resizebox{0.3\textwidth}{!}{ \includegraphics{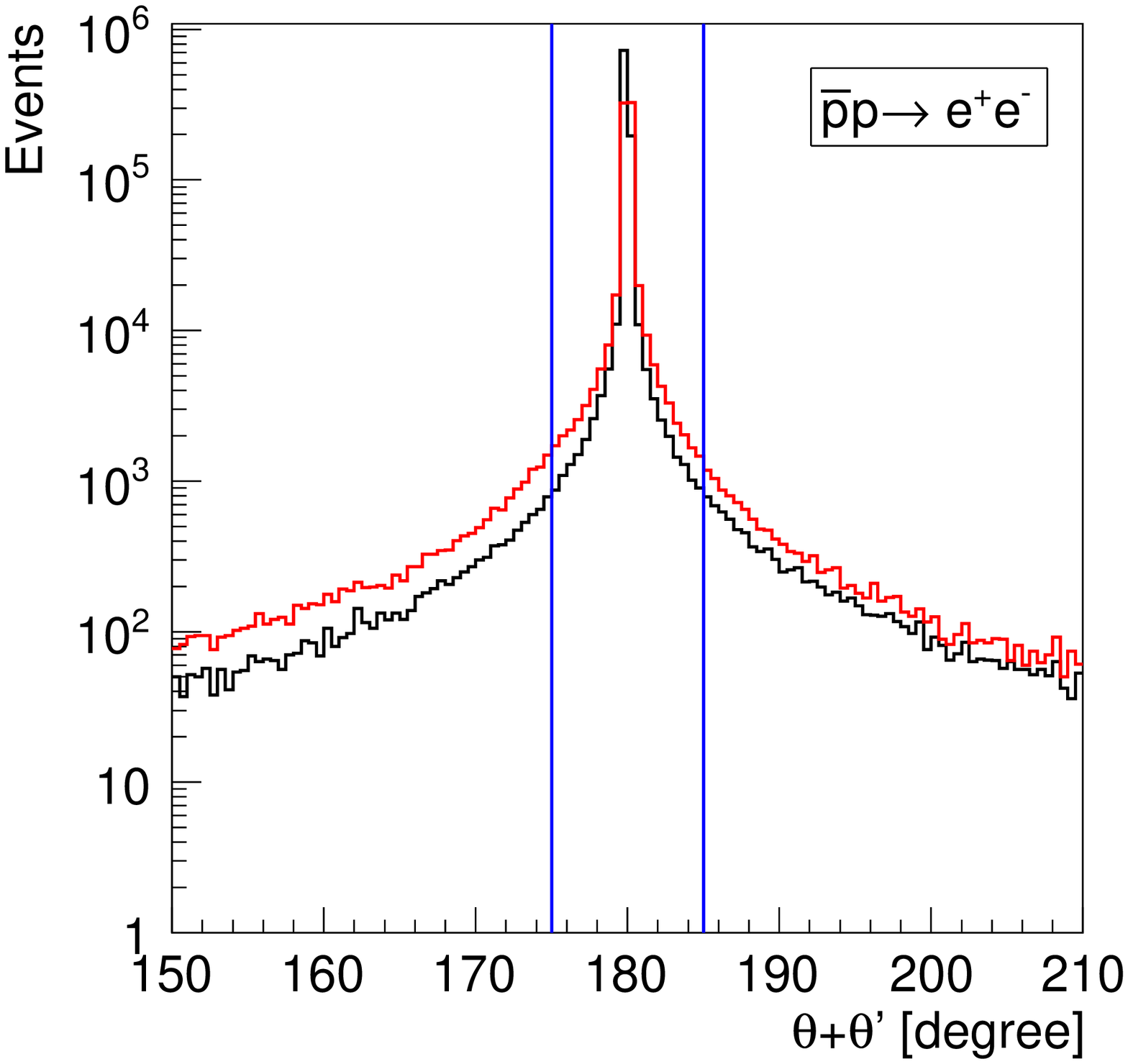} } }
  \subfloat[\label{fig:epem_phi_diff}]{\resizebox{0.3\textwidth}{!}{ \includegraphics{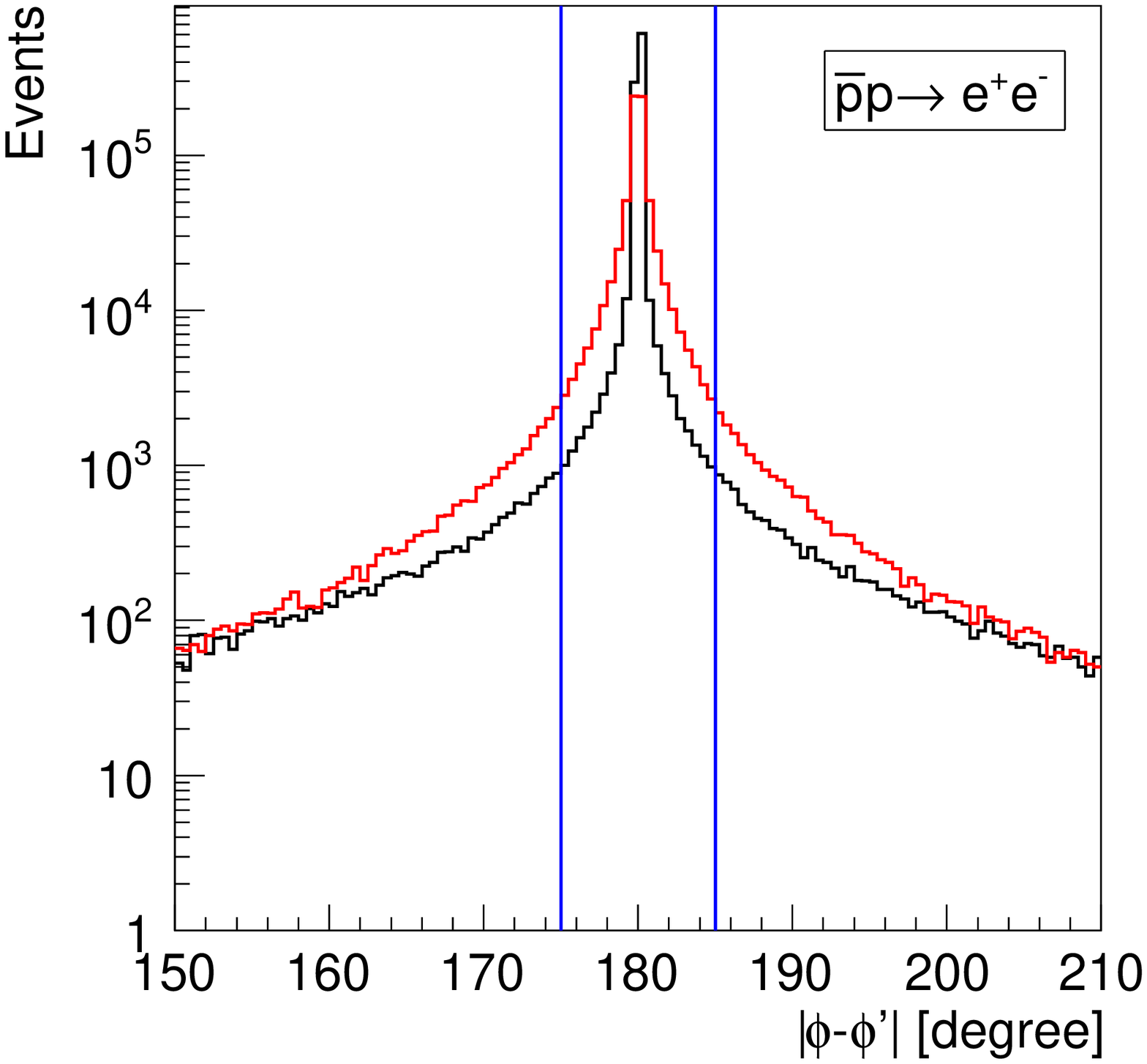} } }
  \subfloat[\label{fig:epem_inv_mass}]{\resizebox{0.3\textwidth}{!}{ \includegraphics{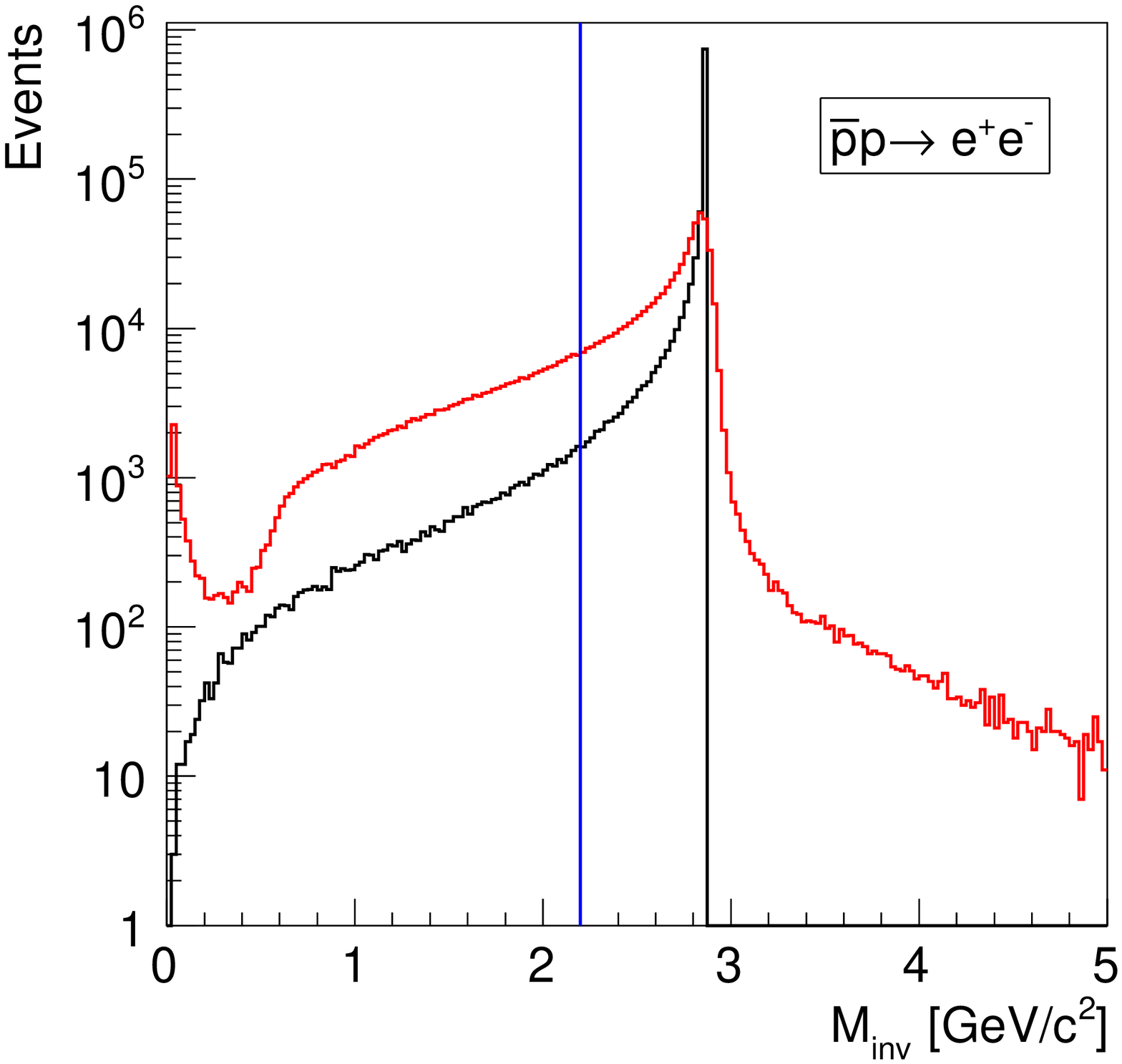} } }  \\
  \subfloat[\label{fig:pipi_theta_diff}]{\resizebox{0.3\textwidth}{!}{ \includegraphics{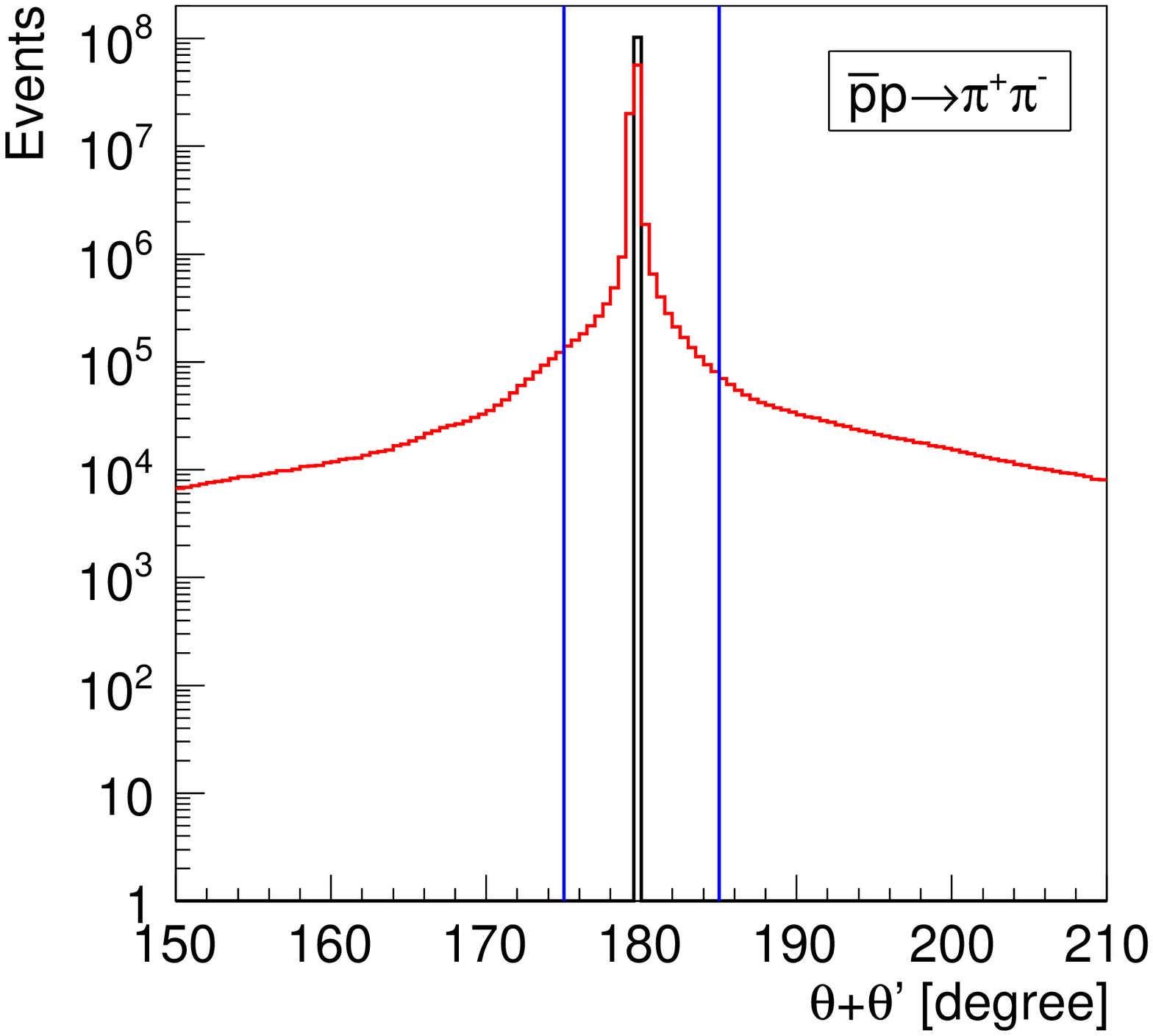} } }
  \subfloat[\label{fig:pipi_phi_diff}]{\resizebox{0.3\textwidth}{!}{ \includegraphics{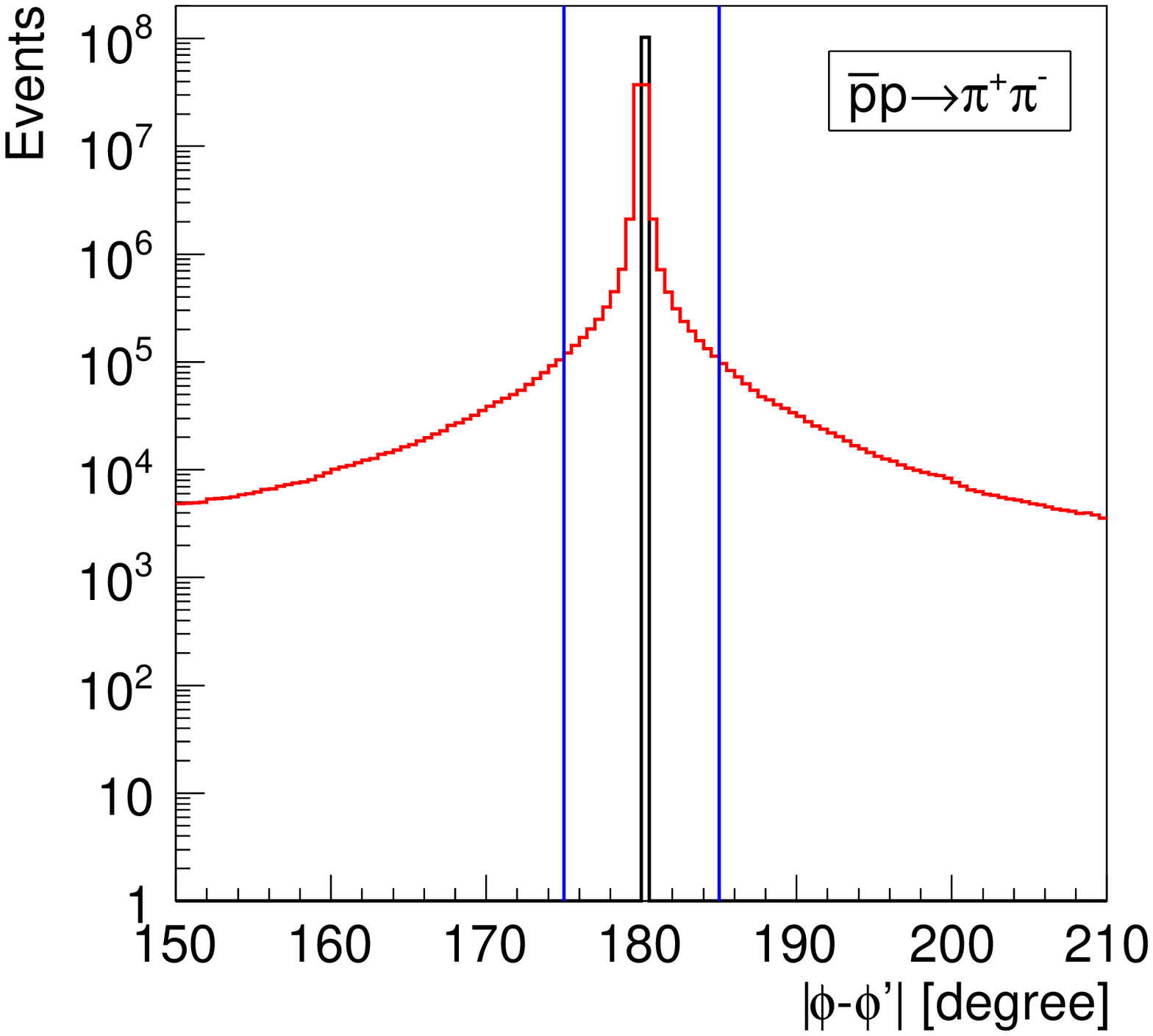} } }
  \subfloat[\label{fig:pipi_inv_mass}]{\resizebox{0.3\textwidth}{!}{ \includegraphics{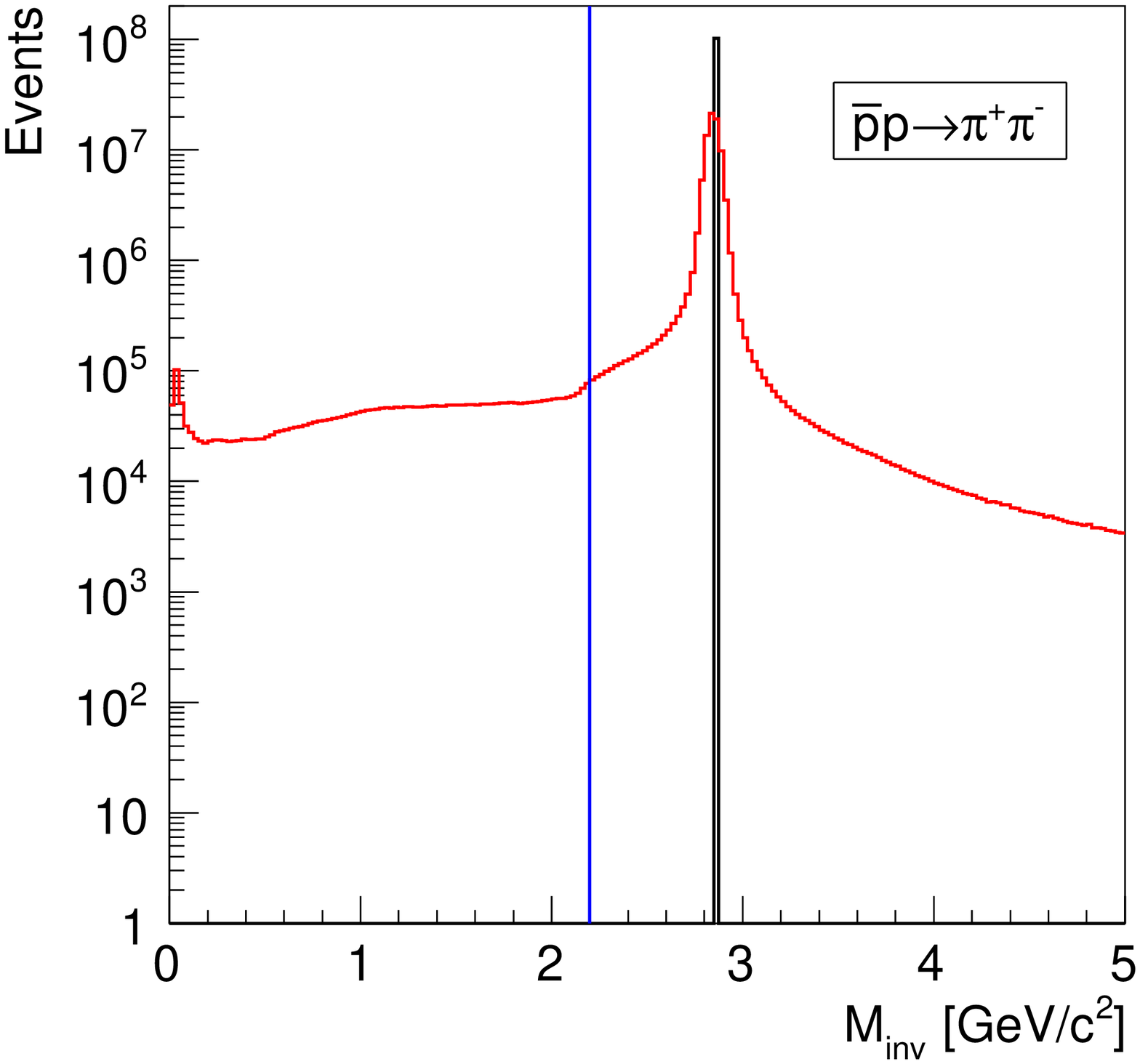} } }
  \caption{Spectra of generated (black) and reconstructed (red) events for different kinematic variables for the signal (top row) and the background (bottom row) at $p_{lab}=3.3$ GeV/$c$: (a, d) the sum of the polar angles in the c.m. frame; (b, e) the difference in the azimuthal angles in the c.m. frame; (c, f) the invariant mass of the reconstructed particles. The blue lines denote the range of the variable accepted for further analysis.}
  \label{fig:varkin}
  \end{center}
\end{figure*}

In Fig.~\ref{fig:varkin}, the relevant kinematic variables are presented for generated and reconstructed MC events, before the selection procedure, at $p_{lab}=3.3$ GeV/$c$.

In contrast to the momentum and energy of a charged particle, the mean polar and azimuthal angles are not affected by the Bremsstrahlung emission during the passage of the particle through matter. Additionally, the sum of the polar angles ($\theta+\theta'$) and the difference of the azimuthal angles ($|\phi-\phi'|$) can be used to reject secondary particles.

\section{Simulation}
\label{sec:analysis}
Two independent simulation studies have been performed for signal and background events. In the following, we will present both studies in detail and clarify the methods in each one. The main differences between them are (i) the angular distribution model used as input for the signal event generator, (ii) the determination of the efficiency and (iii) the fit of the reconstructed events after efficiency correction. By comparing the two simulations we can estimate the effect of the statistical fluctuations, the efficiency determination, and the extraction of the proton FFs using different fit functions. The two approaches are denoted Method I and Method II. Both methods use the same background samples.

\subsection{Background events}
The generator from Ref.~\cite{Zambrana-note:2014} is used for the $ \bar p p \to \pi^+ \pi^-$ background simulation. $10^8$ background events are generated at three incident antiproton beam momenta, $p_{lab}=1.7$, 3.3 and 6.4~GeV/$c$ ($s=5.40$, 8.21, and 13.90~GeV$^2$, respectively). Each method uses an unique set of criteria for the background suppression (see Sections \ref{sec:methodi} and \ref{sec:methodii}).

\subsection[Method I]{Method I\footnote{This work is a part of D. Khaneft's Ph.D. thesis.}}
\label{sec:methodi}

All signal events are generated using the differential cross section parameterized in terms of proton electromagnetic FFs, according to Eq.~(\ref{eq:eqdsigma}). Eq.~(\ref{eq:eqff}) is used as a parameterization of $|G_M|$, together with the hypothesis that $|G_E|=|G_M|$.

Assuming an integrated luminosity of $\mathcal{L}=2$ fb$^{-1}$ (four months of data taking) and 100\% efficiency, the expected number of the produced $e^+e^-$ pairs in the range $-0.8<\cos\theta<0.8$ can be calculated. Table~\ref{table:kincount} shows the total cross section and the expected number of events for different values of the beam momentum. The signal events are generated using Eq.~(\ref{eq:eqdsigma}) and Table~\ref{table:kincount}.

\subsubsection{Particle identification}
\label{sec:a2_pid}
\begin{figure}
  \resizebox{0.5\textwidth}{!}{ \includegraphics{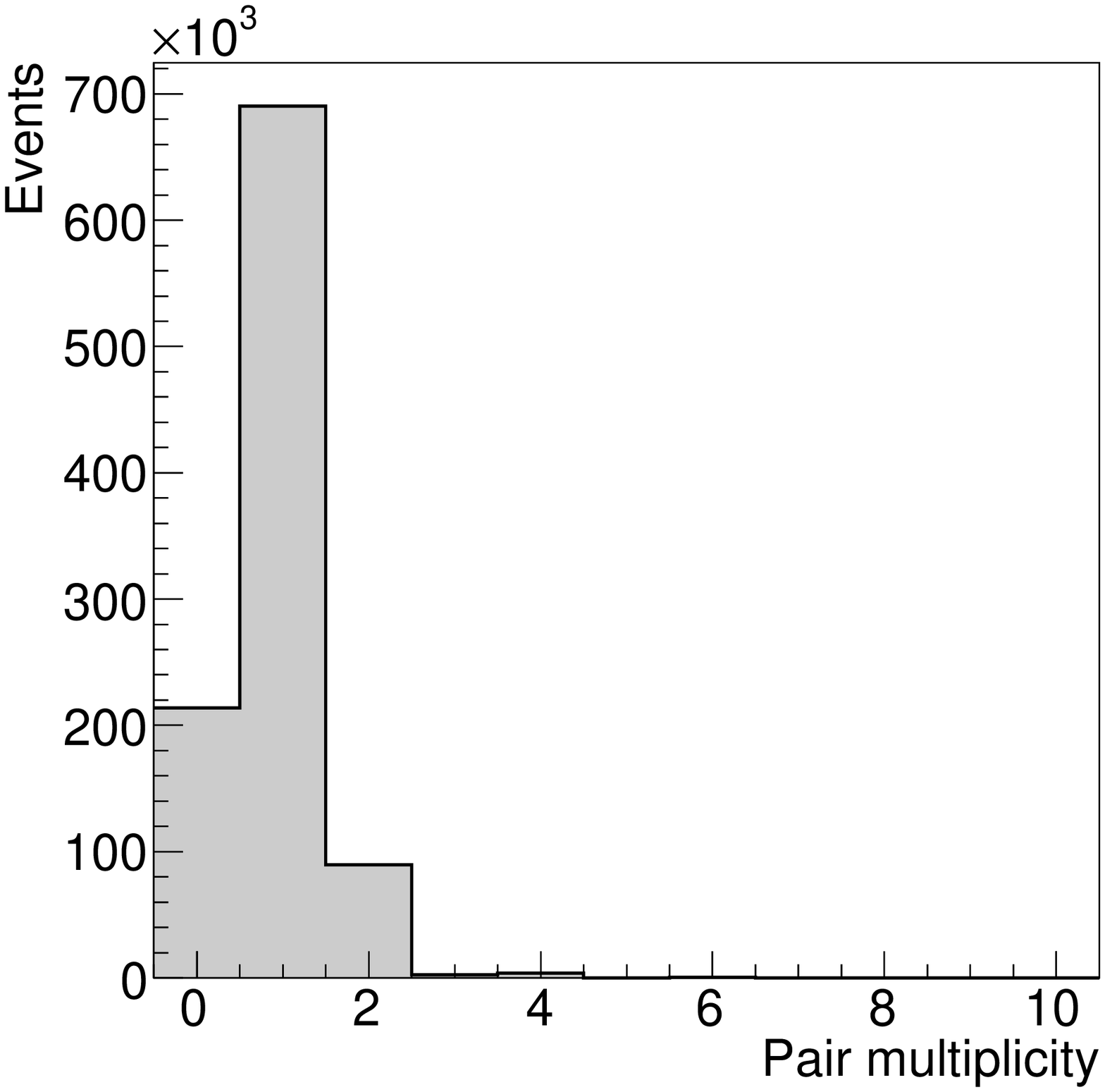} }
  \caption{Multiplicity distribution of the number of reconstructed pairs of particles. $10^6$ signal events were generated at a beam momentum of $p_{lab}=3.3$ GeV/$c$.}
  \label{fig:a2_multiplicity}
\end{figure}

The event selection is performed in two steps. First, events having exactly one positive and one negative reconstructed charged track are selected for further analysis. The number of reconstructed pairs of particles with an opposite charge are shown in Fig.~\ref{fig:a2_multiplicity}. Note that only in $10\%$ of the cases, the multiplicity is larger than one. If an event has \textit{e.g.} one positive and two negative particles, it is considered to have a multiplicity of two, because the positive particle could be associated with either of the two negative particles.

\begin{table*}[th!]
  \begin{center}
  \begin{tabular}{ l l l l l l l l } \hline
    $p_{lab}    $          & [GeV/$c$]  & $1.70$ & $2.78$ & $3.30$ & $4.90$ & $5.90$ & $6.40$ \\ \hline
    PID$_c$ & [\%]         & $>$99      & $>$99   & $>$99   & $>$99    & $>$99    & $>$99    \\
    PID$_s$ & [\%]         & $>$10      & $>$10   & $>$10   & $>$10    & $>$10    & $>$10    \\
    $dE/dx_{STT}$ & [a.u.] & $>$5.8     & $>$5.8  & $>$5.8  & $>$5.8   & $>$5.8   & $>$6.5   \\
    $E_{EMC}/p_{reco}$     & [GeV/(GeV/$c$)] & $>$0.8  & $>$0.8  & $>$0.8  & $>$0.8   & $>$0.8   & $>$0.8   \\
    EMC LM      & -        & $<$0.75 &  $<$0.75 & $<$0.75 & $<$0.75  & -    & -        \\
    EMC E1 & [GeV]         & $>$0.35 &  $>$0.35 & $>$0.35 & $>$0.35  & $>$0.35  & $>$0.35  \\
    $|\theta+\theta'-180|$ & [degree]   & \multicolumn{6}{ c }{<5}   \\
    $|\phi-\phi'-180|$     & [degree]   & \multicolumn{6}{ c }{<5}     \\
    $M_{inv}$              & [GeV/$c$$^2$]         & -     & -     & $>$2.2  & $>$2.2   & $>$2.2   & $>$2.7   \\ \hline
  \end{tabular}
  \caption{Criteria used to select the signal ($e^+e^-$) and suppress the background ($\pi^+\pi^-$) events for each $p_{lab}$ value (Method I).}
  \label{tab:a2_pid}
  \end{center}
\end{table*}

Next, all events passing the selection scheme mentioned above are are filtered through a set of additional criteria listed in Table~\ref{tab:a2_pid}. These criteria are chosen in order to maximize signal reconstruction efficiency while suppressing as many background events as possible. Some cuts are fixed for all values of beam momenta, whereas others are optimized to fit the response of the detector at each energy.

Table~\ref{tab:a2_eff} shows the reconstruction efficiency for the signal ($e^+e^-$) selection and the background ($\pi^+\pi^-$) suppression for each value of $p_{lab}$. 
\begin{table}[th!]
  \begin{center}
  \begin{tabular}{ l l l l} \hline
    $p_{lab}$ [GeV/$c$] & $e^+e^-$ & $\pi^+\pi^-$       \\ \hline
    1.70        & 0.51   & $6.8\times10^{-8}$ \\
    2.78        & 0.54   & -                  \\
    3.30        & 0.46   & $2.0\times10^{-8}$ \\
    4.90        & 0.46   & -                  \\
    5.90        & 0.47   & -                  \\
    6.40        & 0.39   & $2.9\times10^{-8}$ \\ \hline
  \end{tabular}
  \caption{Reconstruction efficiency achieved with the criteria described in Section~\ref{sec:a2_pid} for the signal and the background suppression for each value of $p_{lab}$ (Method I).}
  \label{tab:a2_eff}
  \end{center}
\end{table}

\subsubsection{Determination of the signal efficiency}
\label{sec:a2_sig_eff}
\begin{figure}
  \resizebox{0.5\textwidth}{!}{ \includegraphics{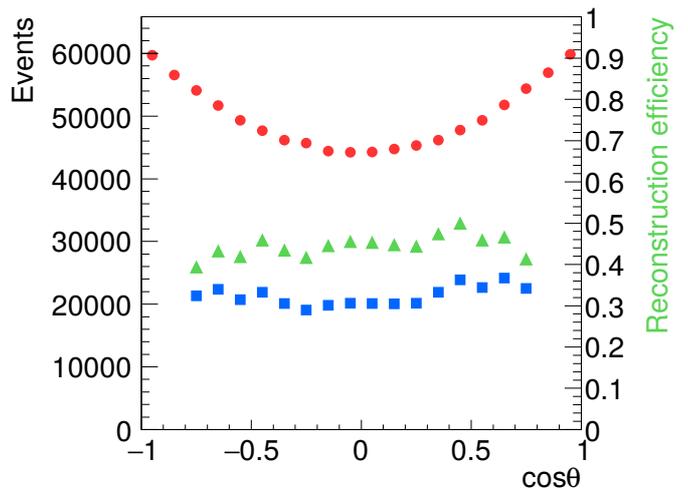} }
  \caption{Angular distribution for $\bar{p}p \rightarrow e^+e^-$ at $p_{lab}=3.3$ GeV/$c$ of generated (red circles) and reconstructed and identified (blue squares) electrons. The reconstruction efficiency (green triangles) corresponds to the $y$-axis scale on the right.}
  \label{fig:a2_eff}
\end{figure}

A significantly larger sample of $e^+e^-$ pairs is simulated for each beam momentum. The signal efficiency is extracted from each sample and equals the ratio between the number of reconstructed events passing the dedicated selection to the number of generated. The uncertainty of the efficiency was calculated in the following way:

\begin{equation}
  \Delta \epsilon_{i}=\sqrt{\epsilon_{i}\frac{(1-\epsilon_{i})}{N^{reco}_{i}}},
\label{eq:a2_eff_err}
\end{equation}
where $\epsilon_{i}$ is the efficiency and $N^{reco}_{i}$ is the number of reconstructed events in the $i$-th bin. The angular distribution of generated electrons, reconstructed and identified events, and the reconstruction efficiency at $p_{lab}=3.3$ GeV/$c$ are presented in Fig.~\ref{fig:a2_eff}.

Thus, the angular distribution of reconstructed and identified electrons can be corrected using the reconstruction efficiency:
\begin{equation}
  N^{corr}_{i} = \frac{N^{reco}_{i}}{\epsilon_{i}},
	\label{eq:a2_ncorr}
\end{equation}
where $N^{corr}_{i}$ is the efficiency corrected number of events in the $i$-th bin.

\subsubsection{Extraction of the ratio R}
\begin{figure*}
  \begin{center}
  \subfloat[\label{fig:events17}]{\resizebox{0.4\textwidth}{!}{ \includegraphics{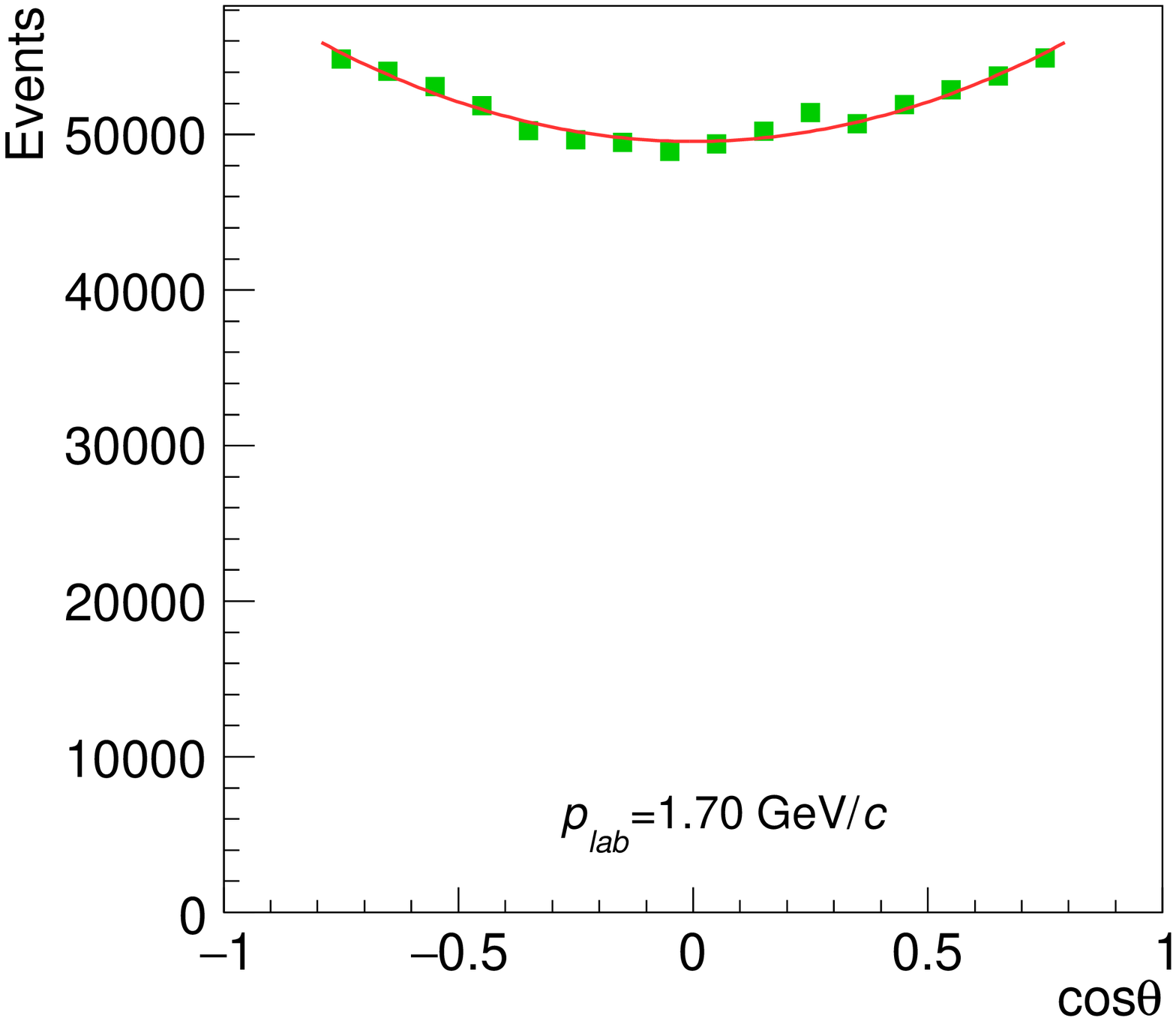} } }
  \subfloat[\label{fig:events278}]{\resizebox{0.4\textwidth}{!}{ \includegraphics{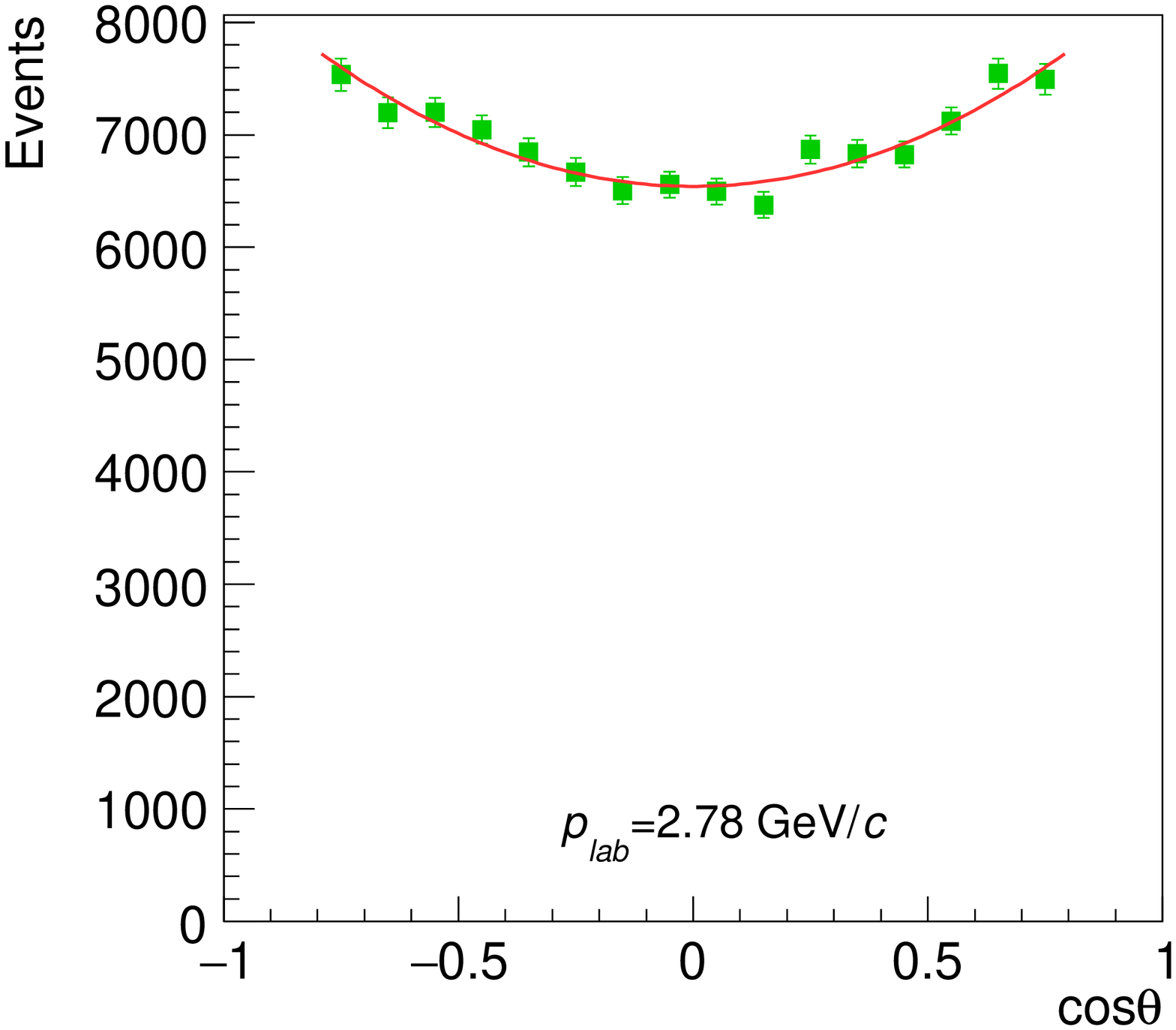} } } \\
  \subfloat[\label{fig:events33}]{\resizebox{0.4\textwidth}{!}{ \includegraphics{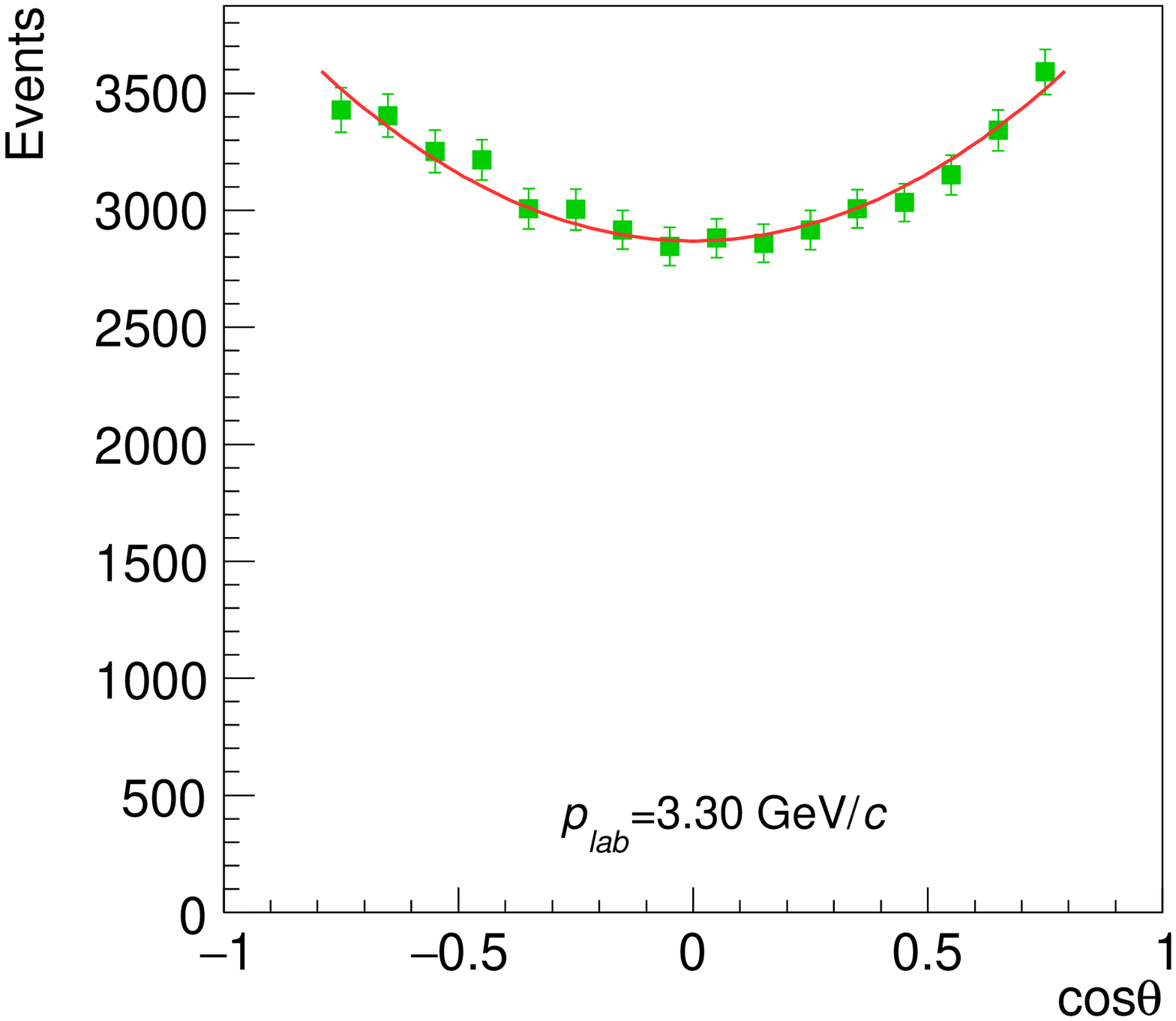} } }
  \subfloat[\label{fig:events49}]{\resizebox{0.4\textwidth}{!}{ \includegraphics{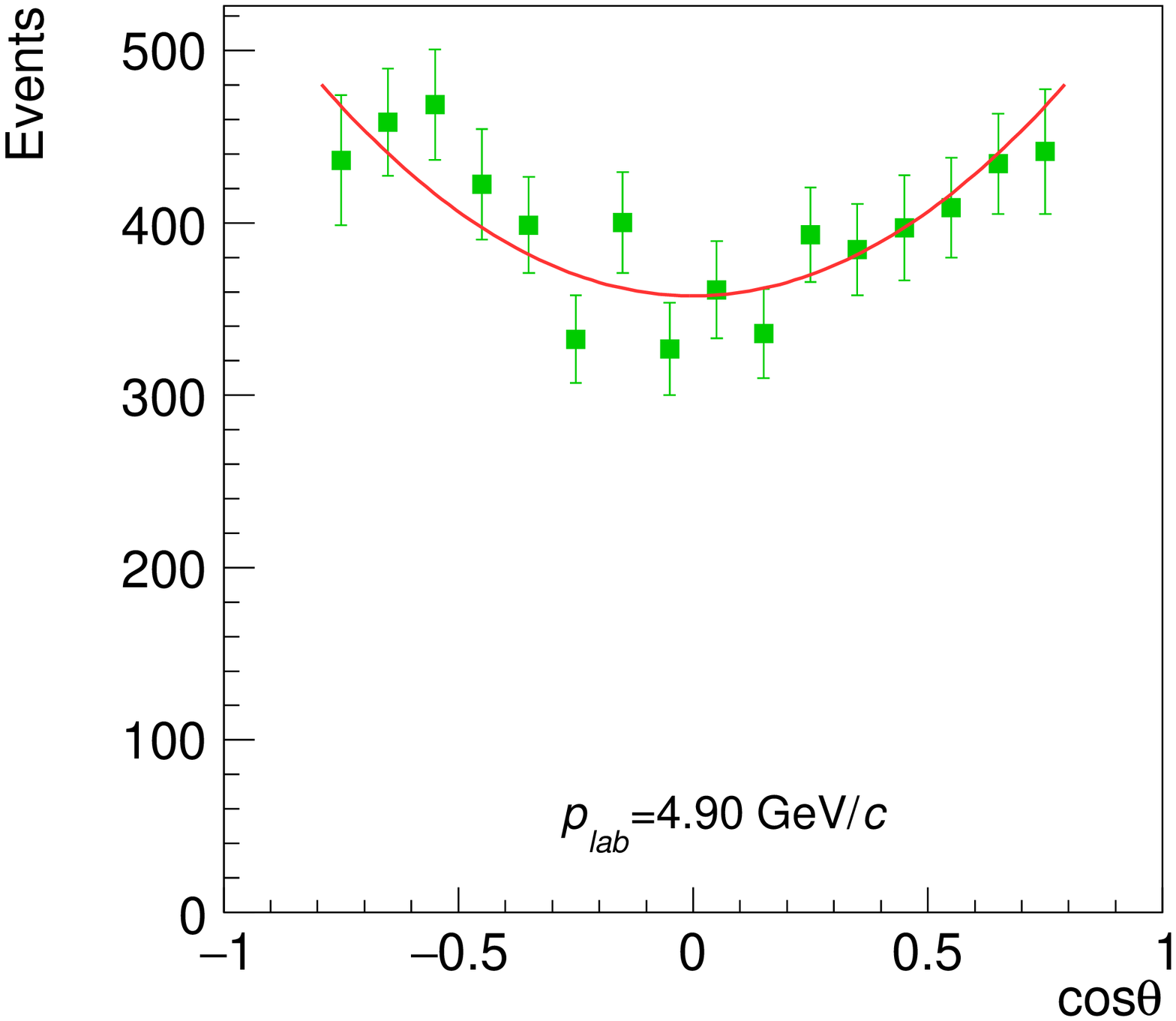} } } \\
  \subfloat[\label{fig:events59}]{\resizebox{0.4\textwidth}{!}{ \includegraphics{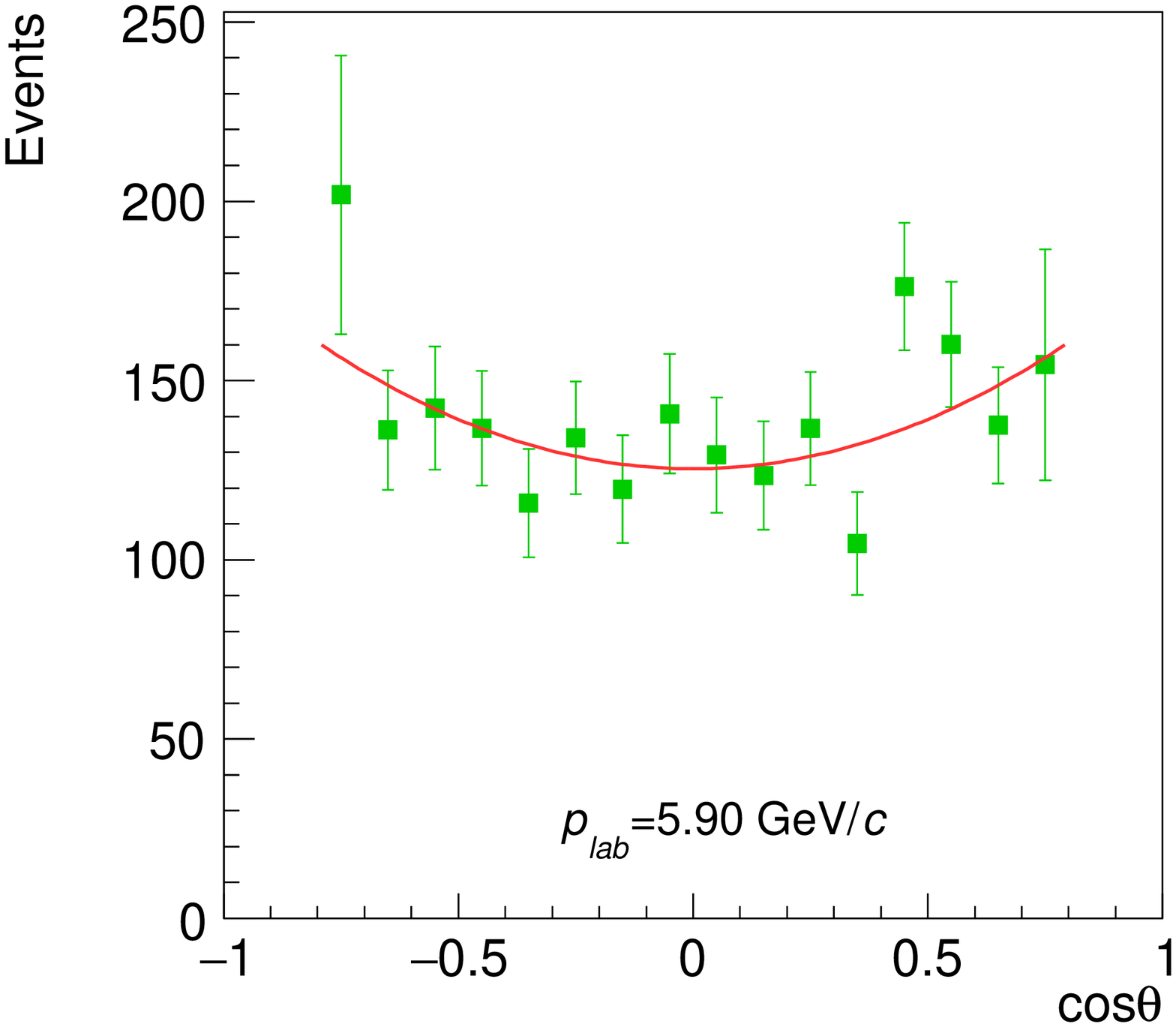} } }
  \subfloat[\label{fig:events64}]{\resizebox{0.4\textwidth}{!}{ \includegraphics{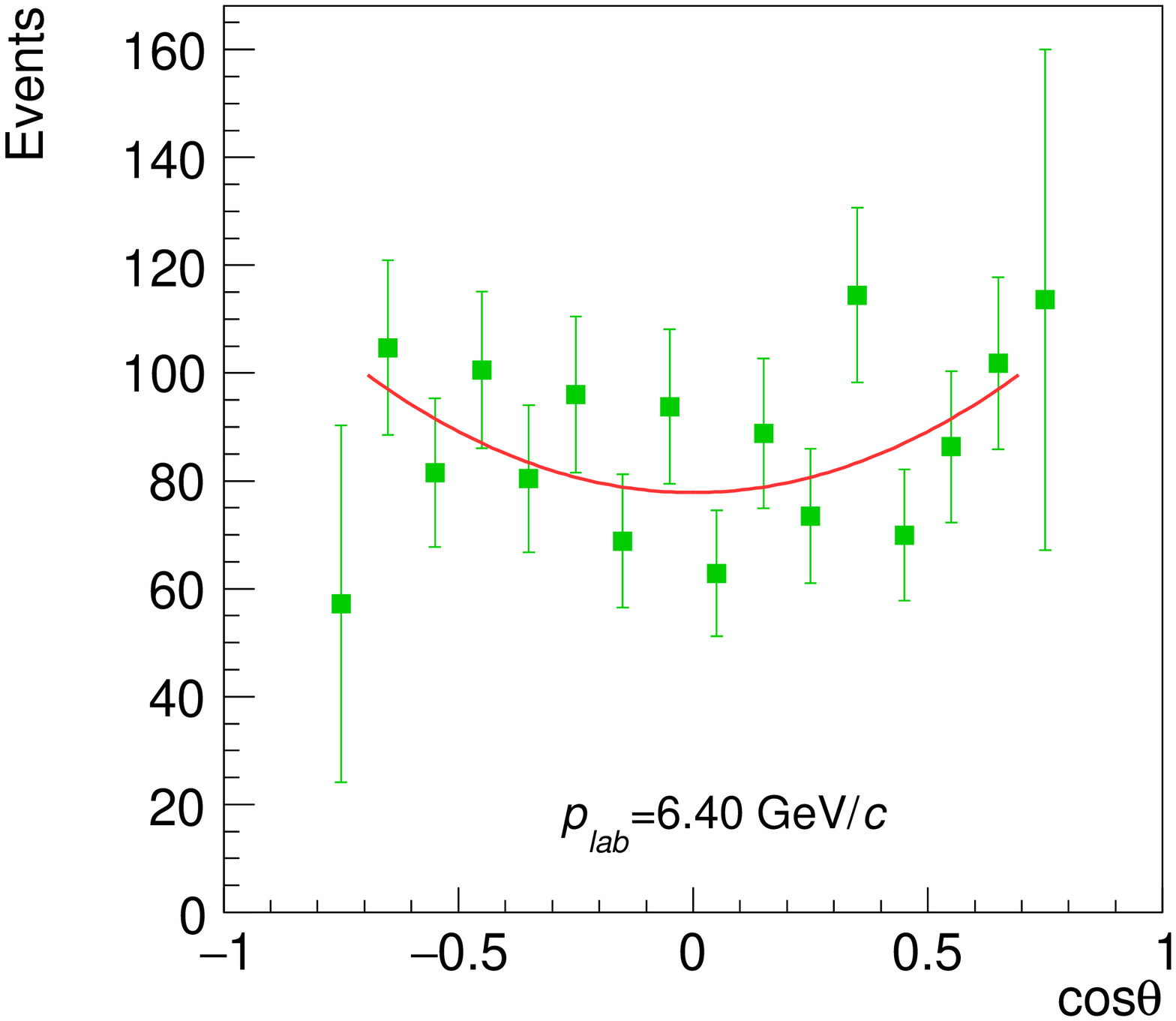} } }
  \caption{Reconstructed and efficiency-corrected angular distributions of generated electrons (green squares) and the fit (red line) for different $p_{lab}$ values: (a) 1.7 GeV/$c$, (b) 2.78 GeV/$c$, (c) 3.3 GeV/$c$, (d) 4.9 GeV/$c$, (e) 5.9 GeV/$c$, and (f) 6.4 GeV/$c$.}
  \label{fig:a2_events_r}
  \end{center}
\end{figure*}

To extract the FF ratio R, the reconstructed angular distributions first need to be corrected using the efficiency correction method described in Section~\ref{sec:a2_sig_eff}. As a second step of this procedure, the corrected angular distribution is fit using the following equation:
\begin{equation}
  \frac{d\sigma}{d\cos\theta} = \frac{\pi\alpha^2}{2\beta s}|G_{M}|^{2}\Big[(1+\cos^{2}\theta)+\frac{\mbox{R}^{2}}{\tau}\sin^{2}\theta\Big],
  \label{eq:a2_zichichi_cs_ratio}
\end{equation}
where R is a free fit parameter. Equation~(\ref{eq:eqff}) is used to calculate the value of $|G_M|$ for each $p_{lab}$. The reconstructed and acceptance corrected angular distribution for the electrons is shown together with the fitted curve in Fig.~\ref{fig:a2_events_r}. For low $p_{lab}$, where the cross section is higher, the fitted curve matches the shape of the angular distribution and the uncertainties are relatively small. At higher $p_{lab}$, the reconstructed angular data points fluctuate and have larger statistical uncertainties. For $p_{lab}=6.4$~(GeV/$c$)$^2$ the fit range is reduced to $|\cos\theta|<0.7$ because of the large uncertainties at $\cos\theta=\pm0.8$. The reduced $\chi^2$, \textit{i.e.} $\chi^2/NDF$ where \footnote{NDF is the number of degrees of freedom.} is close to unity for all $p_{lab}$, except the largest one where $\chi^2/NDF$ approaches 2.

\subsubsection{Individual extraction of $|G_E|$ and $|G_M|$}
To extract $|G_E|$ and $|G_M|$ individually, the differential cross sections are calculated assuming an integrated luminosity of $\mathcal{L}=2$ fb$^{-1}$, and using:
\begin{equation}
  \sigma_i=\frac{N^{corr}_{i}}{\mathcal{L}}\cdot\frac{1}{W_i},
  \label{eq:a2_counts_to_cs}
\end{equation}
where $W_i$ is the width of the $i$-th bin. The cross section uncertainty $\Delta\sigma$ is calculated in the following way:
\begin{equation}
  \Delta\sigma_i=\frac{1}{W_i}\frac{\Delta N^{corr}_{i}}{\mathcal{L}}.
  \label{eq:a2_counts_to_cs_uncertainties}
\end{equation}
Each differential cross section is fit using Eq.~(\ref{eq:eqdsigma}), which includes $|G_E|$ and $|G_M|$ as free parameters.

\subsubsection{Results}
\begin{figure}
  \resizebox{0.5\textwidth}{!}{ \includegraphics{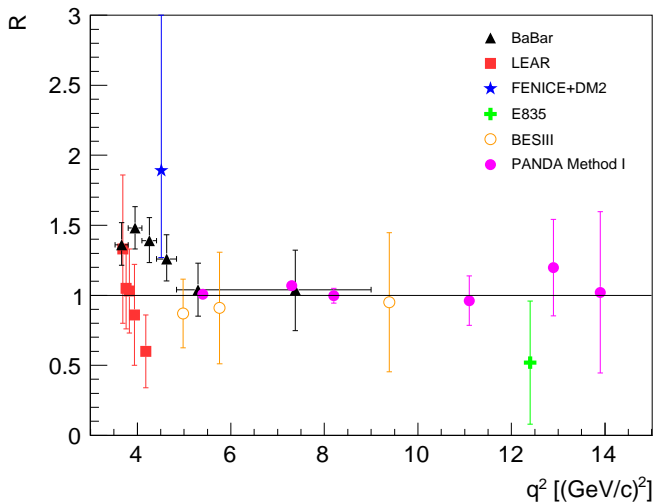} }
  \caption{Form factor ratio extracted from the present simulation (magenta circles) as a function of $q^2$, compared with the existing data. Data are from Ref.~\protect\cite{Bardin:1994am} (red squares), Ref.~\protect\cite{Lees:2013uta} (black triangles), Ref.~\protect\cite{PhysRevD.91.112004} (open orange circles), and Ref.~\protect\cite{Baldini:2005xx} (green cross and blue star).}
  \label{fig:a2_ff_ratio}
\end{figure}

After the fitting procedure, the ratio R and the individual values and the uncertainties of $|G_E|$ and $|G_M|$ are extracted from the fit. The extracted FF ratio is shown in Fig.~\ref{fig:a2_ff_ratio} as a function of $q^2$ together with results of other experiments. From this we conclude that the \PANDA experiment will be able to measure the FF ratio with a high statistical precision of around 1$\%$ at lower $q^2$. Furthermore, \PANDA will provide new measurements in the high $q^2$ domain with a statistical precision of up to 50\%.

\begin{figure*}
  \resizebox{0.5\textwidth}{!}{ \includegraphics{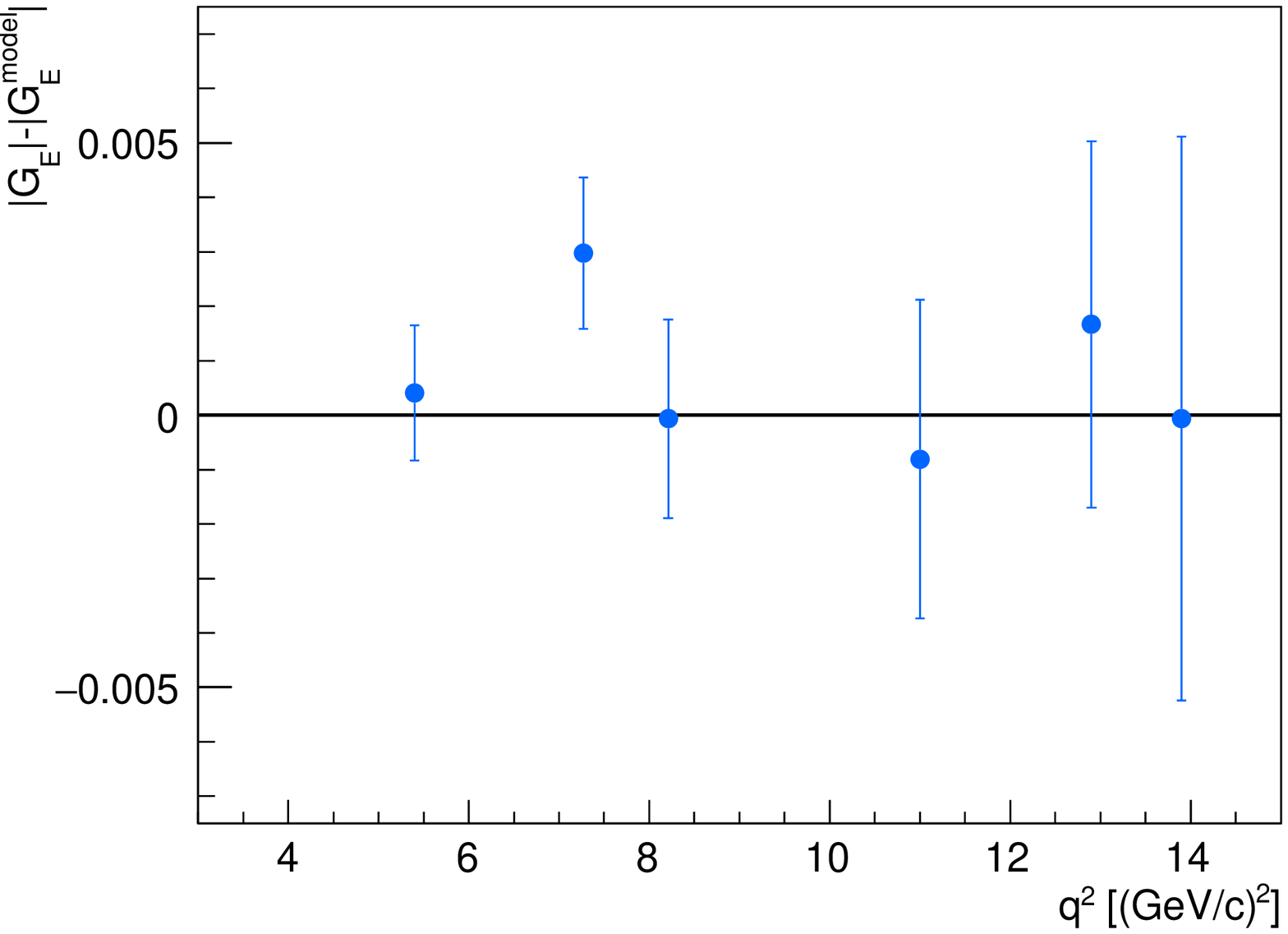} }
  \resizebox{0.5\textwidth}{!}{ \includegraphics{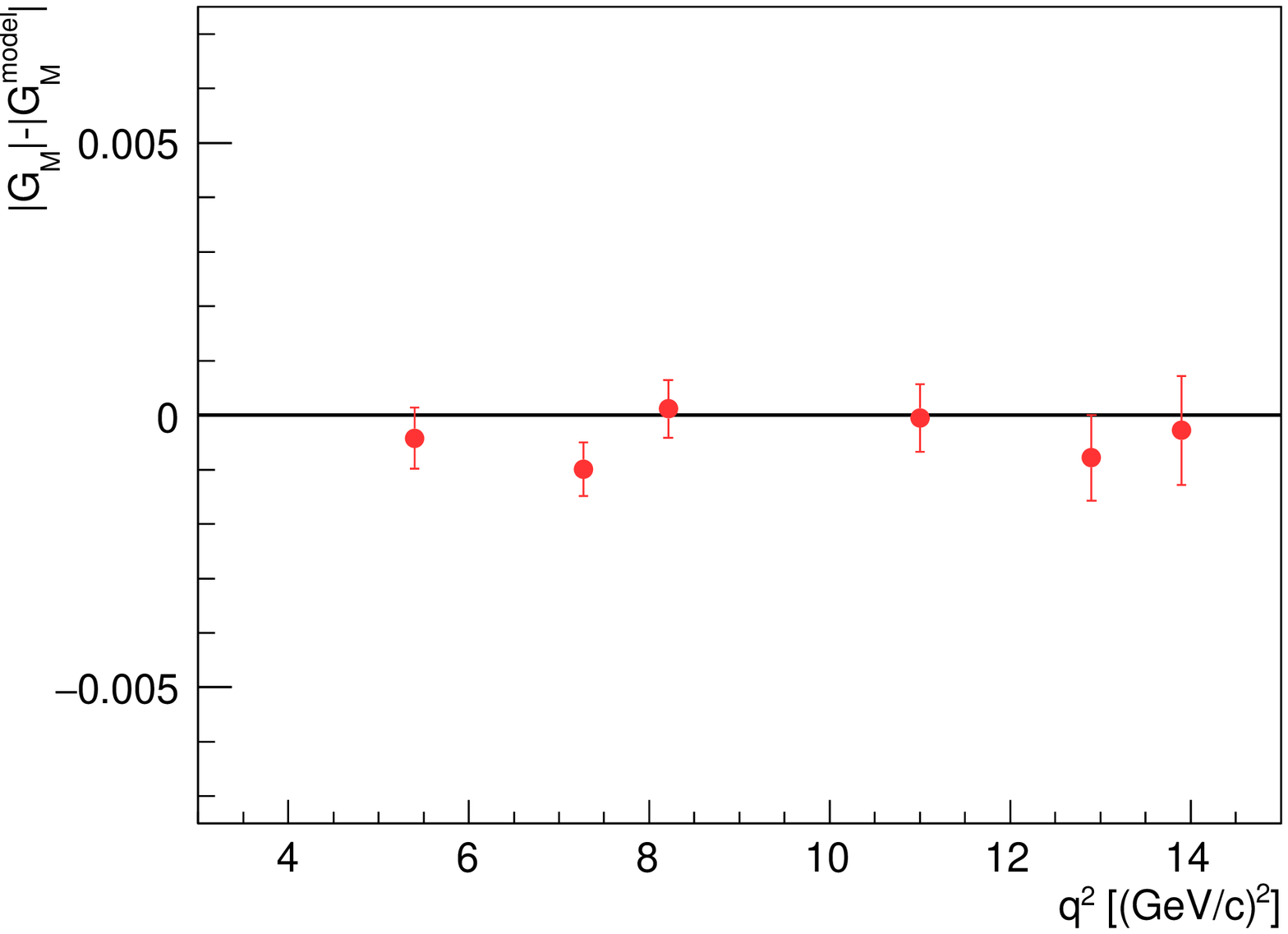} }
  \caption{Residual values of $|G_E|$ (left panel) and $|G_M|$ (right panel) for different $q^2$ values with statistical uncertainties only (Method I).}
  \label{fig:a2_ff_individual}
\end{figure*}

The difference between the expected values and the extracted values of $|G_{E}|$ and $|G_{M}|$ are shown in Fig.~\ref{fig:a2_ff_individual} along with their statistical uncertainties. $|G_M|$ can be measured with uncertainties within the range of $2\%$-$9\%$, whereas $|G_E|$ has uncertainties of about $3\%$-$45\%$. The difference in precision between $|G_E|$ and $|G_{M}|$ is due to the factor $\tau$ in the fit function. Table~\ref{tab:a2_FFs} shows the expected values and uncertainties of the extracted $|G_E|$, $|G_M|$, and R.

\begin{table}
  \setlength{\tabcolsep}{3.0pt} 
  \begin{center}
  \begin{tabular}{ l l l l} \hline
    $q^2$ & $\mbox{R}\pm\Delta \mbox{R}$ & $|G_E|\pm\Delta |G_E|$ & $|G_M|\pm\Delta |G_M|$ \\
    {[}(GeV/$c$)$^2${]} & & & \\ \hline
	  5.40        & 1.0065$\pm$0.0129 & 0.1216$\pm$0.0010 & 0.1208$\pm$0.0004 \\
	  7.27        & 1.0679$\pm$0.0315 & 0.0620$\pm$0.0013 & 0.0580$\pm$0.0004 \\
	  8.21        & 0.9958$\pm$0.0523 & 0.0435$\pm$0.0017 & 0.0437$\pm$0.0005 \\
	  11.12       & 0.9617$\pm$0.1761 & 0.0189$\pm$0.0029 & 0.0197$\pm$0.0006 \\
	  12.97       & 1.1983$\pm$0.3443 & 0.0148$\pm$0.0033 & 0.0123$\pm$0.0007 \\
	  13.90       & 1.0209$\pm$0.5764 & 0.0108$\pm$0.0051 & 0.0106$\pm$0.0010 \\ \hline
  \end{tabular}
  \caption{Expected values and uncertainties of the extracted R, $|G_E|$, and $|G_M|$ (Method I).}
  \label{tab:a2_FFs}
  \end{center}
\end{table}

\subsection{Method II}
\label{sec:methodii}
The second method enables the verification of the results and the investigation of systematic effects. The signal ($\bar p p \to e^+e^-$) is generated with the EvtGen generator using phase space (PHSP) angular distribution. One million events were generated at each of the three incident antiproton momentum values $p_{lab}=1.7$, 3.3, and 6.4 GeV/$c$, respectively. This enables an optimization of the event selection with respect to background suppression, predominately from $\bar p p \to \pi^+\pi^-$.

The signal and background ($\bar p p \to \pi^+\pi^-$) are analyzed in two steps. First, events with one positive and one negative particle are selected. If the event contains more than one positive or negative track, \textit{e.g.} secondary particles produced by the interaction between generated primary particles and detector material, the best pair is identified by selecting one positive and one negative track such that they are emitted closest to back-to-back in the c.m. reference frame. Second, reconstructed variables \textit{e.g.} momentum, deposited energy, and PID probabilities are studied for the selected events in order to achieve the most effective pion rejection. The latter step is explained below.

\subsubsection{PID probability and kinematic cuts}

The applied selection criteria are listed in Table~\ref{tabcutt}, for $p_{lab}=1.7$, 3.3, and 6.4 GeV/$c$. The sequential effects of all cuts, \textit{i.e.}, when applied in a sequence one after the other, as well as the individual impact of each cut, are reported for $p_{lab}=3.3$ GeV/$c$ in Table~\ref{tabcut}.

\begin{table}
\setlength{\tabcolsep}{3.5pt} 
  \begin{center}
  \begin{tabular}{ l l l l l } \hline
    $p_{lab}$                         & [GeV/$c$]  & $1.70$ & $3.30$ & $6.40$ \\ \hline
    PID$_c$ & [\%]                    & $>$99      & $>$99      & $>$99.5    \\
    PID$_s$ & [\%]                    & $>$10      & $>$10      & $>$10    \\
    $dE/dx_{STT}$ & [a.u.]            & $>$6.5     & $>$5.8     & 0 or $>$6.5   \\
    $E_{EMC}/p_{reco}$                & [GeV/(GeV/$c$)]   & $>$0.8     & $>$0.8     & $>$0.8   \\
    EMC LM      & -                   & $<$0.66    & $<$0.75     & $<$0.66  \\
    EMC E1 & [GeV]                    & $>$0.35    & $>$0.35    & $>$0.35  \\
    $|\theta+\theta'-180|$ & [degree] & \multicolumn{3}{ c }{<5}   \\
    $|\phi-\phi'-180|$ & [degree]     & \multicolumn{3}{ c }{<5}     \\
    $M_{inv}$ & [GeV/$c$$^2$]         & -          & $>$2.2     & $>$2.7   \\  \hline
  \end{tabular}
  \caption{Criteria used to select the signal ($e^+e^-$) and suppress the background ($\pi^+\pi^-$) events for each $p_{lab}$ value (Method II).}
  \label{tabcutt}
  \end{center}
\end{table}

\begin{table}[h!]\centering
\setlength{\tabcolsep}{2.5pt} 
\begin{tabular}{ l l l l l }
\hline
Cut & \multicolumn{2}{ l }{individual $\epsilon$} &  \multicolumn{2}{ l }{sequential $\epsilon$} \\ \hline
                    & background           & signal & background          & signal \\
Acceptance/tracking & 0.84                 & 0.86   & 0.84                & 0.86   \\
PID$_c$ +PID$_s$    & $0.35\times 10^{-5}$ & 0.70   & 0.29$\times 10^{-5}$ & 0.61   \\
$dE/dx_{STT}$       & 0.16                 & 0.95   & 0.19$\times 10^{-5}$ & 0.59   \\
$E_{EMC}/p_{reco}$  & 0.43$\times 10^{-3}$ & 0.94   & 0.11$\times 10^{-5}$ & 0.59   \\
EMC E1 + LM         & 0.02                 & 0.84   & 0.19$\times 10^{-6}$ & 0.53   \\
Kinematic cuts      & 0.95                 & 0.73   & 0.98$\times 10^{-8}$ & 0.45   \\ \hline
\end{tabular}
\caption{Individual and sequential efficiency for the signal and the background after the cuts, for $p_{lab}=3.3$ GeV/$c$ and $|\cos\theta|\leq 0.8$ (Method II).}
\label{tabcut}
\end{table}

\begin{figure}
	\centering
	\resizebox{0.50\textwidth}{!}{
	\includegraphics{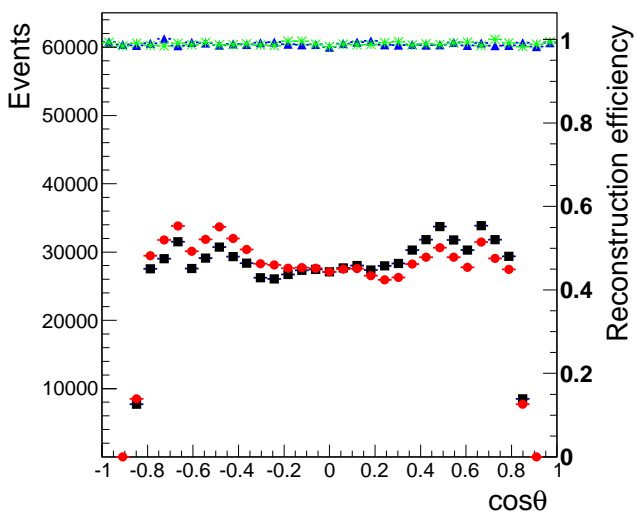}}
	\caption{Angular distribution in the c.m. system of the Monte Carlo events for $e^-$ (green asterisks) and $e^+$ (blue triangles) generated according to the PHSP, for $p_{lab}=3.3$ GeV/$c$. The reconstructed events after the cuts are also shown for $e^-$ (black squares) and $e^+$ (red circles). The right $y$-axis represents the efficiency values.}
	\label{mcemc}
\end{figure}

The selected reconstructed signal events, as well as the undistorted generated events, are shown in Fig.~\ref{mcemc} for $p_{lab}=3.3$ GeV/$c$. The intensity drop at $\cos\theta=0.65$ ($\theta_{lab}\sim 22.3^\circ$) is due to the transition region between the forward and the barrel EMC.

\subsubsection{Extraction of the signal efficiency}
The signal efficiency was extracted for each $\cos\theta$ bin from the events generated according to PHSP. The ratio between the number of PHSP events after cuts $R(\cos\theta)$ to the number of simulated MC events $M(\cos\theta)$ represents the signal efficiency as a function of the angular distribution. It can be written as
\be
\epsilon(\cos\theta)=\frac{R(\cos\theta)}{M(\cos\theta)}.
\label{eff}
\ee
The uncertainty $\Delta R(\cos\theta)=\sqrt{R(\cos\theta)}$ is attributed to the center of each bin of $R(\cos\theta)$. The uncertainty in the efficiency is calculated as follows:
\be
\Delta \epsilon(\cos\theta)= \sqrt{\epsilon(\cos\theta)\frac{(1-\epsilon(\cos\theta))}{R(\cos\theta)}}.
\label{efferr}
\ee

\begin{figure*}
  \resizebox{0.5\textwidth}{!}{ \includegraphics{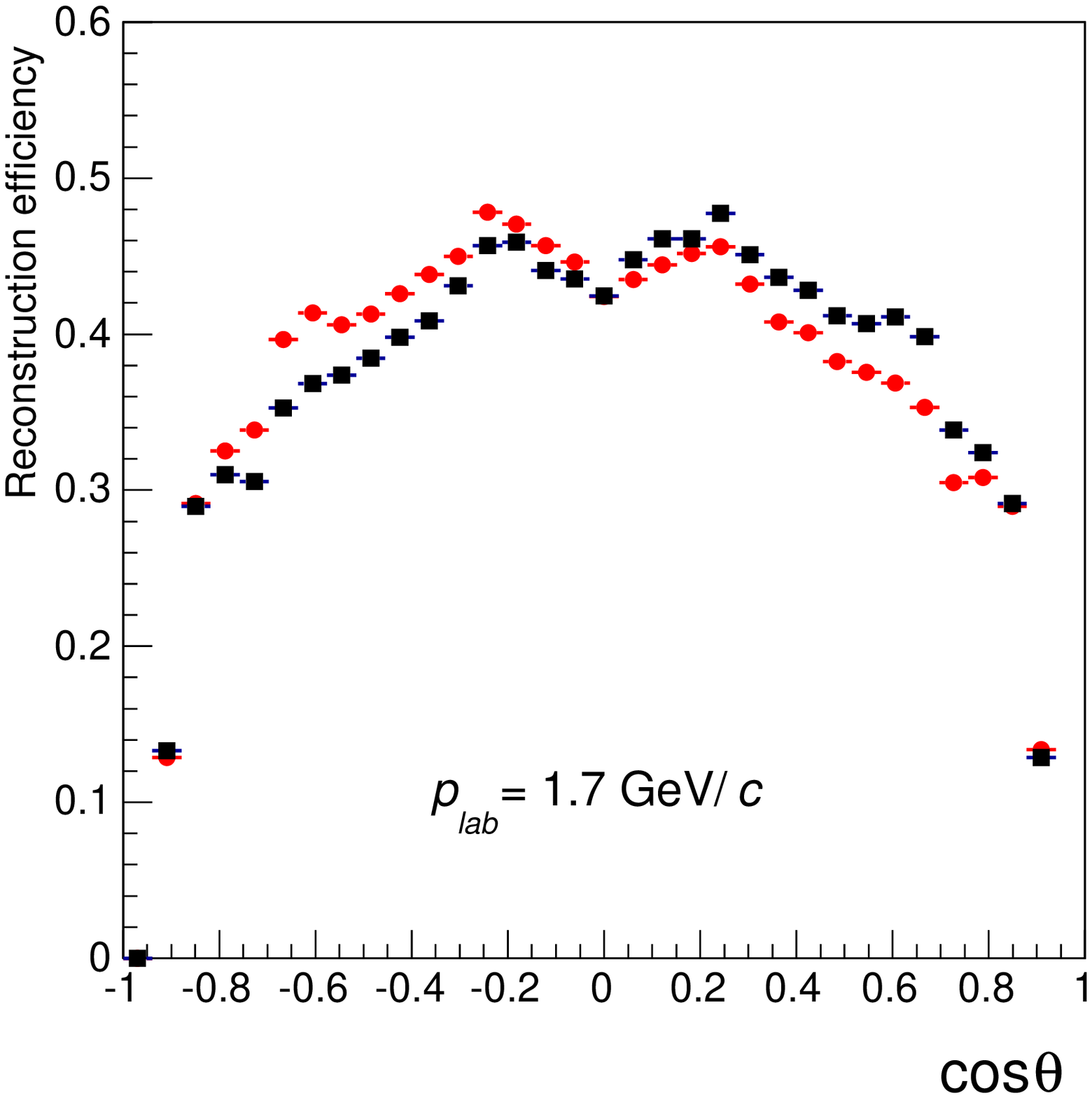} }
  \resizebox{0.5\textwidth}{!}{ \includegraphics{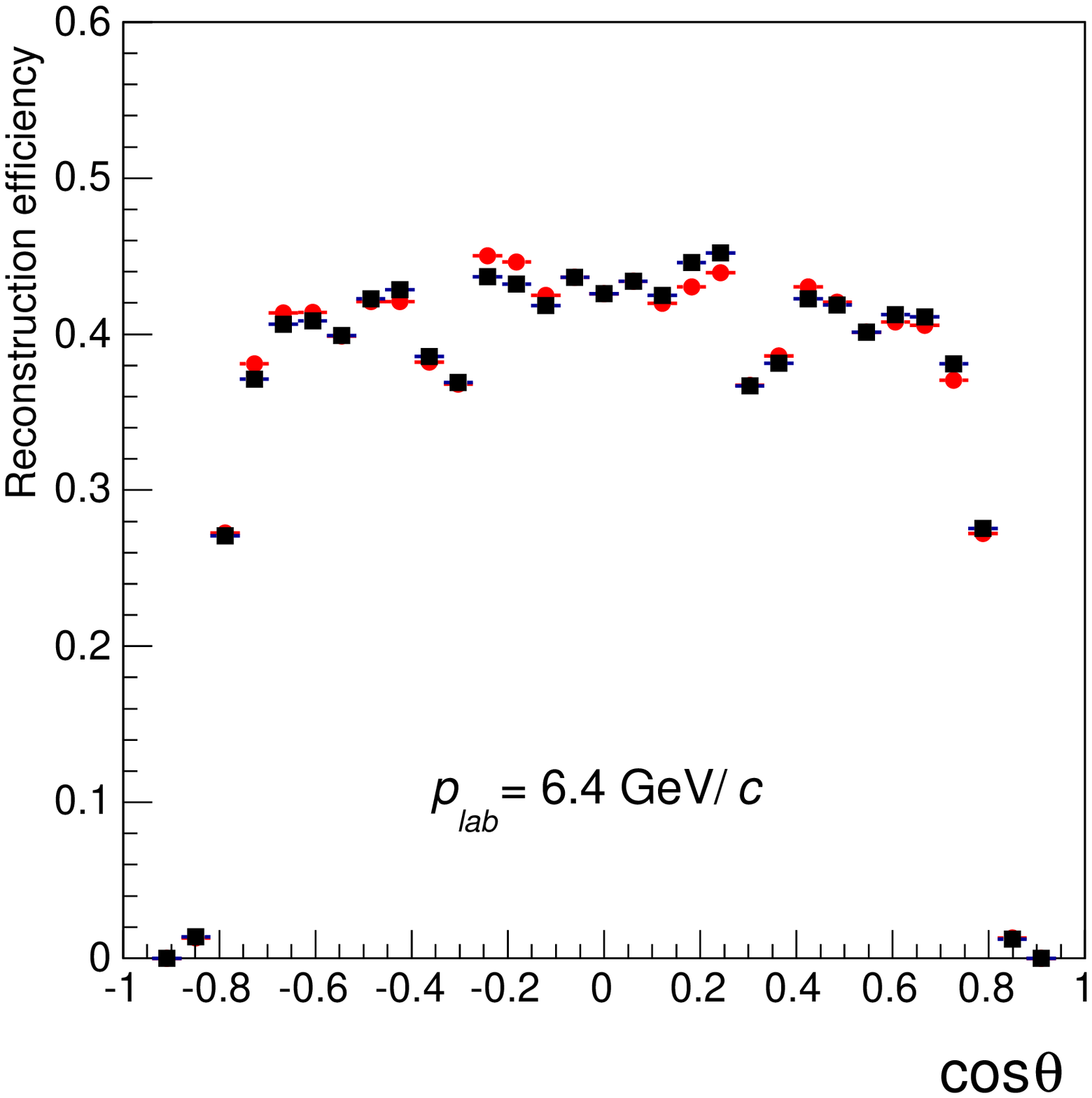} }
  \caption{Signal efficiency for $e^+$ (red circles) and $e^-$ (black squares) as a function of $\cos\theta$ in the c.m. system: (left panel) $p_{lab}=1.7$ and (right panel) 6.4 GeV/$c$.}
  \label{fig:effal}
\end{figure*}

The efficiency distribution of the signal as a function of $\cos\theta$ is shown in Figs.~\ref{mcemc} and \ref{fig:effal} for $p_{lab}=1.7$, 3.3, and 6.4 GeV/$c$. As mentioned above, the reaction mechanism for pion pair production changes as a function of energy and the shape of the efficiency varies with the energy. Due to the drop of the efficiency in the region $|\cos\theta|>0.8$, the analysis for the proton FF measurements is limited to the angular range $\cos\theta\in[-0.8,0.8]$. The integrated efficiency in this region is given in Table~\ref{table:eff}.

\begin{table}
  \begin{center}
  \begin{tabular}{ l l l l} \hline
    $p_{lab}$ [GeV/$c$] & $\epsilon(e^+e^-)$   & $\epsilon(\pi^+\pi^-)$ \\ \hline
    1.7                 & 0.41     & $1.9\times 10^{-8}$ \\
    3.3                 & 0.45     & $9.8\times 10^{-9}$ \\
    6.4                 & 0.41     & $1.9\times 10^{-8}$ \\ \hline
  \end{tabular}
  \caption{Efficiency of the signal ($e^+e^-$) and the background ($\pi^+\pi^-$) integrated over the angular range $|\cos\theta| \leq 0.8$ for each value of $p_{lab}$ (Method II).}
  \label{table:eff}
  \end{center}
\end{table}

\subsubsection{Simulation of events with a realistic angular distribution}
\label{sec:physevent}
The PHSP events have a flat $\cos\theta$ distribution in the c.m. system and do not contain the physics of the proton FFs. In reality, the angular distribution can be described according to Eq.~(\ref{eq:eqsth}). Therefore, the generated $\cos\theta$ histograms are rescaled by the weight $\omega(\cos\theta)=1+{\cal A}\,\cos\theta^2$, where ${\cal A}$ is given according to a model for $G_E$ and $G_M$. In the following, these events will be referred to as \textit{realistic events}.

Note that we do not expect any difference between the simulated electrons and positrons. In a one-photon exchange case, the angular distribution in the c.m. system is described by a forward-backward symmetric even function in $\cos\theta$. Possible contributions from radiative corrections (\textit{i.e.} two-photon exchange diagrams and interference between initial and final state photon emissions) can in principle introduce odd contributions in $\cos\theta$, leading to a forward-backward asymmetry. As it is a binary process, in the absence of odd contributions in the amplitudes, the detection of an electron at a definite value of $\cos\theta$ is equivalent to the detection of a positron at $\cos(\pi-\theta$). This is the case in the present simulation: a one-photon exchange is implied, and the photon emission, calculated with the PHOTOS~\cite{Was:2008zu} package, does not induce any asymmetry.

Once the \PANDA experiment is in operation, the efficiency will be validated using experimental data. In this analysis we have corrected for $e^+$ and $e^-$ MC generated events separately, since some asymmetry appears at the level of reconstruction. This is attributed to the different interaction of positive and negative particles with matter.

To determine the number of the realistic undistorted MC events ($P(\cos\theta)$) and the realistic reconstructed events ($W(\cos\theta)$), three other samples for the reaction $\bar p p \to e^+ e^-$ are generated using the PHSP angular distribution at each $p_{lab}$. The $P(\cos\theta)$ and $W(\cos\theta)$ histograms are rescaled by the weight $w(\cos\theta)$ for the case $\mbox{R}=1$,~${\cal A}=(\tau-1)/(\tau+1)$.

\subsubsection{Normalization: observed events}

The reconstructed realistic events, $W(\cos\theta)$, are normalized according to the integrated count rate $N_{int}(e^+e^-)$ given in Table~\ref{table:kincount}. The integrated count rate depends on the energy of the system and on the luminosity. The number of observed events, $O(\cos\theta)$, is expected to be
\be
O(\cos\theta)=W(\cos\theta)\cdot\frac{N_{int}(e^+ e^-)}{\int_{-0.8}^{0.8}P(\cos\theta) d\cos\theta},
\ee
with an uncertainty $\Delta O(\cos\theta)= \sqrt{O(\cos\theta)}$ since the experimental uncertainty will finally be given by the accumulated statistics of the detected events.

\subsubsection{Efficiency correction and fit}
\label{sec:fit}
\begin{figure}
	\centering
	\resizebox{0.50\textwidth}{!}{
	\includegraphics{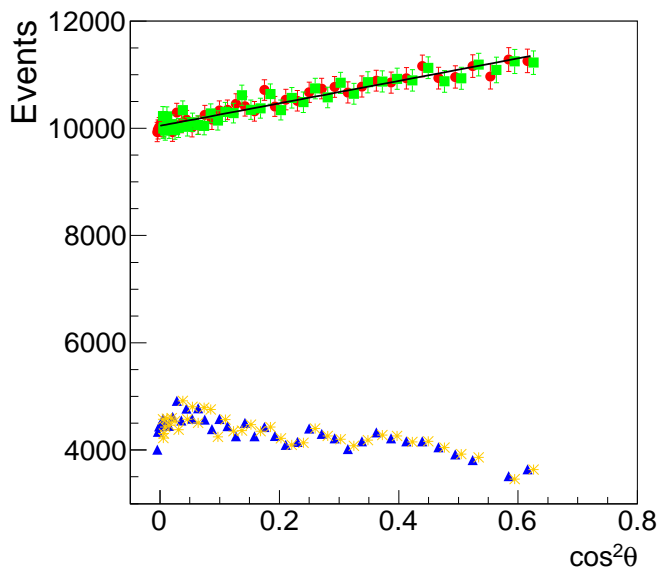}}
	\caption{Observed number of events before efficiency correction $O(\cos\theta)$ (forward events (blue triangles) and backward events (orange stars)) and after efficiency correction $F(\cos\theta)$ (forward events (red circles) and backward events (green squares)), as a function of $\cos^2\theta$ for $p_{lab}=1.7$ GeV/$c$, assuming $\mbox{R}=1$. The solid black line represents the linear fit. The abscissa for backward (forward) events is shifted by +0.005 (-0.005) for better visualization.}
\label{fit1p}
\end{figure}

The fit procedure was applied to the observed events after the correction by the efficiency, $F(\cos\theta)$:
\be
	F(\cos\theta)=O(\cos\theta)/\epsilon(\cos\theta),
\ee
\be
	\frac{\Delta F(\cos\theta)}{F(\cos\theta)}= \sqrt{\left(\frac{\Delta O(\cos\theta)}{O(\cos\theta)}\right)^2+\left(\frac{\Delta \epsilon(\cos\theta)}{\epsilon(\cos\theta)}\right)^2},
\ee
For each $p_{lab}$ value, the distribution $F(\cos\theta)$ as a function of $\cos^2\theta$ was fit with a two-parameter function. The linear fit function is:
\be
	y=a+b x,~\mbox{ with } x=\cos^2\theta,~a\equiv \sigma_0,~b\equiv \sigma_0 {\cal A},
	\label{eq:eqlin}
\ee 
where $a$ and $b$ are the parameters to be determined by minimization. They are related to the FFs, through Eq.~(\ref{eq:eqsth}) and Eq.~(\ref{eq:eqsa}). The number of observed events, before and after the efficiency correction are shown in Fig.~\ref{fit1p} for $p_{lab}=1.7$ GeV/$c$.

From the measured angular asymmetry, the FF ratio R can be calculated using:
\be
	\mbox{R}=\sqrt{\tau\frac{1-{\cal A}}{1+{\cal A}}}.
	\label{eqr}
\ee
In the limit of small uncertainties, provided that first order statistical methods work, the uncertainty in R can be obtained from standard uncertainty propagation on ${\cal A}$:
\be
	\Delta \mbox{R}=\frac{1}{\mbox{R}}\frac{\tau}{(1+{\cal A})^2} \Delta {\cal A}.
	\label{eqrd}
\ee
The results of the fit are reported as a function of $q^2$ in Table~\ref{tab:results}. The relative uncertainty on the proton FF ratio increases from about $1.4\%$ at the low energy point to $40\%$ at $q^2=13.90$ (GeV/$c$)$^2$.

\begin{center}
	\begin{table}[th]\centering
	\begin{tabular}{ l l l l l }
	\hline
  $q^2$  [(GeV/$c$)$^2$] &  $\mbox{R}$ & $ \Delta  \mbox{R}$ &  ${\cal A}$   & $ \Delta {\cal A}$ \\ \hline
  5.40                   & 1 &  0.014 & 0.210  & 0.014 \\
  8.21                   & 1 &  0.050 & 0.400  & 0.042 \\
  13.9                   & 1 &  0.407 & 0.590  & 0.264 \\ \hline
	\end{tabular}
	\caption{Expected statistical uncertainties on the angular asymmetry and the proton FF ratio, for different $q^2$ values assuming an integrated luminosity of 2 fb$^{-1}$ (Method II). The second and the fourth columns are the theoretical values (simulation inputs). The third and the fifth columns are the results of the fit on the uncertainties. The statistical uncertainties are extracted in the angular range $|\cos\theta|\leq0.8$.}
	\label{tab:results}
	\end{table}
\end{center}

\begin{figure}\centering
	\resizebox{0.50\textwidth}{!}{
	\includegraphics{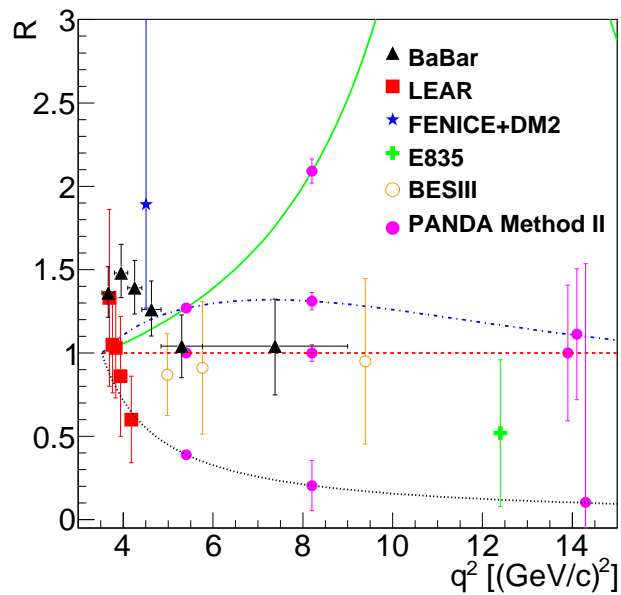}}
	\caption{Expected statistic precision on the determination of the proton FF ratio from the present simulation (magenta circles) as a function of $q^2$, compared with the existing data. Data are from Ref.~\protect\cite{Bardin:1994am} (red squares), Ref.~\protect\cite{Lees:2013uta} (black triangles), Ref.~\protect\cite{PhysRevD.91.112004} (open orange circles), and Ref.~\protect\cite{Baldini:2005xx} (green cross and blue stars). The statistical uncertainties are extracted in the angular range $|\cos\theta|\leq 0.8$. Curves are the graphic representations related to the theoretical predictions, as explained in the text.}
	\label{fig:ratio13mod}
\end{figure}

The advantage of the PHSP based generator method is that the same simulation can be used to test different models, by weighting each event according to the applied model. The points obtained from the present simulations and the published world data on the proton FF ratio are shown in Fig.~\ref{fig:ratio13mod} for the different values of R predicted by theoretical models. The curves are theoretical predictions from vector dominance model (solid green line), extended Gary-Kr\"{u}mpelmann (dash-dotted blue line), and a naive quark model (dashed red line) where the parameters have been adjusted as in Ref.~\cite{TomasiGustafsson:2005kc}, and Ref.~\cite{Kuraev:2011vq} (dotted black line). For a fixed energy point, the relative uncertainty in the proton FF ratio increases when the the ratio approaches zero, giving a meaningless value at the highest beam momentum $p_{lab}=6.4$ GeV/$c$.

\subsubsection{Statistical uncertainties on $|G_E|$ and $|G_M|$}

The individual determination of $|G_E|$ and $|G_M|$ was obtained from a two-parameter fit to the efficiency corrected histograms, as defined in Section \ref{sec:fit}:
\be
	y = a+b\cos^2\theta,
	\label{eq:eqa2}
\ee
where $a$ and $b$ are the two fit parameters. Based on Eq.~(\ref{eq:eqsth}), $|G_E|$ and $|G_M|$ are extracted from $a=\sigma_0 {\cal L}$ and $b=\sigma_0 {\cal A} {\cal L}$ by:
\ba
	|G_M|^2&=&\displaystyle\frac{a+b}{2\cal N},~
	|G_E|^2=\tau\displaystyle\frac{a-b}{2\cal N},\\\nn
	{\cal N}&=&\frac{\mathrm{\pi} \alpha^2}{2\beta s} {\cal L}.
	\label{eq:eqAME}
\ea
Their uncertainties are obtained by:

\ba
	\Delta |G_M|^2&=& \frac{1}{2\cal N} \sqrt{ (\Delta a)^2 +(\Delta b)^2},\nn\\
	\Delta |G_E|^2&=& \frac{\tau}{2\cal N} \sqrt{ (\Delta a)^2 +(\Delta b)^2},
\ea
where $\Delta a$ and $\Delta b$ are the statistical uncertainties of $a$ and $b$, obtained from the fit, respectively.
The results of the fits are reported in Table~\ref{tab:resultsgegm}. The simulation input values of $|G_E|$ and $|G_M|$ are reproduced within the uncertainty ranges.

\begin{table}
	\begin{center}
	\begin{tabular}{ l l l l }
	\hline
	$q^2$  [(GeV/$c$)$^2$]  & $|G_M|=|G_E|$ & $ \Delta |G_M|$ & $ \Delta |G_E|$ \\ \hline
	5.40                    & 0.1212 & 0.0007  &  0.0010 \\
	8.21                    & 0.0438 & 0.0007  &  0.0002 \\
	13.9                    & 0.0109 & 0.0010  &  0.0035 \\\hline
	\end{tabular}
	\caption{Expected statistical uncertainties in the proton FFs, for different $q^2$ values (Method II). The second column is the theoretical value (simulation input). The third and fourth columns are the results of the fit. The statistical uncertainties are extracted in the angular range $|\cos\theta|\leq0.8$.}
	\label{tab:resultsgegm}
\end{center}
\end{table}

\subsubsection{Results with the full and reduced luminosity mode}

FAIR is designed to provide instantaneous luminosities up to $2 \times 10^{32}$ cm$^{-2}$ s$^{-1}$. The results presented in this work correspond to an integrated luminosity of 2 fb$^{-1}$ per beam momentum setting, which can be accumulated in about 4 months of data taking at the design luminosity. With this integrated luminosity, the proton FF ratio ($\mbox{R}=1$) can be determined with a relative statistical precision of 1.4\%, 5\% and 40.7\% at $q^2=5.40$, 8.21, and 13.90 (GeV/$c$)$^2$, respectively (Table~\ref{tab:results}). A separate measurement of $|G_E|$ and $|G_M|$ is possible. The relative uncertainty on $|G_E|$ ($|G_M|$) increases from about 0.8\% (0.6\%) at $q^2=5.40$ (GeV/$c$)$^2$ to 37\% (9\%) at $q^2=13.90$ (GeV/$c$)$^2$ (Table~\ref{tab:resultsgegm}). The proton effective FF, $|F_p|$ (see Eq.~(\ref{eq:Fp1})), can be determined in the region below $q^2=13.90$ (GeV/$c$)$^2$ with a few percent statistical error. The measurement of $|F_p|$ can be extended to higher $q^2$ values according to the experimental efficiency.

In the startup phase of the \PANDA experiment, a luminosity about 20 times lower than that proposed in this paper is expected. With an integrated luminosity of 0.2 fb$^{-1}$ (8 months of data taking with the reduced scenario), the statistical uncertainty on the proton FF ratio increases by a factor of $\sim\sqrt{10}$ compared to the full luminosity mode. As a consequence, the upper limit of the measurable FF range will be reduced to $\sim$10 (GeV/$c$)$^2$. At $q^2=10$ (GeV/$c$)$^2$ the expected precision of the FF ratio will be around $40\%$ (estimated using Method II).

\subsection{Comparison}
Both methods demonstrate consistent results at $q^2=5.4$ and 8.2 (GeV/$c$)$^2$ where the expected statistics is relatively high. At $q^2=13.9$ (GeV/$c$)$^2$ Method I gives a larger statistical uncertainty than Method II: At this energy point, the number of events is small and, as a consequence, the statistical fluctuations are important. This is not taken into account in Method II, where about $10^6$ events have been generated and rescaled. Indeed, the statistical fluctuations are arbitrary, and by repeating Method I multiple times one can obtain a Gaussian distribution of the statistical uncertainty, where the mean, the most probable value, is equal to the uncertainty value obtained with Method II. The uncertainty distribution of the angular asymmetry ${\cal{A}}$ is verified to be Gaussian symmetric up to $q^2=13.9$ (GeV/$c$)$^2$.

\section{Systematic uncertainties}
\label{sec:systematic}

Since a full systematic study requires both experimental data and MC, we are limited in our ability to estimate every possible source. Therefore, in the following we will discuss some of the sources of systematic uncertainties which can be tested with MC only. A more precise estimation of systematic uncertainties will not be feasible until the design and construction of the detector is completed.

\subsection{Luminosity measurement}
 
The \PANDA experiment will use $\bar{p}p$ elastic scattering for the luminosity measurement. Based on Ref.~\cite{LumiTDR} the systematic uncertainty on the luminosity measurement might vary from 2\% to 5\%, depending on the beam energy, the $\bar{p}p$ elastic scattering parameterization, and $\bar{p}p$ inelastic background contamination. We considered the relative systematic luminosity uncertainty $\Delta{\cal L}/{\cal L}$ to be 4.0\% for all beam momenta. Table~\ref{tab:systematic} shows the impact of the luminosity uncertainty on the precise extraction of $|G_E|$ and $|G_M|$.

\subsection{Detector alignment}

Thanks to the almost 4$\pi$ acceptance of the \PANDA detector, misalignments of its different components will not affect the determination of proton FFs. Known displacements will be corrected for during the reconstruction of the raw data. The effect of small displacements up to a few hundred micrometers will give rise to spatial uncertainties that are much smaller than the foreseen uncertainties from the tracking resolution.

\subsection{Pion background}

Using the achieved background rejection factor listed in Table~\ref{tab:a2_eff}, we can estimate the effect of misidentification of background events contaminating the signal. The analysis in Section~\ref{sec:analysis} was repeated with signal and background events mixed together. The number of added background events was calculated in accordance with the achieved background suppression.

In addition, the cross section of the background channel $\bar{p}p \rightarrow\pi^+\pi^-$ will be measured at \PANDA with a very high precision due to its large cross section. Therefore, systematic uncertainties due to the model of the background differential cross section used in simulations are expected to be negligible. The impact of the background on the FFs precision is shown in Table~\ref{tab:systematic}. 

\subsection{Sensitivity to odd \mbox{\boldmath$\cos\theta$} contributions}
The analysis above assumes even $\cos\theta$ angular distributions, as expected from the one-photon exchange mechanism. However, odd contributions may be present in the data. One example is the presence of the two-photon exchange (TPE) mechanism, which may play a role at large $p_{lab}$ and give rise to odd $\cos\theta$ terms in the angular distribution (see Refs.~\cite{Guttmann:2011ze,Pacetti:2015iqa}).

In the presence of TPE, the matrix element of the reaction $\bar p p \to e^+ e^-$ contains three complex amplitudes: $\tilde{G}_E$, $\tilde{G}_M$ and $F_3$ \cite{Gakh:2005wa}, instead of two FFs. The differential cross section of $\bar p p \to e^+ e^-$ including the TPE contributions can be approximated as:
\be
	\frac{d\sigma}{d\cos\theta} = \frac{\pi\alpha^2}{2 q^2} \sqrt{\frac{\tau}{\tau-1}}D,
	\label{eq:tpe}
\ee
with radiative corrections or TPE
\ba
D=&(&1+\cos^2\theta)|G_M|^2+\displaystyle\frac{1}{\tau}\sin^2\theta|G_E|^2 \nn\\
&+& 2\sqrt{\tau(\tau-1)} \left(\frac{G_E}{\tau}-G_M\right)F_3 \cos\theta\sin^2 \theta.
\ea
The three amplitudes, denoted $G_E$, $G_M$ and $F_3$, are considered as real functions of $q^2$ since their relative phases are not known. In the following, we investigate the limit of a detectable odd $\cos\theta$ contribution. If present in the data, the source of the asymmetry may be either more complicated underlying physics, \textit{e.g.} radiative corrections, TPE, or experimental artifacts, \textit{e.g.} non-symmetric detection of leptons which haven't been properly corrected for. The MC histograms (produced according to the PHSP model) are rescaled according to Eq.~(\ref{eq:tpe}) with $\mbox{R}=1$ and $F_3/G_M=0$, 0.02, 0.05 and 0.20. We fit the angular distributions by the function:
\be
y=a_0 +a_1 \cos^2\theta+a_2 \cos\theta(1-\cos^2\theta),
\label{eq:method2fit}
\ee
where $a_2$, directly related to the ratio $F_3/G_M$, gives the relative size of odd contributions. The results of the fit are reported in Table~\ref{tab:tpe}. Below $F_3/G_M=0.05$, $a_2$ is compatible with zero, indicating a sensitivity to an asymmetry larger than 5\% at $q^2=5.4$ (GeV/$c$)$^2$. The extraction of $\mbox{R}$ and ${\cal{A}}$ is not affected by the relative size of $F_{3}/G_{M}$.

In the case of a charge symmetric detection of electrons and positrons, the interference term between the one- and two-photon-exchange channels will not contribute to the differential cross section \cite{Gakh:2005wa}. Since \PANDA will be able to detect both electrons and positrons (exclusive processes) and the contribution of TPE is symmetric between them, the TPE contribution can be eliminated by adding electron and positron angular distributions.

\begin{table}\centering
	\setlength{\tabcolsep}{3.5pt} 
	\begin{tabular}{ l l l l l l} \hline
    	                                  & $q^2$ & Stat    & \multicolumn{2}{ c }{Systematic} \\
  	                                    & [(GeV/$c$)$^2$] &       & Bg    & Lumi  & Total \\ \hline
	\multirow{3}{*}{$\Delta |G_E|/|G_E|$} & 5.40            & 0.9\% & 0.3\% & 2.0\% & 2.2\% \\
    	                                  & 8.21            & 4.1\% & 2.9\% & 2.0\% & 5.4\% \\
  	                                    & 13.9           & 48\%  & 3.1\% & 2.0\% & 48\% \\ \hline
	\multirow{3}{*}{$\Delta |G_M|/|G_M|$} & 5.40            & 0.4\% & 2.8\% & 2.0\% & 3.5\% \\
    	                                  & 8.21            & 1.2\% & 1.1\% & 2.0\% & 2.6\% \\
  	                                    & 13.9           & 9.4\% & 1.0\% & 2.0\% & 9.7\% \\ \hline
	\multirow{3}{*}{$\Delta$ R/R}         & 5.40            & 1.3\% & 2.9\% & n/a   & 3.3\% \\
	                                      & 8.21            & 5.3\% & 4.0\% & n/a   & 6.6\% \\
	                                      & 13.9           & 56\%  & 4.1\% & n/a   & 57\% \\ \hline
	\end{tabular}
	\caption{Effect of systematic and statistical uncertainties, as well as their total contribution, on the precision of $|G_E|$, $|G_M|$, and R (Method I).}
	\label{tab:systematic}
\end{table}

\begin{table*}[th!]\centering
\begin{tabular}{ l l l l l l l}
\hline
$q^2$  [(GeV/$c$)$^2$]  & $F_{3}/G_{M}$ [\%] & $a_{0}$ &  $a_{1}$ & $a_{2}$ & $\Delta R$ &  $\Delta{\cal{A}}$ \\ \hline
5.40  & 0  & 10045$\pm$34 & 2080$\pm$130 & 76$\pm$83     & 0.014 & 0.01 \\
5.40  & 2  & 10045$\pm$34 & 2080$\pm$130 & -0.6$\pm$83   & 0.014 & 0.01 \\
5.40  & 5  & 10045$\pm$34 & 2080$\pm$130 & -115$\pm$83   & 0.014 & 0.01 \\
5.40  & 20 & 10045$\pm$34 & 2080$\pm$130 & -687$\pm$83   & 0.014 & 0.01 \\ \hline
8.21  & 0  & 579$\pm$6    & 231$\pm$24   & -3.5$\pm$14.8 & 0.05  & 0.04 \\
8.21  & 2  & 579$\pm$6    & 231$\pm$24   & -19.9$\pm$15  & 0.05  & 0.04 \\
8.21  & 5  & 579$\pm$6    & 231$\pm$24   & -44.3$\pm$15  & 0.05  & 0.04 \\
8.21  & 20 & 579$\pm$6    & 231$\pm$24   & -166$\pm$15   & 0.05  & 0.04 \\ \hline
13.9  & 0  & 17$\pm$1     & 10$\pm$4.2   & -0.1$\pm$2.6  & 0.4   & 0.26 \\
13.9  & 2  & 17$\pm$1     & 10$\pm$4.2   & -1.4$\pm$2.6  & 0.4   & 0.26 \\
13.9  & 5  & 17$\pm$1     & 10$\pm$4.2   & -3.4$\pm$2.6  & 0.4   & 0.26 \\
13.9  & 20 & 17$\pm$1     & 10$\pm$4.2   & -13.4$\pm$2.6 & 0.4   & 0.25 \\ \hline
\end{tabular}
\caption{The results from the fit of the angular distributions using Eq.~(\ref{eq:method2fit}), for $q^2=5.40$, 8.21 and 13.90 (GeV/$c$)$^2$.}
\label{tab:tpe}
\end{table*}

\subsection{Contribution to FFs}

The contributions of the luminosity and background to the precision of extracted values of FFs are reported in Table~\ref{tab:systematic} together with the statistical contribution. The background contamination is on the level of a few percent for all values of $q^2$. The luminosity uncertainty affects only $|G_E|$ and $|G_M|$, since the luminosity measurement is not needed for R determination. At lower $q^2$ values, where the number of signal events is relatively large, the total uncertainty is dominated by the background contamination and luminosity contributions. In the intermediate energy domain, the amount of statistics decreases and affects the total uncertainty on the same level as systematic uncertainties. At higher $q^2$, the main contribution to the total uncertainty is given by the statistical uncertainty due to the small signal cross section.

\section{Competitiveness of the \PANDA experiment}
\label{sec:competitiveness}

The moduli of the individual FFs, $|G_{E}|$ and $|G_{M}|$, will be measured for the first time at BESIII using the data collected at 20 different $q^2$ values between 4.0 and 9.5 (GeV/$c$)$^2$ \cite{ProposalHuang}. A statistical precision on the FF ratio between $9\%$ and $35\%$ is expected. Based on these numbers, it is clear that \PANDA will extend these measurements up to about $q^2=14$ (GeV/$c$)$^2$, with a precision better than that expected at BESIII or comparable in the case of the reduced luminosity mode.

The modulus of the proton FFs ratio can also be measured at Belle \cite{Abashian:2000cg}, using the initial state radiation (ISR) technique, with a comparable accuracy to the BABAR data. The Belle detector was operating on the KEKB $e^+e^-$ collider \cite{Akai:2001pf}. An integrated luminosity of about 1040 fb$^{-1}$ was collected at KEKB between 1999 and 2010. Most of the data were taken at the $\Upsilon(4S)$ resonance. The upgraded facility of the KEKB collider (SuperKEKB) aims to accumulate 50 ab$^{-1}$ by about 2025 \cite{Ferber:2015}. The Belle II experiment may provide the most accurate data on the proton FF ratio. So far, no estimation has been presented by the collaboration for the FF measurement at Belle and Belle II. One disadvantage of the ISR technique is that it only allows extraction of FFs in wide bins of $q^2$. This is in contrast to the formation reaction that \PANDA will use, where  the precision of $q^2$ is given, in general, by the very precise beam momentum resolution.

\section{Conclusion}

\begin{figure}\centering
\resizebox{0.50\textwidth}{!}{
\includegraphics{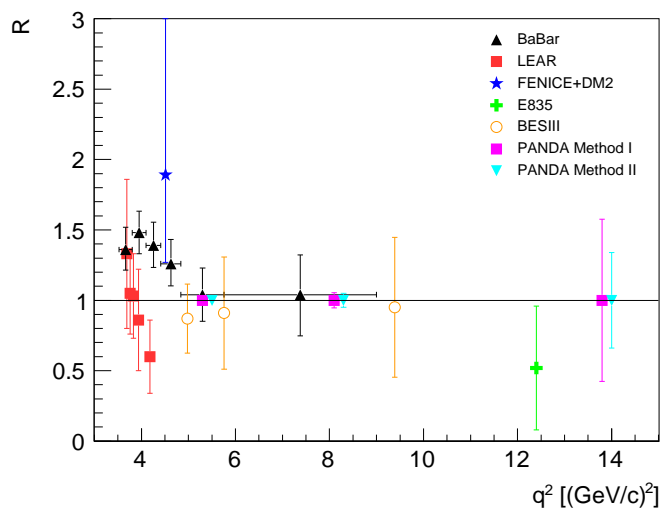}}
\caption{Expected statistical precision on the determination of the proton FF ratio from the present simulations (magenta squares and blue down triangles) for $\mbox{R}=1$ as a function of $q^2$, compared with the existing data. Data are from Ref.~\protect\cite{Bardin:1994am} (red squares), Ref.~\protect\cite{Lees:2013uta} (black up triangles), Ref.~\protect\cite{PhysRevD.91.112004} (open orange circles), and Ref.~\protect\cite{Baldini:2005xx} (green cross and blue star). The statistical uncertainties are extracted in the angular range $|\cos\theta|\leq 0.8$.}
\label{fig:ratiocomb}
\end{figure}

Feasibility studies for the measurement of the process $\bar p p \to e^+ e^-$ at \PANDA have been performed and reported in this work. Full simulations for the processes $\bar p p \to e^+ e^-$ have been carried out with the PandaRoot software. A total of 300 million events of the main background process $\bar p p \to \pi^+ \pi^-$ have been generated and reconstructed at three energy points. A background suppression factor of the order of $\sim10^8$ has been achieved, keeping a large and sufficient signal efficiency for the proton FF measurements at \PANDA. Two independent simulations have been performed for the signal using i) two different models in the event generator, ii) a different number of generated events iii) two sets of event selection criteria and iv) two fit functions to extract the proton electromagnetic FFs. The results from the two simulations, assuming $\mbox{R}=1$, are shown in Fig.~\ref{fig:ratiocomb}, together with the existing experimental data. For $q^2=5.4$ and 8.2~(GeV/$c$)$^2$, the results are consistent with each other.

The determination of the statistical uncertainties on R has been extended with Method II to different models of the proton FF ratio. A larger relative uncertainty has been obtained for $\mbox{R}<1$ than for the case $\mbox{R}=1$.

Compared to the previous analysis \cite{Sudol:2009vc}, the GEANT4 description of the detector is more realistic. Furthermore, the MC data are digitized and reconstructed using realistic pattern recognition and tracking algorithms, which was not the case in the old software framework. The total signal efficiency is improved by 5--10$\%$ in comparison with the previous studies \cite{Sudol:2009vc} thanks to the new selection procedure and new PID capabilities available in the PandaRoot framework. However, the experimental cuts have to be fine-tuned using measured data.

The \PANDA experiment at FAIR will extend the knowledge of the TL electromagnetic proton FFs in a large kinematic range. The present results show that the statistical uncertainty at $q^2 \geq 14$ (GeV/$c$)$^2$ will be comparable to the one obtained by BABAR at $\sim$ 7 (GeV/$c$)$^2$.

The study of the systematic uncertainties shows that the background misidentification and luminosity uncertainty dominate the total uncertainty at lower $q^2$, while in the high energy domain the total uncertainty is dominated by the statistical fluctuations due to the smaller $\bar{p}p\rightarrow e^+e^-$ cross section. The total relative uncertainty of individual FFs is expected to be in the range 2--48\% and 3--57\% for the ratio R.

The absolute cross section measurement depends essentially on the precision achieved in the luminosity measurement, which is expected to be around 4\%. The \PANDA experiment at FAIR will allow the individual determination of the proton FFs in the TL region, from 5.4 (GeV/$c$)$^2$ to 13.9 (GeV/$c$)$^2$. This is essential for a global analysis of the FFs in the SL and TL regions, to test the models which apply in the whole kinematic range and require analytical continuation of FFs.

\section{Acknowledgments}
\sloppy
The success of this work relies critically on the expertise and dedication of the computing organizations that support \PANDA. We acknowledge financial support from 
the Science and Technology Facilities Council (STFC), British funding agency, Great Britain;
the Bhabha Atomic Research Center (BARC) and the Indian Institute of Technology, Mumbai, India; 
the Bundesministerium f\"ur Bildung und Forschung (BMBF), Germany; 
the Carl-Zeiss-Stiftung 21-0563-2.8/122/1 and 21-0563-2.8/131/1, Mainz, Germany;
the Center for Advanced Radiation Technology (KVI-CART), Groningen, the Netherlands;
the CNRS/IN2P3 and the Universit\'e Paris-Sud, France;
the Deutsche Forschungsgemeinschaft (DFG), Germany;
the Deutscher Akademischer Austauschdienst (DAAD), Germany;
the Forschungszentrum J\"ulich GmbH, J\"ulich, Germany;
the FP7 HP3 GA283286, European Commission funding;
the Gesellschaft f\"ur Schwerionenforschung GmbH (GSI), Darmstadt, Germany;
the Helmholtz-Gemeinschaft Deutscher Forschungszentren (HGF), Germany;
the INTAS, European Commission funding;
the Institute of High Energy Physics (IHEP) and the Chinese Academy of Sciences, Beijing, China;
the Istituto Nazionale di Fisica Nucleare (INFN), Italy;
the Ministerio de Educaci\'{o}n y Ciencia (MEC) under grant FPA2006-12120-C03-02, Spain;
the Polish Ministry of Science and Higher Education (MNiSW) grant No. 2593/7, PR UE/2012/2, and the National Science Center (NCN) DEC-2013/09/N/ST2/02180, Poland;
the State Atomic Energy Corporation Rosatom, National Research Center Kurchatov Institute, Russia;
the Schweizerischer Nationalfonds zur Forderung der wissenschaftlichen Forschung (SNF), Switzerland;
the Stefan Meyer Institut f\"ur Subatomare Physik and the \"Osterreichische Akademie der Wissenschaften, Wien, Austria;
the Swedish Research Council, Sweden.
\bibliographystyle{unsrt}
\bibliography{Biblio.bib}

\begin{thebibliography}{10}

\bibitem{Lutz:2009ff}
W.~Erni et~al.
\newblock 2009.
\newblock arXiv:0903.3905 [hep-ex].

\bibitem{Wiedner:2011mf}
W.~Ulrich.
\newblock {\em Prog. Part. Nucl. Phys.}, 66:477--518, 2011.

\bibitem{Kuraev:2011vq}
E.~A. Kuraev, E.~Tomasi-Gustafsson, and A.~Dbeyssi.
\newblock {\em Phys. Lett. B}, 712:240--244, 2012.

\bibitem{Hofstadter:1956qs}
R~Hofstadter.
\newblock {\em Rev. Mod. Phys.}, 28:214--254, 1956.

\bibitem{Akhiezer:1968ek}
A.~I. Akhiezer and M.P. Rekalo.
\newblock {\em Sov. Phys. Dokl.}, 13:572, 1968.
\newblock [Dokl. Akad. Nauk Ser. Fiz. 180,1081(1968)].

\bibitem{Akhiezer:1974em}
A.~I. Akhiezer and M.P. Rekalo.
\newblock {\em Sov. J. Part. Nucl.}, 4:277, 1974.
\newblock [Fiz. Elem. Chast. Atom. Yadra4,662(1973)].

\bibitem{Perdrisat:2006hj}
C.~F. Perdrisat, V.~Punjabi, and M.~Vanderhaeghen.
\newblock {\em Prog. Part. Nucl. Phys.}, 59:694, 2007.

\bibitem{Jones:1999rz}
M.~K. Jones et~al.
\newblock {\em Phys. Rev. Lett.}, 84:1398, 2000.

\bibitem{Punjabi:2005wq}
V.~Punjabi et~al.
\newblock {\em Phys. Rev. C}, 71:055202, 2005.

\bibitem{Gayou:2001qd}
O.~Gayou et~al.
\newblock {\em Phys. Rev. Lett.}, 88:092301, 2002.

\bibitem{Puckett:2010ac}
A.~Puckett et~al.
\newblock {\em Phys. Rev. Lett.}, 104:242301, 2010.

\bibitem{Bardin:1994am}
G.~Bardin et~al.
\newblock {\em Nucl. Phys. B}, 411:3--32, 1994.

\bibitem{Lees:2013uta}
J.~P. Lees et~al.
\newblock {\em Phys. Rev. D}, 88(7):072009, 2013.

\bibitem{Ablikim:2015vga}
M.~Ablikim et~al.
\newblock {\em Phys. Rev. D}, 91(11):112004, 2015.

\bibitem{Kivel:2012zz}
N.~Kivel and M.~Vanderhaeghen.
\newblock {\em Prog. Part. Nucl. Phys.}, 67:491, 2012.

\bibitem{TomasiGustafsson:2008gq}
E.~Tomasi-Gustafsson and M.~P. Rekalo.
\newblock 2008.
\newblock arXiv:0810.4245 [hep-ph].

\bibitem{Sudol:2009vc}
M.~Sudol et~al.
\newblock {\em Eur. Phys. J. A}, 44:373--384, 2010.

\bibitem{SpataroPandaRoot}
S.~Spataro.
\newblock {\em Journal of Physics: Conference Series}, 396(2):022048, 2012.

\bibitem{tosca}
{\em TOSCA 9.0 Reference Manual.}, 2003.

\bibitem{Agostinelli:2002hh}
S.~Agostinelli et~al.
\newblock {\em Nucl. Instrum. Meth. A}, 506:250--303, 2003.

\bibitem{Zichichi:1962ni}
A.~Zichichi, S.~M. Berman, N.~Cabibbo, and R.~Gatto.
\newblock {\em Nuovo Cim.}, 24:170--180, 1962.

\bibitem{Eisenhandler:1975kx}
E.~Eisenhandler et~al.
\newblock {\em Nucl. Phys. B}, 96:109--154, 1975.

\bibitem{VandeWiele:2010kz}
J.~Van~de Wiele and S.~Ong.
\newblock {\em Eur. Phys. J. A}, 46:291--298, 2010.

\bibitem{MoraEspi12}
M.C.~Mora Espi.
\newblock {\em Feasibility studies for accessing nucleon structure observables
  with the PANDA experiment at the future FAIR facility}.
\newblock PhD thesis, Johannes Gutenberg-Universit\"at, 2012.

\bibitem{TomasiGustafsson:2001za}
E.~Tomasi-Gustafsson and M.~P. Rekalo.
\newblock {\em Phys. Lett. B}, 504:291--295, 2001.

\bibitem{Belushkin:2006qa}
M.~A. Belushkin, H.-W. Hammer, and U.-G. Meissner.
\newblock {\em Phys. Rev. C}, 75:035202, 2007.

\bibitem{Denig:2012by}
A.~68 and G.~Salm\`{e}.
\newblock {\em Prog. Part. Nucl. Phys.}, 68:113--157, 2013.

\bibitem{Pacetti:2015iqa}
S.~Pacetti, R.~Baldini~Ferroli, and E.~Tomasi-Gustafsson.
\newblock {\em Phys. Rept.}, 550-551:1--103, 2015.

\bibitem{Haidenbauer:2014kja}
J.~Haidenbauer, X.~W. Kang, and U.-G. Meissner.
\newblock {\em Nucl. Phys. A}, 929:102, 2014.

\bibitem{Lorenz:2015pba}
I.~T. Lorentz, H.-W. Hammer, and U.-G. Meissner.
\newblock {\em Phys. Rev. D}, 92:034018, 2015.

\bibitem{Bianconi:2015vva}
A.~Bianconi and E.~Tomasi-Gustafsson.
\newblock {\em Phys. Rev. C}, 93:035201, 2016.

\bibitem{Lees:2013rqd}
J.~P. Lees et~al.
\newblock {\em Phys. Rev. D}, 88:032011, 2013.

\bibitem{Lees:2013ebn}
J.~P. Lees et~al.
\newblock {\em Phys. Rev. D}, 87(9):092005, 2013.

\bibitem{Haidenbauer:2006dm}
J.~Haidenbauer, H.-W. Hammer, and U.-G. Meissner.
\newblock {\em Phys. Lett. B}, 643:29, 2006.

\bibitem{Andreotti:2003bt}
M.~Andreotti et~al.
\newblock {\em Phys. Lett. B}, 559:20--25, 2003.

\bibitem{Ambrogiani:1999bh}
M.~Ambrogiani et~al.
\newblock {\em Phys. Rev. D}, 60:032002, 1999.

\bibitem{Antonelli:1998fv}
A.~Antonelli et~al.
\newblock {\em Nucl. Phys. B}, 517:3--35, 1998.

\bibitem{Armstrong:1992wq}
T.~A. Armstrong et~al.
\newblock {\em Phys. Rev. Lett.}, 70:1212--1215, 1993.

\bibitem{Delcourt:1979ed}
B.~Delcourt et~al.
\newblock {\em Phys. Lett. B}, 86:395--398, 1979.

\bibitem{Bisello:1983at}
D.~Bisello et~al.
\newblock {\em Nucl. Phys. B}, 224:379, 1983.

\bibitem{Bisello:1990rf}
D.~Bisello et~al.
\newblock {\em Z. Phys.}, C48:23--28, 1990.

\bibitem{Ablikim:2005nn}
M.~Ablikim et~al.
\newblock {\em Phys. Lett. B}, 630:14--20, 2005.

\bibitem{Pedlar:2005sj}
T.~K. Pedlar et~al.
\newblock {\em Phys. Rev. Lett.}, 95:261803, 2005.

\bibitem{Shirkov:1997wi}
D.~V. Shirkov and I.~L. Solovtsov.
\newblock {\em Phys. Rev. Lett.}, 79:1209--1212, 1997.

\bibitem{Dbeyssi:2013}
A.~Dbeyssi.
\newblock {\em Study of the internal structure of the proton with the PANDA
  experiment at FAIR}.
\newblock PhD thesis, Universit\'e Paris-sud, 2013.

\bibitem{Zambrana-note:2014}
{Zambrana, M. and others}.
\newblock 2014.
\newblock Internal note.

\bibitem{Dover:1992vj}
C.~B. Dover, T.~Gutsche, M.~Maruyama, and A.~Faessler.
\newblock {\em Prog. Part. Nucl. Phys.}, 29:87--174, 1992.

\bibitem{Moussallam:1984zj}
B.~Moussallam.
\newblock {\em Nucl. Phys. A}, 429:429--444, 1984.

\bibitem{Eide:1973tb}
A.~Eide et~al.
\newblock {\em Nucl. Phys. B}, 60:173--220, 1973.

\bibitem{Buran:1976wc}
T.~Buran et~al.
\newblock {\em Nucl. Phys. B}, 116:51--64, 1976.

\bibitem{Stein:1977en}
N.~A. Stein et~al.
\newblock {\em Phys. Rev. Lett.}, 39:378--381, 1977.

\bibitem{White:1994tj}
C.~White et~al.
\newblock {\em Phys. Rev. D}, 49:58--78, 1994.

\bibitem{Armstrong:1986ng}
T.~A. Armstrong et~al.
\newblock {\em Nucl. Phys. B}, 284:643, 1987.

\bibitem{Ward:1980cw}
D.~R. Ward et~al.
\newblock {\em Nucl. Phys. B}, 172:302, 1980.

\bibitem{Chen:1977ik}
C.~K. Chen, T.~Fields, D.~S. Rhines, and J.~Whitmore.
\newblock {\em Phys. Rev. D}, 17:42, 1978.

\bibitem{Eastman:1973va}
P.~S. Eastman et~al.
\newblock {\em Nucl. Phys. B}, 51:29--56, 1973.

\bibitem{Mandelkern:1972ba}
M.~A. Mandelkern, R.~R. Burns, P.~E. Condon, and J.~Schultz.
\newblock {\em Phys. Rev. D}, 4:2658--2666, 1971.

\bibitem{Domingo:1967gra}
V.~Domingo, G.~P. Fisher, L.~Marshall~Libby, and R.~Sears.
\newblock {\em Phys. Lett. B}, 25:486--488, 1967.

\bibitem{Erni:2009pt}
W.~Erni et~al.
\newblock 2009.
\newblock arXiv:0907.0169 [physics.ins-det].

\bibitem{Erni:2012kva}
W.~Erni et~al.
\newblock 2012.
\newblock arXiv:1207.6581v2 [physics.ins-det].

\bibitem{Erni:2013ita}
W.~Erni et~al.
\newblock {\em Eur. Phys. J. A}, 49:25, 2013.

\bibitem{Merle:2014wra}
O.~Merle et~al.
\newblock {\em Nucl. Instrum. Meth. A}, 766:96--100, 2014.

\bibitem{Gruber2016104}
L.~Gruber et~al.
\newblock {\em Nuclear Instruments and Methods in Physics Research Section A:
  Accelerators, Spectrometers, Detectors and Associated Equipment}, 824:104 --
  105, 2016.
\newblock Frontier Detectors for Frontier Physics: Proceedings of the 13th Pisa
  Meeting on Advanced Detectors.

\bibitem{Erni:2008uqa}
W.~Erni et~al.
\newblock 2008.
\newblock arXiv:0810.1216v1 [physics.ins-det].

\bibitem{Kavatsyuk:2011zz}
M.~Kavatsyuk et~al.
\newblock {\em Nucl. Instrum. Meth. A}, 648:77--91, 2011.

\bibitem{Bohmer:2011zz}
F.~V. Bohmer et~al.
\newblock {\em Nucl. Phys. Proc. Suppl.}, 215:278--280, 2011.

\bibitem{MuonTDR}
The~PANDA Collaboration.
\newblock {Muon Detectors Technical Design Report}.
\newblock Unpublished.

\bibitem{Schwarz:2011zzc}
C.~Schwarz et~al.
\newblock {\em Nucl. Instrum. Meth. A}, 639:169--172, 2011.

\bibitem{fairroot}
M.~Al-Turany et~al.
\newblock {\em Journal of Physics: Conference Series}, 396(2):022001, 2012.

\bibitem{Brun:1997pa}
R.~Brun and F.~Rademakers.
\newblock {\em Nucl. Instrum. Meth. A}, 389:81--86, 1997.

\bibitem{Lange:2001uf}
D.~J. Lange.
\newblock {\em Nucl. Instrum. Meth. A}, 462:152--155, 2001.

\bibitem{PhysRevD.91.112004}
M.~Ablikim et~al.
\newblock {\em Phys. Rev. D}, 91:112004, Jun 2015.

\bibitem{Baldini:2005xx}
R.~Baldini et~al.
\newblock {\em Eur. Phys. J. C}, 46:421--428, 2006.

\bibitem{Was:2008zu}
Z.~Was, P.~Golonka, and G.~Nanava.
\newblock {\em Nucl. Phys. Proc. Suppl.}, 181-182:269--274, 2008.

\bibitem{TomasiGustafsson:2005kc}
E.~Tomasi-Gustafsson, F.~Lacroix, C.~Duterte, and G.~I. Gakh.
\newblock {\em Eur. Phys. J. A}, 24:419--430, 2005.

\bibitem{LumiTDR}
The~PANDA Collaboration.
\newblock {Luminosity Detector Technical Design Report}.
\newblock Unpublished.

\bibitem{Guttmann:2011ze}
J.~Guttmann, N.~Kivel, and M.~Vanderhaeghen.
\newblock {\em Phys. Rev. D}, 83:094021, 2011.

\bibitem{Gakh:2005wa}
G.~I. Gakh and E.~Tomasi-Gustafsson.
\newblock {\em Nucl. Phys. A}, 761:120--131, 2005.

\bibitem{ProposalHuang}
G.~Huang et~al.
\newblock {Precision measurement of R values, high mass charmonium states and
  QCD studies with BESIII at BEPCII}.
\newblock {BESIII Proposal}.

\bibitem{Abashian:2000cg}
A.~Abashian et~al.
\newblock {\em Nucl. Instrum. Meth. A}, 479:117--232, 2002.

\bibitem{Akai:2001pf}
K.~Akai et~al.
\newblock {\em Nucl. Instrum. Meth. A}, 499:191--227, 2003.

\bibitem{Ferber:2015}
T.~Ferber.
\newblock 2015.
\newblock Towards First Physics at Belle II, DPG 2015.

\end{thebibliography}

\end{document}